\pdfoutput=1
\documentclass[12pt,a4paper]{article}
\usepackage{ifthen} %
\newboolean{pdflatex}
\setboolean{pdflatex}{true} %

\newboolean{articletitles}
\setboolean{articletitles}{true} %

\newboolean{uprightparticles}
\setboolean{uprightparticles}{false} %

\def\paperauthors{LHCb collaboration} %
\def\paperasciititle{Measurement of charm production asymmetries at 13.6 TeV with the LHCb detector} %
\def\papertitle{Measurements of charmed meson and antimeson production asymmetries at $\sqrt{s}=$13.6\tev} %
\def\paperkeywords{{High Energy Physics}, {LHCb}} %
\def\papercopyright{\the\year\ CERN for the benefit of the LHCb collaboration} %
\def\paperlicence{CC BY 4.0 licence}
\def\paperlicenceurl{https://creativecommons.org/licenses/by/4.0/}

\newif\ifEnableSectionTOCLinks
\EnableSectionTOCLinksfalse %

\usepackage[top=1in, bottom=1.25in, left=1in, right=1in]{geometry}

\usepackage{microtype}
\usepackage{lineno}  %
\usepackage{xspace} %
\usepackage{caption} %

\usepackage{graphicx}  %
\usepackage{color}
\usepackage{colortbl}
\graphicspath{{./figs/}} %
\usepackage{amsmath} %
\usepackage{amssymb}
\usepackage{amsfonts}
\usepackage{upgreek} %

\newcommand*\patchAmsMathEnvironmentForLineno[1]{%
\expandafter\let\csname old#1\expandafter\endcsname\csname #1\endcsname
\expandafter\let\csname oldend#1\expandafter\endcsname\csname
end#1\endcsname
 \renewenvironment{#1}%
   {\linenomath\csname old#1\endcsname}%
   {\csname oldend#1\endcsname\endlinenomath}%
}
\newcommand*\patchBothAmsMathEnvironmentsForLineno[1]{%
  \patchAmsMathEnvironmentForLineno{#1}%
  \patchAmsMathEnvironmentForLineno{#1*}%
}
\AtBeginDocument{%
\patchBothAmsMathEnvironmentsForLineno{equation}%
\patchBothAmsMathEnvironmentsForLineno{align}%
\patchBothAmsMathEnvironmentsForLineno{flalign}%
\patchBothAmsMathEnvironmentsForLineno{alignat}%
\patchBothAmsMathEnvironmentsForLineno{gather}%
\patchBothAmsMathEnvironmentsForLineno{multline}%
\patchBothAmsMathEnvironmentsForLineno{eqnarray}%
}

\usepackage{hyperref}
\usepackage{hyperxmp}
\hypersetup{pdftex,
pdfauthor={\paperauthors},
pdftitle={\paperasciititle},
pdfkeywords={\paperkeywords},
pdfcopyright={Copyright (C) \papercopyright},
pdflicenseurl={\paperlicenceurl}}

\usepackage[colorinlistoftodos,textsize=scriptsize]{todonotes}

\usepackage[bottom,flushmargin,hang,multiple]{footmisc}

\usepackage[all]{hypcap} %

\usepackage{dcolumn}
\usepackage{booktabs}          %

\newcolumntype{d}[1]{D{.}{.}{#1}}
\newcommand\mc[1]{\multicolumn{1}{c}{#1}} %

\usepackage{xspace} 
\usepackage{upgreek}

\def\lhcb   {\mbox{LHCb}\xspace}

\def\velo   {VELO\xspace}

\def\scifi     {SciFi\xspace}

\def\MagUp {magnet up\xspace}
\def\MagDown {magnet down\xspace}

\def\hltone {HLT1\xspace}
\def\hlttwo {HLT2\xspace}

\ifthenelse{\boolean{uprightparticles}}%
{

 \def\Ppi         {\ensuremath{\uppi}\xspace}

 \def\Pphi        {\ensuremath{\upphi}\xspace}

 \def\PDelta      {\ensuremath{\Delta}\xspace}                 
 \def\PXi         {\ensuremath{\Xi}\xspace}                 
 \def\PLambda     {\ensuremath{\Lambda}\xspace}                 
 \def\PSigma      {\ensuremath{\Sigma}\xspace}                 
 \def\POmega      {\ensuremath{\Omega}\xspace}                 
 \def\PUpsilon    {\ensuremath{\Upsilon}\xspace}
 \let\oldPi\Pi
 \def\PPi         {\ensuremath{\oldPi}\xspace}

 \def\PB      {\ensuremath{\mathrm{B}}\xspace}                 
 \def\PD      {\ensuremath{\mathrm{D}}\xspace}                 

 \def\PK      {\ensuremath{\mathrm{K}}\xspace}                 
 \def\Pb      {\ensuremath{\mathrm{b}}\xspace}                 
 \def\Pc      {\ensuremath{\mathrm{c}}\xspace}

 \def\Pp      {\ensuremath{\mathrm{p}}\xspace}                 

 \def\Ps      {\ensuremath{\mathrm{s}}\xspace}

 \def\thebaroffset{0.0em}
}
{

 \def\Ppi         {\ensuremath{\pi}\xspace}

 \def\Pphi        {\ensuremath{\phi}\xspace}

 \mathchardef\PDelta="7101
 \mathchardef\PXi="7104
 \mathchardef\PLambda="7103
 \mathchardef\PSigma="7106
 \mathchardef\POmega="710A
 \mathchardef\PUpsilon="7107
 \mathchardef\PPi="7105
 \def\PB      {\ensuremath{B}\xspace}                 
 \def\PD      {\ensuremath{D}\xspace}                 

 \def\PK      {\ensuremath{K}\xspace}                 
 \def\Pb      {\ensuremath{b}\xspace}                 
 \def\Pc      {\ensuremath{c}\xspace}

 \def\Pp      {\ensuremath{p}\xspace}                 

 \def\Ps      {\ensuremath{s}\xspace}

 \def\thebaroffset{0.18em}
}
\newcommand{\offsetoverline}[2][\thebaroffset]{\kern #1\overline{\kern -#1 #2}}%

\makeatletter
\ifcase \@ptsize \relax%
  \newcommand{\miniscule}{\@setfontsize\miniscule{4}{5}}%
\or%
  \newcommand{\miniscule}{\@setfontsize\miniscule{5}{6}}%
\or%
  \newcommand{\miniscule}{\@setfontsize\miniscule{5}{6}}%
\fi
\makeatother

\DeclareRobustCommand{\optbar}[1]{\shortstack{{\miniscule (\rule[.5ex]{1.25em}{.18mm})}
  \\ [-.7ex] $#1$}}

\def\squark    {{\ensuremath{\Ps}}\xspace}

\def\cquark    {{\ensuremath{\Pc}}\xspace}

\def\bquark    {{\ensuremath{\Pb}}\xspace}

\def\pion   {{\ensuremath{\Ppi}}\xspace}

\def\pip    {{\ensuremath{\pion^+}}\xspace}
\def\pim    {{\ensuremath{\pion^-}}\xspace}
\def\pipm   {{\ensuremath{\pion^\pm}}\xspace}

\def\kaon    {{\ensuremath{\PK}}\xspace}

\def\KorKbar {\kern \thebaroffset\optbar{\kern -\thebaroffset \PK}{}\xspace}

\def\Kp      {{\ensuremath{\kaon^+}}\xspace}
\def\Km      {{\ensuremath{\kaon^-}}\xspace}

\def\KS      {{\ensuremath{\kaon^0_{\mathrm{S}}}}\xspace}

\def\Dbar    {{\ensuremath{\offsetoverline{\PD}}}\xspace}
\def\D       {{\ensuremath{\PD}}\xspace}

\def\DorDbar {\kern \thebaroffset\optbar{\kern -\thebaroffset \PD}\xspace}
\def\Dz      {{\ensuremath{\D^0}}\xspace}
\def\Dzb     {{\ensuremath{\Dbar{}^0}}\xspace}
\def\Dp      {{\ensuremath{\D^+}}\xspace}
\def\Dm      {{\ensuremath{\D^-}}\xspace}
\def\Dpm     {{\ensuremath{\D^\pm}}\xspace}

\def\DpDm    {\ensuremath{\Dp {\kern -0.16em \Dm}}\xspace}

\def\Dstarp  {{\ensuremath{\D^{*+}}}\xspace}
\def\Dstarm  {{\ensuremath{\D^{*-}}}\xspace}

\def\theDstarp{{\ensuremath{\D^{*}(2010)^{+}}}\xspace}

\def\Dsp     {{\ensuremath{\D^+_\squark}}\xspace}

\def\Dspm    {{\ensuremath{\D^{\pm}_\squark}}\xspace}

\def\DporDsp {{\ensuremath{\D_{(\squark)}^+}}\xspace}

\def\B       {{\ensuremath{\PB}}\xspace}

\def\BorBbar {\kern \thebaroffset\optbar{\kern -\thebaroffset \PB}\xspace}

\def\Bd      {{\ensuremath{\B^0}}\xspace}

\def\BdorBdbar {\kern \thebaroffset\optbar{\kern -\thebaroffset \Bd}\xspace}
\def\Bu      {{\ensuremath{\B^+}}\xspace}

\def\Bp      {{\ensuremath{\Bu}}\xspace}

\def\Bpm     {{\ensuremath{\B^\pm}}\xspace}

\def\Bs      {{\ensuremath{\B^0_\squark}}\xspace}

\def\BsorBsbar {\kern \thebaroffset\optbar{\kern -\thebaroffset \Bs}\xspace}

\def\Y#1S{\ensuremath{\PUpsilon{(#1S)}}\xspace}

\def\proton      {{\ensuremath{\Pp}}\xspace}

\def\Lz          {{\ensuremath{\PLambda}}\xspace}
\def\Lbar        {{\ensuremath{\offsetoverline{\PLambda}}}\xspace}
\def\LorLbar     {\kern \thebaroffset\optbar{\kern -\thebaroffset \PLambda}\xspace}

\def\Lb           {{\ensuremath{\Lz^0_\bquark}}\xspace}
\def\Lbbar        {{\ensuremath{\Lbar{}^0_\bquark}}\xspace}

\newcommand{\decay}[2]{\mbox{\ensuremath{#1\!\to #2}}\xspace} 

\def\to                 {\ensuremath{\rightarrow}\xspace}

\def\CP                {{\ensuremath{C\!P}}\xspace}

\newcommand{\ACP}{{\ensuremath{A_{\CP}}}\xspace}

\def\AT#1     {\ensuremath{A_{\mathrm{T}}^{#1}}\xspace}           %

\def\C#1      {\ensuremath{\mathcal{C}_{#1}}\xspace}                       %
\def\Cp#1     {\ensuremath{\mathcal{C}_{#1}^{'}}\xspace}                    %
\def\Ceff#1   {\ensuremath{\mathcal{C}_{#1}^{\mathrm{(eff)}}}\xspace}        %
\def\Cpeff#1  {\ensuremath{\mathcal{C}_{#1}^{'\mathrm{(eff)}}}\xspace}       %
\def\Ope#1    {\ensuremath{\mathcal{O}_{#1}}\xspace}                       %
\def\Opep#1   {\ensuremath{\mathcal{O}_{#1}^{'}}\xspace}                    %

\newcommand{\nospaceunit}[1]{\ensuremath{\text{#1}}}       
\newcommand{\aunit}[1]{\ensuremath{\text{\,#1}}}       
\newcommand{\tev}{\aunit{Te\kern -0.1em V}\xspace}
\newcommand{\gev}{\aunit{Ge\kern -0.1em V}\xspace}
\newcommand{\mev}{\aunit{Me\kern -0.1em V}\xspace}
\newcommand{\kev}{\aunit{ke\kern -0.1em V}\xspace}
\newcommand{\ev}{\aunit{e\kern -0.1em V}\xspace}
 
\newcommand{\mevc}{\ensuremath{\aunit{Me\kern -0.1em V\!/}c}\xspace}
\newcommand{\gevc}{\ensuremath{\aunit{Ge\kern -0.1em V\!/}c}\xspace}
\newcommand{\mevcc}{\ensuremath{\aunit{Me\kern -0.1em V\!/}c^2}\xspace}
\newcommand{\gevcc}{\ensuremath{\aunit{Ge\kern -0.1em V\!/}c^2}\xspace}

\def\cm   {\aunit{cm}\xspace}

\def\mm   {\aunit{mm}\xspace}

\def\mum  {\ensuremath{\,\upmu\nospaceunit{m}}\xspace}

\def\pb {\aunit{pb}\xspace}
\def\invpb {\ensuremath{\pb^{-1}}\xspace}
\def\fb   {\ensuremath{\aunit{fb}}\xspace}
\def\invfb   {\ensuremath{\fb^{-1}}\xspace}

\def\sec  {\ensuremath{\aunit{s}}\xspace}

\newcommand{\stat}{\aunit{(stat)}\xspace}
\newcommand{\syst}{\aunit{(syst)}\xspace}

\newcommand{\chisqip}{\ensuremath{\chi^2({\text{IP})}}\xspace}

\def\gsim{{~\raise.15em\hbox{$>$}\kern-.85em
          \lower.35em\hbox{$\sim$}~}\xspace}
\def\lsim{{~\raise.15em\hbox{$<$}\kern-.85em
          \lower.35em\hbox{$\sim$}~}\xspace}

\def\sPlot{\mbox{\em sPlot}\xspace}

\def\sqs   {\ensuremath{\protect\sqrt{s}}\xspace}

\def\pt         {\ensuremath{p_{\mathrm{T}}}\xspace}

\def\evtgen     {\mbox{\textsc{EvtGen}}\xspace}

\def\geant      {\mbox{\textsc{Geant4}}\xspace}

\def\herwig     {\mbox{{Herwig}}\xspace}

\def\photos     {\mbox{\textsc{Photos}}\xspace}

\def\pythia     {\mbox{\textsc{Pythia}}\xspace}

\def\roofit     {\mbox{\textsc{RooFit}}\xspace}

\def\tell1  {TELL1\xspace}
\def\ukl1   {UKL1\xspace}

\newcommand{\ie}{\mbox{\itshape i.e.}\xspace}

\newcommand{\cf}{\mbox{\itshape cf.}\xspace}

\newcommand{\lhcborcid}[1]{\href{https://orcid.org/#1}{\hspace*{0.1em}\raisebox{-0.45ex}{\includegraphics[width=1em]{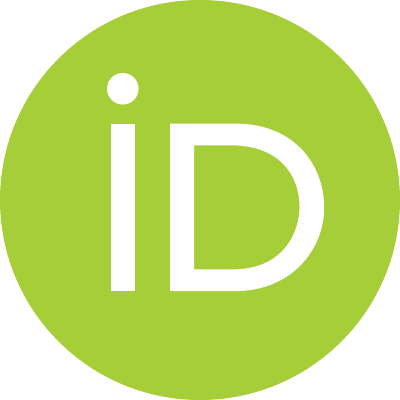}}}}

\newcommand{\Araw}{\ensuremath{A_\mathrm{raw}}\xspace}
\newcommand{\Adet}{\ensuremath{A_\mathrm{det}}}

\newcommand{\Arec}{\ensuremath{A_\mathrm{rec}}}
\newcommand{\Apid}{\ensuremath{A_\mathrm{PID}}}

\newcommand{\adet}{\Adet}
\newcommand{\arec}{\Arec}
\newcommand{\apid}{\Apid}

\newcommand{\Aprod}{\ensuremath{A_{\text{prod}}}}
\def\Ap    {\Aprod}
\def\AP    {\Ap}

\newcommand{\Asec}{\ensuremath{A_{\text{sec}}}}
\newcommand{\fsec}{\ensuremath{f_{\text{sec}}}}
\newcommand{\DsDp}{\DporDsp}

\newcommand{\Xc}{\ensuremath{X_{c}}\xspace}
\newcommand{\Xcbar}{\ensuremath{\offsetoverline{X}_{c}}\xspace}

\newcommand{\herwigseven}{\mbox{\herwig}~7}
\hypersetup{
  colorlinks   = true, %
  urlcolor     = blue, %
  linkcolor    = blue, %
  citecolor    = red   %
}

\ifEnableSectionTOCLinks
    \usepackage[explicit]{titlesec} %

    \let\oldcontentsline\contentsline
    \renewcommand

    \titleformat{\section}{\normalfont\Large\bf}{\hyperlink{tocsection.\thesection}{{\thesection} \parbox[t]{\dimexpr\textwidth-1pc}{#1}}}{1pc}{}

    \titleformat{\subsection}{\normalfont\bf}{\hyperlink{tocsubsection.\thesubsection}{{\thesubsection} \parbox[t]{\dimexpr\textwidth-1pc}{#1}}}{1pc}{}

    \titleformat{name=\section,numberless}[display]{}{}{0pt}{\normalfont\Huge\bfseries #1}
\fi

\usepackage{cite} %
\usepackage{makecell}
\usepackage{mciteplus}
 \usepackage{longtable} %

\begin{document}

\renewcommand{\thefootnote}{\fnsymbol{footnote}}
\setcounter{footnote}{1}

%
%
%
%
%
%
%
%
%
%
%
%

%
%
%
\begin{titlepage}
\pagenumbering{roman}

\vspace*{-1.5cm}
\centerline{\large EUROPEAN ORGANIZATION FOR NUCLEAR RESEARCH (CERN)}
\vspace*{1.5cm}
\noindent
\begin{tabular*}{\linewidth}{lc@{\extracolsep{\fill}}r@{\extracolsep{0pt}}}
\ifthenelse{\boolean{pdflatex}}%
{\vspace*{-1.5cm}\mbox{\!\!\!\includegraphics[width=.14\textwidth]{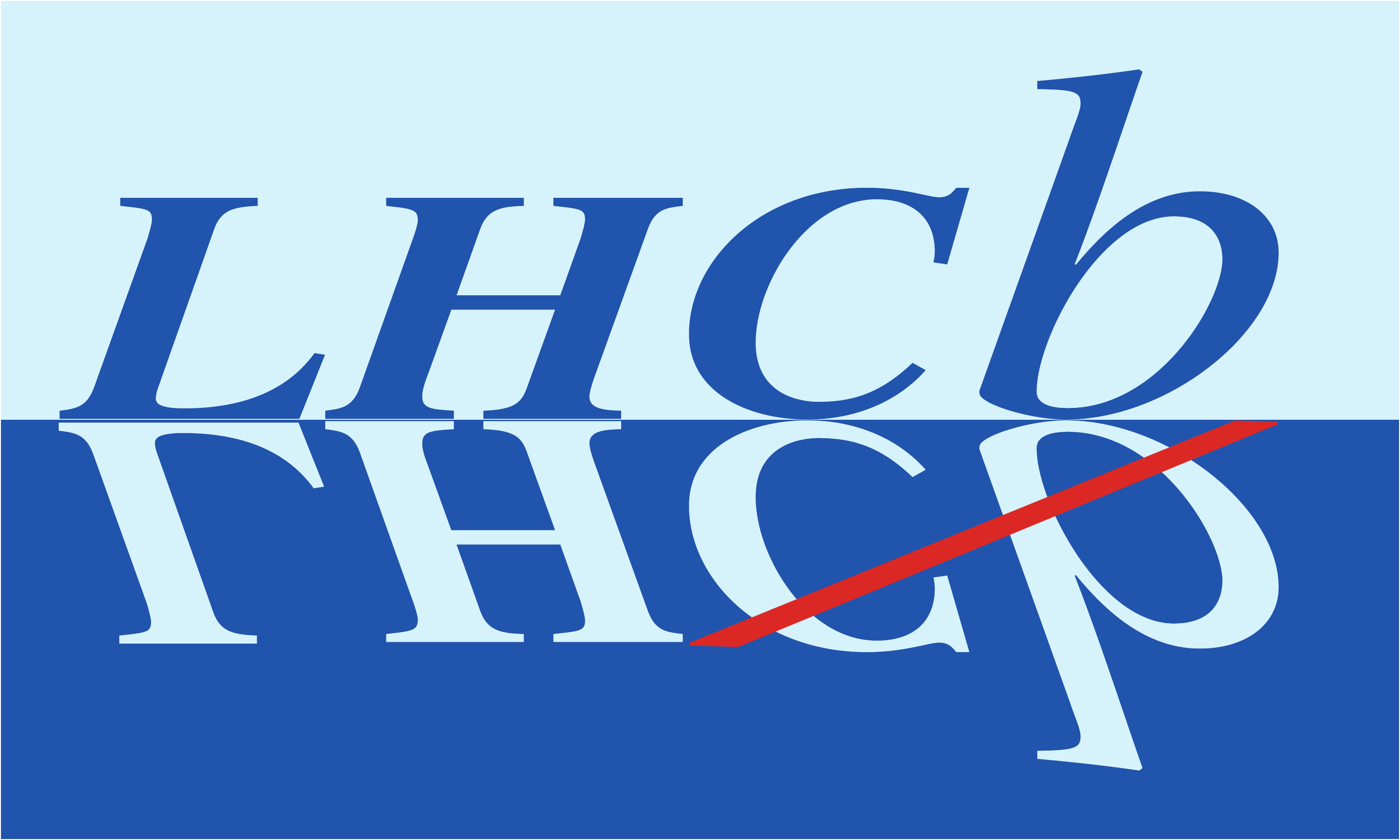}} & &}%
{\vspace*{-1.2cm}\mbox{\!\!\!\includegraphics[width=.12\textwidth]{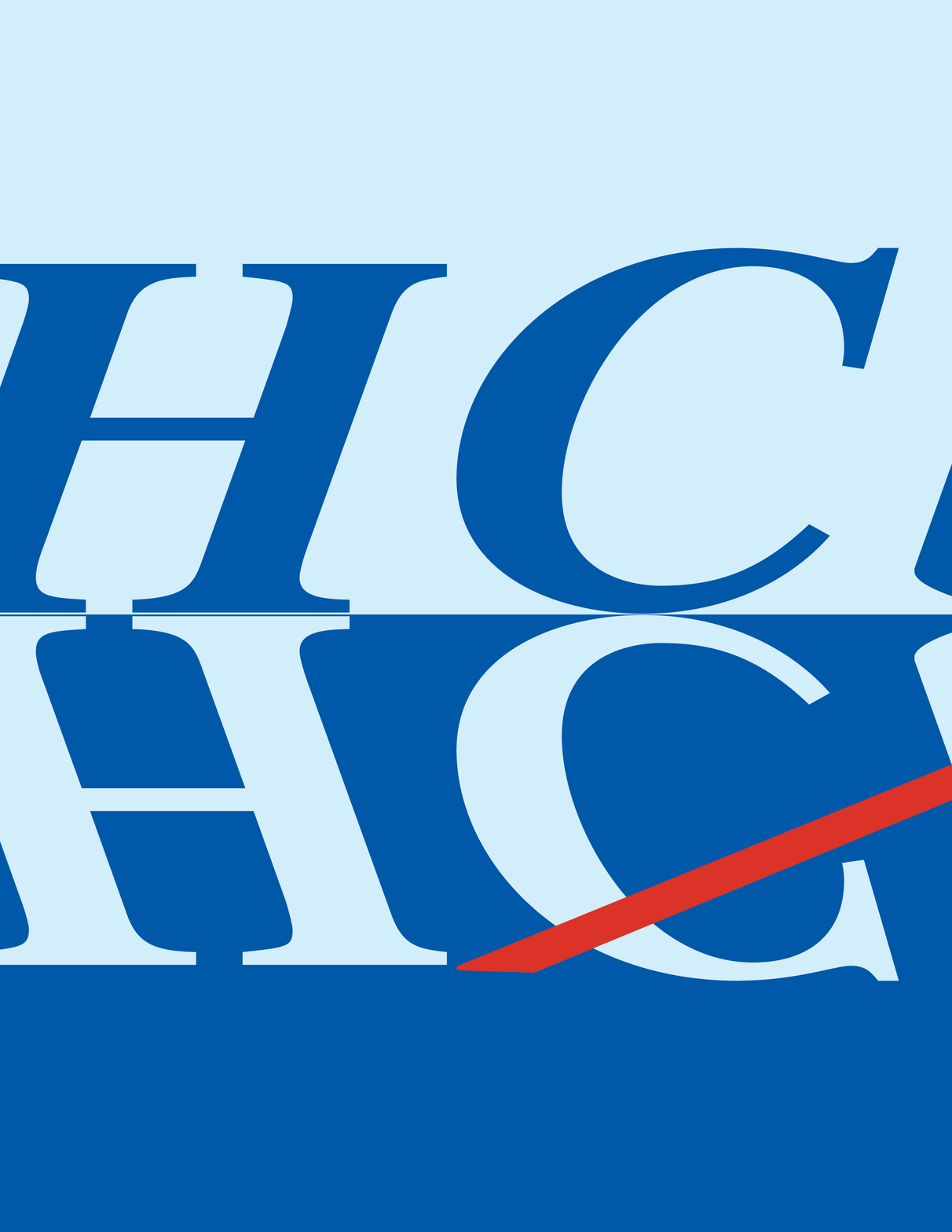}} & &}%
\\
 & & CERN-EP-2025-087 \\  %
 & & LHCb-PAPER-2024-052 \\  %
 & & October 10, 2025 \\ %
 & & \\
\end{tabular*}

\vspace*{4.0cm}

{\normalfont\bfseries\boldmath\huge
\begin{center}
  \papertitle 
\end{center}
}

\vspace*{2.0cm}

\begin{center}
\paperauthors\footnote{Authors are listed at the end of this paper.}
\end{center}

\vspace{\fill}

\begin{abstract}
  \noindent
  This article presents doubly differential measurements of the asymmetries in production rates between mesons containing a charm quark and those containing an anti\-charm quark in proton-proton collisions at a centre-of-mass energy of $\sqs=13.6\tev$ using data recorded by the \lhcb experiment. 
  The asymmetries of \Dz, \Dp and \Dsp mesons are measured for two-dimensional intervals in transverse momentum and pseudorapidity, within the range $2.5<\pt<25.0\gevc$ and $2.0<\eta<4.5$. 
  No significant production asymmetries are observed. 
  Comparisons to the \pythia~8 and \herwigseven~event generators are also presented, and their agreement with the data is evaluated.
  These measurements constitute the first measurements of production asymmetries at this centre-of-mass energy of colliding beams, and the first measurements with the \lhcb Run 3 detector.
\end{abstract}

\vspace*{2.0cm}

\begin{center}
  Published in JHEP 10 (2025) 050

\end{center}

\vspace{\fill}

{\footnotesize 
\centerline{\copyright~\papercopyright. \href{\paperlicenceurl}{\paperlicence}.}}
\vspace*{2mm}

\end{titlepage}

\newpage
\setcounter{page}{2}
\mbox{~}
%
%
%
%
%
%
%
%
%
%
%
%
%
%
%
%
%
%
%
%

 %
%

\renewcommand{\thefootnote}{\arabic{footnote}}
\setcounter{footnote}{0}

\cleardoublepage

\pagestyle{plain} %
\setcounter{page}{1}
\pagenumbering{arabic}

%
%
%

%
%
%
\section{Introduction}
\label{sec:Introduction}
The initial production of charm-anticharm pairs in proton-proton ($pp$) collisions at the LHC is governed primarily by quantum chromodynamics (QCD). 
To the leading order, heavy quarks and antiquarks are produced in equal amounts~\cite{Norrbin:2000zc,Gauld2019}. 
However, this symmetry is altered during hadronisation due to the influence of the valence quarks in the colliding protons, resulting in asymmetries between particle and antiparticle production.
These asymmetries are expected to be more pronounced near the beam axis, making them particularly relevant for the \lhcb experiment. 
Designed to study the decays of charmed and bottom hadrons, \lhcb is instrumented in the forward region, and can observe such effects with high precision. 
The production asymmetry is defined as
\begin{equation}
    \Aprod(\Xc) \equiv \frac{\sigma(\Xc) - \sigma(\Xcbar)}{\sigma(\Xc) + \sigma(\Xcbar)},
\end{equation}
where $\sigma(\Xc)$ denotes the inclusive production cross-section of a charmed meson $\Xc$ (specifically $\Xc = \Dz, \Dp, \Dsp$) in $pp$ collisions. This includes mesons produced either directly or via decays of excited charm resonances. Charmed mesons produced through either mechanism are collectively referred to as prompt.
The corresponding cross-section for the charge-conjugate meson, $\Xcbar$, is denoted by $\sigma(\Xcbar)$. 

Various theoretical frameworks have explored and discussed charmed meson-antimeson production asymmetries, notably in the context of the Lund string model~\cite{Norrbin:2000zc}, cluster-hadronisation model~\cite{Bellm:2015jjp}, meson-cloud model~\cite{Cazaroto:2013wy} and heavy-quark recombination~\cite{Das:1977cp,Lai:2014xya}.
These studies show that measuring production asymmetries is crucial for enhancing our understanding of production processes.
Their relevance in estimating the atmospheric neutrino flux has been highlighted~\cite{Maciula:2017wov,Goncalves:2018zzf}, along with their relation to the intrinsic charm content of the proton~\cite{Sufian:2020coz,Maciula:2020dxv}. 
In addition, precise knowledge of the production asymmetries of charmed mesons in $pp$ collisions is an important input to precision measurements of \CP-violating observables at the \lhcb experiment, as production asymmetries contribute to background asymmetries in such measurements~\cite{LHCb-PAPER-2022-024}.

Evidence for production asymmetries at the subpercent level has been reported by the LHCb collaboration for $\DporDsp$ mesons, using $pp$ collision data at the LHC at centre-of-mass energies of $\sqs=7$ and $8\tev$~\cite{LHCb-PAPER-2012-026,LHCb-PAPER-2018-010}.
The measured $\Dsp$ production asymmetry revealed some discrepancies~\cite{LHCb-PAPER-2018-010} with respect to \pythia8. 
Production asymmetries are expected to be sensitive to the implementation of the hadronisation models within the event generators\cite{LHCb-PAPER-2021-016}. 
Following recent efforts to improve such models~\cite{Fieg:2023kld}, further comparisons with new data are desirable. 
Although production asymmetries are anticipated to diminish at larger centre-of-mass energies for fixed rapidity~\cite{Cazaroto:2013wy}, experimental evidence for this dependence has not yet been reported, with only measurements at $\sqrt{s} = 7$ and $8\tev$ currently available.

This paper presents the first measurements of production asymmetries for $\Dz$, $\Dp$, and $\Dsp$ mesons in $\sqs=13.6\tev$ $pp$ collisions. These asymmetries are measured using \mbox{$\Dz\to\Km\pip$}, \mbox{$\Dp\to\phi\pip$}, and \mbox{$\Dsp\to\phi\pip$} decays, as a function of both the
component of the momentum transverse to the beam ($\pt$) and 
the pseudorapidity ($\eta$) of the charmed mesons, in two-dimensional intervals spanned by
\begin{align}
  \label{eq:binning}
  \pt \; [\!\gevc] : &\;[2.5 , 4.7]\, ;\, [4.7 , 6.5]\,;\, [6.5 , 8.5]\,;\, [8.5 , 25.0] \, . \nonumber \\
  \eta: &\; [2.0 , 3.0]\,;\, [3.0 , 3.5]\,;\, [3.5 , 4.0] \,;\, [4.0, 4.5] \, .
\end{align}

The analysis is performed using the first $pp$ collision data collected by the upgraded \lhcb detector~\cite{LHCb-DP-2022-002} during Run 3 of the LHC.
The achieved experimental precision is comparable to existing measurements at lower centre-of-mass energies, despite a significantly lower integrated luminosity. 

\section{\lhcb Run 3 detector}
\label{sec:Detector}
The \lhcb Run 3 detector~\cite{LHCb-DP-2022-002,LHCb-DP-2008-001,LHCb-DP-2014-002} is a single-arm forward spectrometer covering the pseudo\-rapidity range $2<\eta <5$, designed for the study of particles containing \bquark or \cquark quarks. 
This detector has been substantially upgraded for the Run 3 data-taking period, which started in 2022. 
The upgraded detector is designed to operate at an instantaneous luminosity of $2 \times 10^{33} \cm^{-2} \sec^{-1}$, approximately five times larger with respect to the Run~1--2 detector.  

The tracking system has been fully replaced and consists of a silicon-pixel vertex detector (\velo) surrounding the $pp$ interaction region~\cite{LHCb-TDR-013}, a large-area silicon-strip detector (UT) located upstream of a dipole magnet with a bending power of about $4{\mathrm{\,T\,m}}$, and three stations of scintillating fibre detectors (\scifi) placed downstream of the magnet~\cite{LHCb-TDR-015}.
Different types of charged hadrons are distinguished using information from two ring-imaging Cherenkov (RICH) detectors~\cite{LHCb-DP-2012-003,LHCb-TDR-014}. 
The whole photon detection system of the Cherenkov detectors has been renewed for the upgraded detector. Photons, electrons and hadrons are identified by a calorimeter system consisting of an electromagnetic
and a hadronic calorimeter~\cite{LHCb-DP-2020-001}. 
Muons are identified by a
system composed of alternating layers of iron and multiwire
proportional chambers~\cite{LHCb-DP-2012-002}.

Readout of all detectors into an all-software trigger~\cite{LHCb-TDR-016, LHCb-TDR-021}
is a central feature of the upgraded detector, enabling the reconstruction of events at the LHC bunch-crossing rate of approximately 30 MHz, and their selection in real time. 
The trigger system is implemented in two stages: a first inclusive stage (\hltone) running on GPUs, based primarily on charged-particle reconstruction, which reduces the data volume by roughly a factor of 20, and a second stage (\hlttwo) running on CPUs, which performs the full offline-quality reconstruction and a selection of physics signatures. 

The implementation of this all-software trigger circumvents the limiting systematic uncertainty in previous measurements of production asymmetries associated with the hardware trigger~\cite{LHCb-PAPER-2018-010}, and significantly improves the selection efficiency for hadronic decays.

In preparation for analysis, physics data undergoes a centralized, offline processing step known as {\tt Sprucing}. Subsequently, highly automated LHCb Analysis Productions deliver physics-analysis-ready data across the entire LHCb physics program~\cite{sprucing, FunTuple}. Analysis specific computational tasks were managed with the tool {\tt Ganga}~\cite{ganga}.

In the simulation, $pp$ collisions are generated using \pythia~\cite{Sjostrand:2007gs,Sjostrand:2006za} with a specific \lhcb configuration~\cite{LHCb-PROC-2010-056}. 
Decays of unstable particles are described by \evtgen~\cite{Lange:2001uf}, in which final-state radiation is generated using \photos~\cite{davidson2015photos}. 
The interaction of the generated particles with the detector, and its response, are implemented using the \geant toolkit~\cite{Allison:2006ve,Agostinelli:2002hh} as described in Ref.~\cite{LHCb-PROC-2011-006}.

\section{Datasets recorded in 2022 and 2023}
The datasets, recorded in 2022 and 2023, are the first from the upgraded experiment. 
During this period, the detector was still in its commissioning phase, with its performance and stability improving continuously.

In 2022, data used in this analysis were recorded in two fills of the LHC. The two fills have different configurations for the dipole magnet's polarity, referred to as \MagUp and \MagDown, swapping the bending direction of the charged particles. 
In these datasets, each bunch crossing had an average of either two or three visible $pp$ interactions. 
The VELO modules, which are moved towards the beam line at the beginning of a fill, operated with an extra clearance with respect to the beam line of $1.5\mm$, compared to the design value, for additional safety. 
The UT detector was not yet installed. 
Instabilities in the front-end electronics led to a data acquisition efficiency of about 70\%. 
At the \hltone stage, data were processed at an input rate of at most 17 MHz, reconstructing tracks with information from the \velo and SciFi detectors, and selecting events based on either the activity in the calorimeters, or using track-based selections. 
Small movements of the VELO detector from the design position along the $x$-direction\footnote{The LHCb coordinate system is a right-handed system centred on the nominal $pp$ collision point, with the $z$-axis pointing along the beam direction towards the detectors downstream, the $y$-axis pointing vertically upwards, and the $x$-axis pointing in the horizontal direction.} were observed, starting with $1 \mum/\mathrm{min}$ after closure and reducing to less than $0.1 \mum/\mathrm{min}$ after 100 minutes.
These movements were corrected for by frequently updating the detector alignment used in the \hltone reconstruction according to the anticipated movement over time.
Due to the limited integrated luminosity, all data after \hltone could be stored, including the raw detector information. This allowed for a delayed processing of the data by \hlttwo, which included improvements in the detector alignment.
This alignment included a fine-grained correction for the movements of the \velo detector and an improved alignment of the \scifi detector.

In 2023, data were recorded with only one magnet-polarity configuration (\MagDown).
After an incident in the LHC-\velo vacuum safety system which deformed the boundary between the detector and beam volumes, the \velo was retracted horizontally (\ie along the $x$-direction) from its nominal position by initially $\pm31.0\mm$ and later $\pm25.5\mm$~\cite{RodriguezRodriguez:2024beh}.
This affected the detector's angular acceptance and impact-parameter (IP) resolution, where the IP is the minimum distance between a particle trajectory and a $pp$ collision vertex (PV).  
To avoid unnecessary radiation damage to the detectors and partially mitigate residual instabilities in the front-end electronics, the experiment operated for the majority of the time at an instantaneous luminosity corresponding to an average of 1.1 visible $pp$ interactions per bunch crossing. 
Moreover, \hltone operated with relaxed track reconstruction and selection requirements (with respect to its operation at the design instantaneous luminosity), at an input rate of the order of 20 MHz. 
The UT was installed, but not included in the regular data taking and therefore tracks were reconstructed using only the information from the \velo and \scifi detectors, as in 2022.
The spatial alignment of all detectors was periodically updated at the time of data taking. %

The $\Dz$ meson production asymmetry is measured using datasets of 2022 and 2023 corresponding to an integrated luminosity of $15$ and $162\invpb$, respectively, while for the $\Dp$ and $\Dsp$ meson production asymmetries, the data samples correspond to $15$ and $41\invpb$, respectively. 
The difference in integrated luminosity used for the charged and neutral mesons is due to the trigger configuration of part of the 2023 data, which did not include the relevant selections for the measurements with $\Dsp$ and $\Dp$ mesons.
The different operating conditions of the LHCb detector necessitate that data recorded under different configurations are analysed separately, with corrections applied accordingly. 
Due to the impact of the open VELO configuration on the angular acceptance, all measurements at $\eta > 4$ are determined using only the data from 2022. 

\section{Methodology}
\label{sec:method}
The production asymmetries are determined from the yields of $\Xc$ and $\Xcbar$ mesons, reconstructed through their decay to a final state $f$ and $\overline{f}$, respectively. The raw asymmetry is defined as
\begin{equation}
	\Araw (\Xc \to f) \equiv \frac{N(\Xc \to f) - N(\Xcbar \to \overline{f})}{N(\Xc \to f) + N(\Xcbar \to \overline{f})},
\end{equation}
in which $N$ denotes the number of decays observed in the dataset. The raw asymmetry contains contributions from detector-induced asymmetries, \Adet, and potential asymmetries due to charge parity ($\CP$) symmetry violation in the considered decays, \ACP, which are corrected as discussed later.

The reconstructed charmed mesons are primarily produced directly from \proton\proton collisions (prompt production), while a fraction, $\fsec$, originates from \bquark-hadron decays (secondary production), with an asymmetry $A_\text{sec}(\Xc)$. 
Given that both \CP and production asymmetries are of order 1\%, and detection asymmetries are well below 10\%, the following equation 
\begin{align}\label{eq:Araw_sec}
    \Araw (\Xc \to f) &=  (1-\fsec) \Aprod(\Xc) + \Adet
    (f) \nonumber \\ &\phantom{=} + \ACP(\Xc \to f) + \fsec A_\text{sec}(\Xc),
\end{align}
is valid up to corrections of order $10^{-4}$, which is at least one order of magnitude below the statistical sensitivity of the presented measurement. 
The different contributions in Eq.~\ref{eq:Araw_sec} are determined from data, with minimal input from simulation.

Detector-induced asymmetries depend on the final-state particles of the charmed-meson decay and their kinematics. 
Accordingly, their determination for the measurement of $\Ap(\Dz)$ follows a different approach with respect to that of the $\Dp$ and $\Dsp$ mesons. 

The production asymmetries $\Aprod(\Dsp)$ and $\Aprod(\Dp)$ are measured using the decays \mbox{$\Dsp\to\phi\pip$} and \mbox{$\Dp\to\phi\pip$}, respectively, with \mbox{$\Pphi\to\Km\Kp$}.
Due to the final states being symmetric in the production of the two charged kaons, the relevant detector-induced asymmetry arises from the different reconstruction efficiencies $\varepsilon(\pipm)$ for positively and negatively charged pions, defined as
\begin{align}
    \label{eq:Adet_pi}
    \adet(\pip) \equiv \frac{\varepsilon(\pip) - \varepsilon(\pim)}{\varepsilon(\pip) + \varepsilon(\pim)}.
\end{align}
Residual contributions from any asymmetric kaon-pair production are considered as systematic uncertainties. 

The asymmetry $\adet(\pip)$ consists of two contributions, $\arec(\pip)$ and $\apid(\pip)$, where the former corresponds to the asymmetry in reconstructing the pion track, and the latter corresponds to the asymmetry related to the pion particle-identification (PID) efficiency.
Each contribution is determined using a specific calibration sample. 
The determination of $\arec(\pip)$ is performed using a large sample of \mbox{$\KS\to\pip\pim$} decays, following the tag-and-probe method introduced in Refs.~\cite{LHCb-DP-2019-003,LHCb-PAPER-2021-016,LaurentThesis}. 
The method relies on the assumption that, due to the negligible magnetic field, VELO tracks are unaffected by instrumental charge asymmetries -- an assumption validated within experimental precision~\cite{LHCb-PAPER-2021-016,LaurentThesis}.
Values for $\apid(\pip)$ are determined using samples of $\Dz\to\Km\pip$ decays,\footnote{The inclusion of charge-conjugated processes is implied throughout, unless stated otherwise.} where the $\Dz$ mesons are produced in $\theDstarp\to\Dz\pip$ decays (henceforth denoted as
$\Dstarp$), and are selected without PID requirements~\cite{LHCb-PUB-2016-021}.

Corrections for possible $\CP$ asymmetries in the decay $\Dp\to\phi\pip$ are applied using the experimental result \mbox{$\ACP(\Dp\to\phi\pip) = (0.3 \pm 4.0 \stat \pm 2.9 \syst)\times10^{-4}$}~\cite{LHCb-PAPER-2019-002}. For the $\Dsp\to\phi\pip$ and $\Dz\to\Km\pip$ decays, which primarily occur via Cabibbo-favoured tree-level amplitudes, the contribution of $\CP$ violation is assumed to be negligible. 

The production asymmetry of \Dz mesons is measured with \mbox{$\Dz\to\Km\pip$} decays. The flavour of the $\Dz$ meson at production can be inferred from the charge of the decay products up to negligible corrections due to neutral-meson mixing and doubly Cabibbo-suppressed decays~\cite{HFLAV24}. 
Here, the measured raw asymmetry receives contributions from detector-induced asymmetry for the combined $\Km \pip$ meson pair, $\adet(\Km\pip)$.
This detection asymmetry is determined using the combination of decays of $\Dz$ mesons produced from $\Dstarp$ meson decays
\begin{align}\label{eq:adetkpi}
    \adet(\Km\pip) &= \Araw( \Dstarp \to \Dz(\to\Km\pip)\pip ) \nonumber \\
     &\phantom{=} \, - \Araw( \Dstarp \to \Dz(\to\Km\Kp)\pip ) \\
     &\phantom{=} \, + \ACP(\Dz\to\Km\Kp), \nonumber
\end{align}
using the charge-symmetric production of the kaons in $\Dz\to\Km\Kp$ decays, and the measured value of $\ACP(\Dz\to\Km\Kp) = (7.7 \pm 5.7)\times 10^{-4}$~\cite{LHCb-PAPER-2022-024}. The $\Dz$ mesons produced from $\Dstarp$ decays are subsequently referred to as tagged $\Dz$ candidates, and represent a subset of the untagged $\Dz$ candidates used as the signal. 

\section{Data selection}
\label{sec:selection}

\subsection{Selection of charged charmed mesons}
\label{sec:selection_dsp_dp}
The decay candidates are selected by the \hltone trigger based on the presence of either one track displaced from all PVs with a high \pt, or a displaced two-track vertex. 
Potential biases are removed by only considering candidates that are triggered by the symmetric kaon pair at this stage.
At the \hlttwo stage, $\DporDsp$ candidates are selected by combining three tracks originating from a common vertex, two of which have opposite charge.
The sum of the tracks' $\pt$ is required to be above $3\gevc$. 
Each of the tracks must be incompatible with originating directly from a PV, which is ensured through requirements on the square of the significance of the IP, $\chisqip$.
The charmed mesons are associated to the PV that has the smallest IP.
The kaon and pion candidates are required to be identified as such using the information from the RICH detectors. 

Offline, candidates are only considered if the invariant mass of the kaon pair, $m(\Kp\Km)$, is in a window of approximately $15\mevcc$ around the known $\Pphi$-meson mass~\cite{PDG2024}. 
Candidates with pion momentum below $7\gevc$ exhibit a very large instrumental asymmetry and are therefore excluded. 
For the data recorded in 2022 (2023), the pion candidate is required to have a maximum $\pt$ of $5\gevc$  ($7\gevc$) to ensure a good overlap with the calibration sample.

The charmed-meson candidates must be compatible with originating directly from their associated PV through a requirement on $\chisqip$. 
The requirement is stricter for the $\Dsp$-meson selection in 2023 data than in 2022 data, due to the worsened IP resolution and the expected larger background contamination. 
Despite this selection, contributions from $\bquark$-hadron decays are expected to remain, which are estimated using those charmed-meson candidates that fail the $\chisqip$ requirement in addition. 
Examples of the $\chisqip$ distributions in data, along with the impact of the selection requirements in different years of data taking, are presented in  Sec.~\ref{sec:secondaries}, in which estimates of the residual secondary component are also presented. 

Due to an imperfect detector alignment, charge-dependent inaccuracies in the determination of a particle's momentum can occur, also known as curvature biases. 
These biases are mostly present for the data recorded in 2023, for which the global alignment procedure was complicated due to the open \velo configuration and the exclusion of the UT, and are mitigated by an offline calibration. No such calibration is needed for the 2022 data, which benefited from an improved alignment in the delayed processing, reducing the curvature biases to a negligible level. 
The offline calibration uses $\Dp\to\Km\pip\pip$ decays to determine the momentum-scale factors for which the invariant-mass distributions of the $\Dp$ and $\Dm$ mesons agree. 
This last calibration primarily impacts the measurements at low charmed-meson $\pt$, where the asymmetries change by about $0.2\%$. 

\begin{figure}[]
\centering
\includegraphics[width=0.49\textwidth]{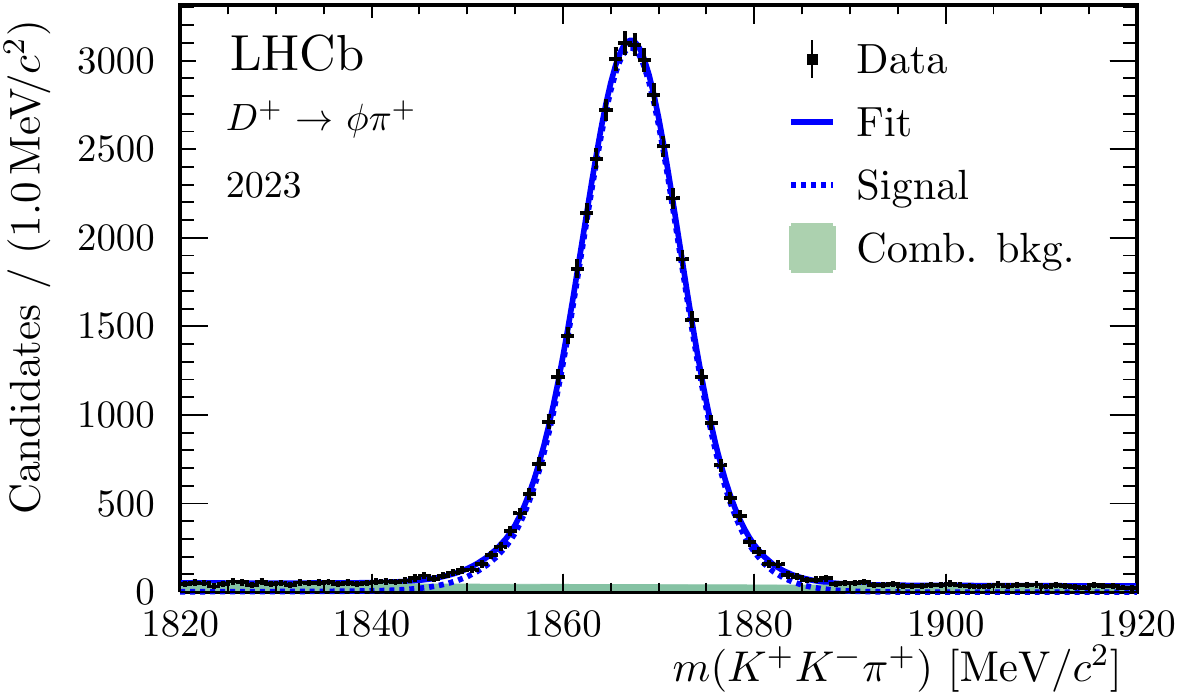}
\includegraphics[width=0.49\textwidth]{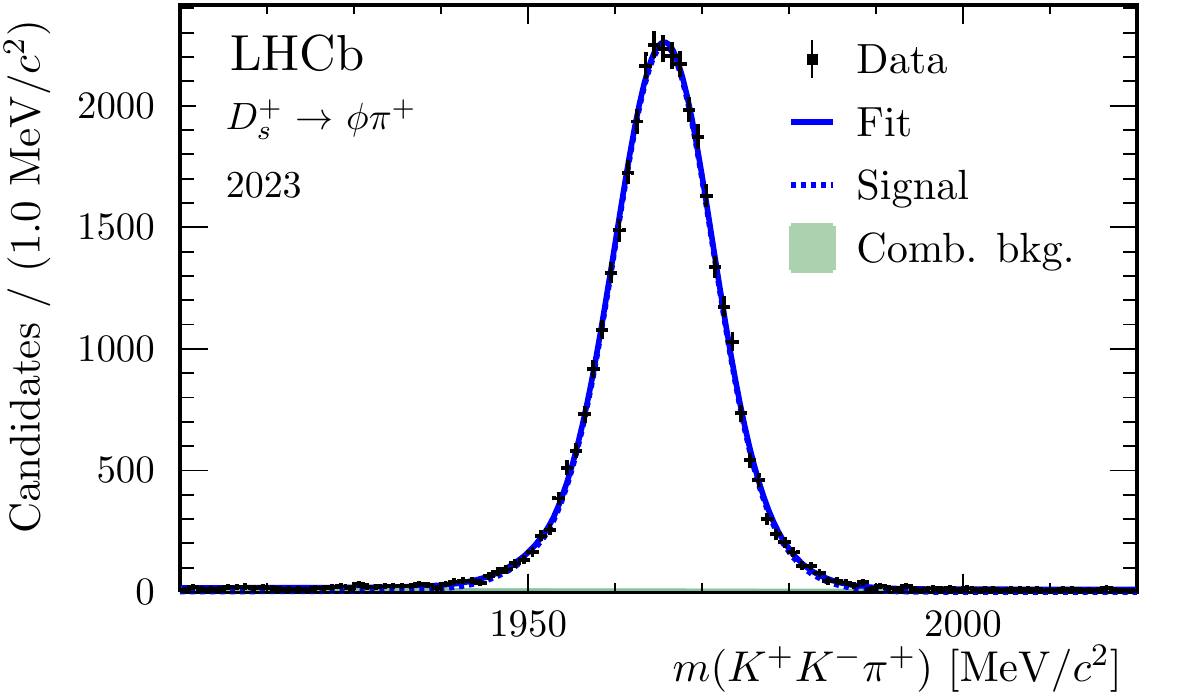} 

\caption{Invariant-mass distributions of $\Kp\Km\pip$ combinations for positively charged (left) $\Dp$ and (right) $\Dsp$ candidates in the 2023 data. Fits to the data points are also shown. 
}\label{fig:selection_dsp_dp_mass_peaks_no_cuts}
\end{figure}

Approximately $107 \times 10^{3}$ and $72 \times 10^{3}$ $\Dsp$ decays are selected in 2022 and 2023, respectively. The signal yields after selection for $\Dp$ decays are approximately $61 \times 10^{3}$ in 2022 and $94 \times 10^{3}$ in 2023. The invariant-mass distribution of the $\Kp\Km\pip$ combination, \mbox{$m(\Kp\Km\pip)$}, which also illustrates the purity of these data, is presented in Fig.~\ref{fig:selection_dsp_dp_mass_peaks_no_cuts} for the 2023 data. 
The distributions for the other datasets are presented in Appendix~\ref{appendix:mass_plots}.

\subsection{Selection of the {\boldmath$\KS\to\pip\pim$} calibration channel}
Decays of the type $\KS\to\pip\pim$, used to determine $\arec(\pip)$, are selected through a set of dedicated \hlttwo trigger selections. 
These selections consider $\pip \pim$ meson pairs, where one pion, the probe, is reconstructed using only information from the \velo, while the other is reconstructed using information from all the tracking detectors.
Since the $\KS$ mesons are produced in abundance at the LHC, especially at low $\pt$, not all events containing $\KS$ mesons can be retained. 
For this purpose, the \hlttwo selections suppress these candidates through random filters, using four intervals in the inferred $\pt$ of the probe pion and favouring those at higher $\pt$ to aid the kinematic overlap with that of signal pions. 
The \hltone selection is independent of the probe pion to avoid possible biases in determining the reconstruction asymmetries.
Exploiting the long lifetime of the $\KS$ meson, the candidates are selected by requiring a significant displacement of the decay vertex with respect to any PV. 

The $\KS$-meson candidates are required to be compatible with originating directly from their associated PV. 
Using this geometric constraint, the $\pip\pim$ invariant mass can be inferred from a kinematic fit and used subsequently to separate signal and background. 
An additional constraint to the known \KS mass~\cite{PDG2024} allows for a more accurate determination of the momentum of the probe pion, used to parametrise $\arec(\pip)$. 

Candidates with an invariant mass in the range 400--600\mevcc, and for which the momentum of the probe pion is at least $7\gevc$ are retained.
Background from $\PLambda \to \proton \pim$ decays, where the proton is mistakenly assumed to be a pion, is rejected by removing candidates with a mass in the vicinity of the known $\PLambda$ baryon mass when the proton hypothesis is assumed, retaining only those with $m({\proton \pim}) > 1180 \mevcc$. 

In total, $5\times10^6$ and $9\times10^6$ $\KS$ mesons are selected in the 2022 and 2023 dataset, respectively. The invariant-mass distribution for the 2023 dataset is presented in Fig.~\ref{fig:selection_ks_mass_peaks_no_cuts}. Additional material is presented in Appendix~\ref{appendix:mass_plots}.

\begin{figure}[]
\centering
\includegraphics[width=0.49\textwidth]{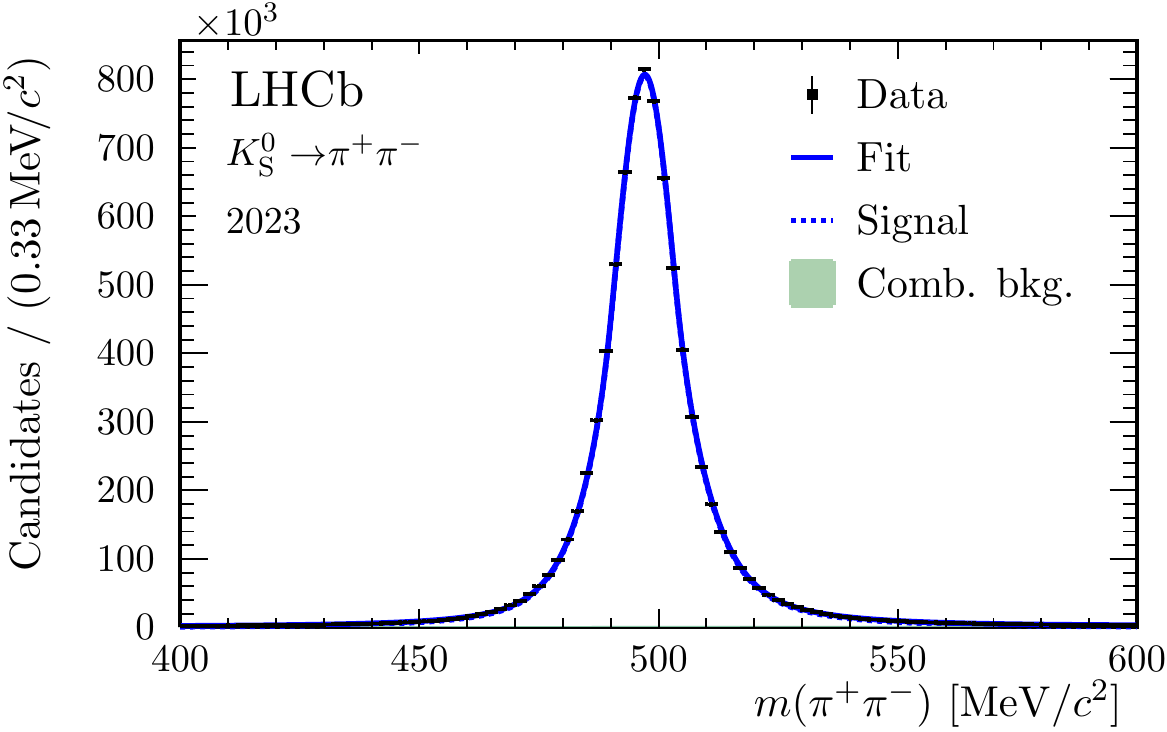}

\caption{Invariant-mass distribution of $\pip\pim$ combinations for $\KS$ candidates, obtained by constraining the origin of the $\KS$ meson to its associated PV, in the 2023 data. A fit to the data points is also shown. 
}\label{fig:selection_ks_mass_peaks_no_cuts}
\end{figure}

\subsection{Selection of untagged neutral charmed mesons}
At the \hltone stage, candidates are required to have either a track displaced from all PVs with high \pt, or a displaced two-track vertex, similar to the selection applied for the charged mesons.
At the \hlttwo stage, $\Dz\to\Km\pip$ candidates are formed by combining two oppositely charged tracks, each with $\pt > 800\mevc$ and $p > 5\gevc$, that do not point to any PV and originate from a common vertex. 
One of the tracks is required to be consistent with the kaon hypothesis, while the other is identified as a pion, using the RICH information. 
The $\Km\pip$ combination is retained if it has a $\pt$ of at least $2\gevc$, and its vertex is significantly displaced from any PV. 

In the offline selection, \Dz candidates originating from interactions with the detector material are removed by requiring the distance transverse to the beam line to be smaller than $13\mm$. 
Similar to the selection of the charged modes, the $\Dz$ candidates are required to be compatible with originating directly from the associated PV through a requirement on $\chisqip$.
Residual $\bquark$-hadron contributions are expected and estimated using candidates failing the $\chisqip$ requirement in addition (\cf Sec.~\ref{sec:secondaries}).

\begin{figure}[]
\centering
\includegraphics[width=0.49\textwidth]{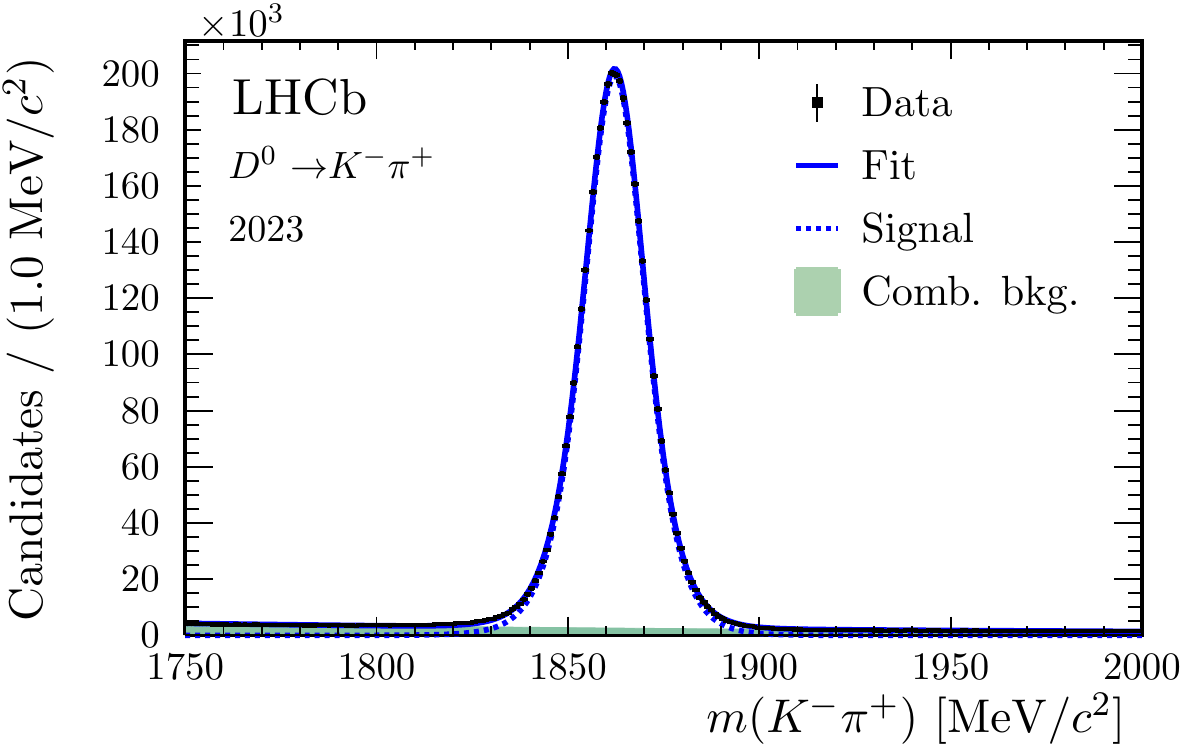}\\

\caption{Invariant-mass distribution of $\Km\pip$ combinations for untagged $\Dz$ candidates in the 2023 data. A fit to the data points is also shown. Candidates of the charge-conjugated decay are not included in this figure.
}\label{fig:selection_dz_mass_peaks}
\end{figure}

Approximately $1.1\times10^6$ and $8.7\times10^6$ $\Dz\to\Km\pip$ decays are selected from the 2022 and 2023 samples, respectively. Figure~\ref{fig:selection_dz_mass_peaks} presents the $m(\Km\pip)$ distribution for the selected candidates in the 2023 data, providing an indication of the sample's purity. 
Additional material is presented in Appendix~\ref{appendix:mass_plots}.

\subsection{Selection of the {\boldmath$\Dstarp$} calibration channels}
At the \hltone stage, the selection of tagged $\Dz$ candidates, used to calibrate the instrumental asymmetries for the $\Dz$ production asymmetry, is identical to that of untagged candidates. At the \hlttwo stage, tagged candidates are selected by first forming $\Dz\to\Km\pip$ and $\Dz\to\Km\Kp$ candidates.
These are created by combining two tracks that originate from a common vertex, with identical kinematic criteria as those in the untagged $\Dz\to\Km\pip$ selection.
Only the $\Dz\to\Km\pip$ candidates with $1.80 < m(\Km\pip) < 1.92\gevcc$ are retained, with a similar requirement for the $\Dz\to\Km\Kp$ candidates.
The mutually exclusive distinction between the $\Dz\to\Km\pip$ and $\Dz\to\Km\Kp$ decays is achieved by requiring the kaons or pions to be identified as such by the RICH detectors. 
These $\Dz$ candidates are then combined with another track, with $\pt > 200\mevc$ and $p> 1\gevc$, to form $\Dstarp\to\Dz\pip$ candidates.

To ensure that these calibration samples are representative of the signal, the offline selection requirements on the $\Dz$ candidate are identical to those of the $\Dz\to\Km\pip$ signal decays.
The deflection by the magnetic field can drastically impact the trajectory of the pion originating directly from the $\Dstarp$ decay, which typically carries little momentum. This leads to a large variation of its reconstruction efficiency depending on its charge and direction, and, subsequently, to large instrumental asymmetries in specific kinematic regions of the pion.
These regions are therefore removed from the data sample through the requirement $|p_x| < 0.32\times(p_z - 1900)$, where $p_x$ and $p_z$ are the projections of the pion momentum onto the LHCb coordinate system expressed in $\mevc$.
To increase the purity of the sample, only events with a single signal candidate are considered.

\begin{figure}[]
\centering
\includegraphics[width=0.49\textwidth]{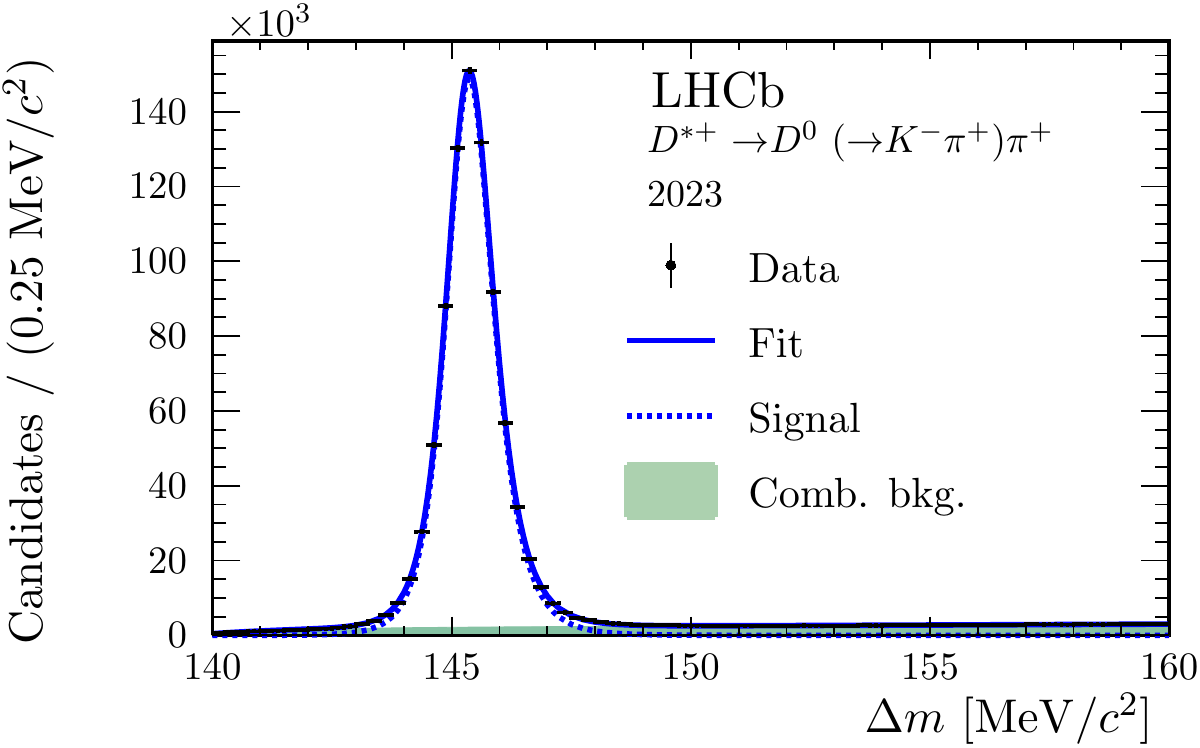}
\includegraphics[width=0.49\textwidth]{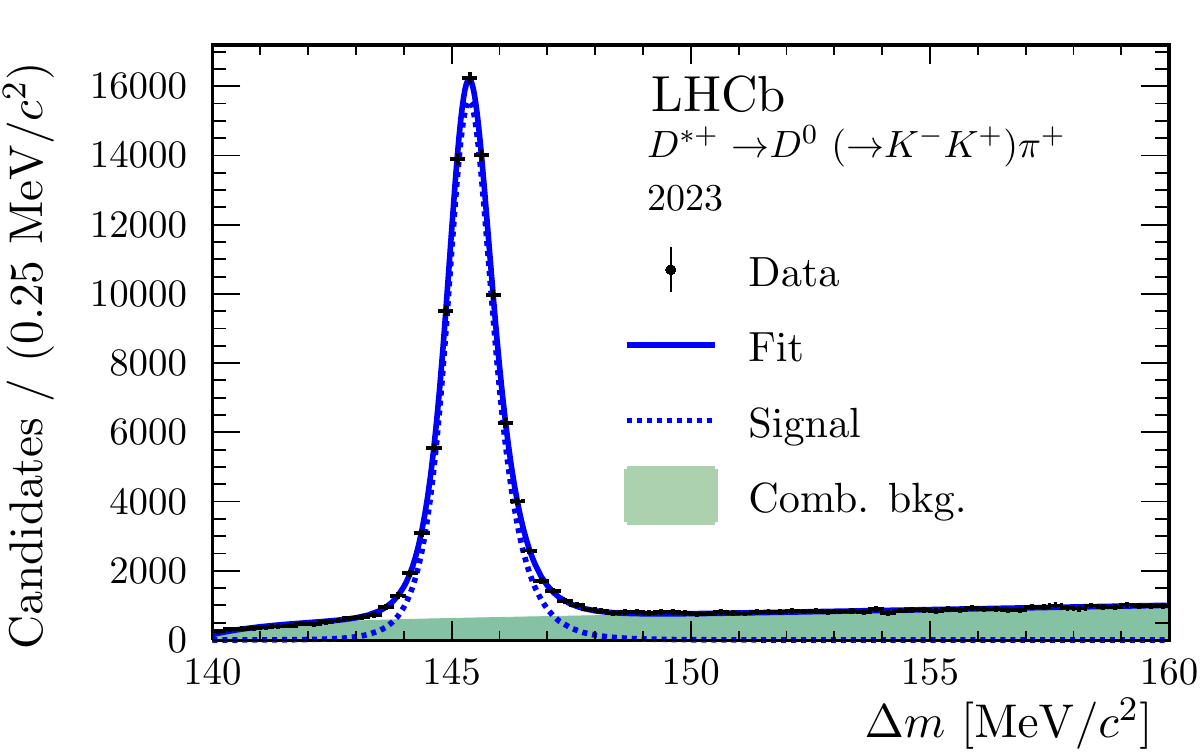}\\

\caption{Distributions of $\Delta m$ for positively charged $\Dstarp \to \Dz\pip$ candidates, used as control modes for the measurement of the $\Dz$ production asymmetry, with (left) $\Dz\to\Km\pip$, and (right) \mbox{$\Dz\to\Km\Kp$} decays. Data collected in 2023 are shown, along with a fit to the data. 
}\label{fig:selection_dstar_mass_peaks}
\end{figure}

For the 2022 and 2023 data samples,  $7.2\times10^5$ and $1.6\times10^6$ \mbox{$\Dstarp\to\Dz(\to\Km\pip)\pip$} are selected, respectively, along with $7.9\times10^4$ and $1.6\times10^5$ \mbox{$\Dstarp\to\Dz(\to\Kp\Km)\pip$} decays. 
Figure~\ref{fig:selection_dstar_mass_peaks} presents the distributions of the difference between the invariant masses of the $\Dstarp$ and $\Dz$ mesons, $\Delta m$, for the selected candidates of both decay modes in the 2023 data.
Additional material is presented in Appendix~\ref{appendix:mass_plots}.

\section{Determination of production asymmetries} 
\subsection{Charged charmed mesons}
After partitioning the data in the kinematic regions of Eq.~\ref{eq:binning}, the raw asymmetries of the $\Dsp$ and $\Dp$ mesons are determined through simultaneous unbinned maximum-likelihood fits to the $\Kp\Km\pip$ and $\Km\Kp\pim$ invariant-mass distributions. 
In these fits, the signal components are described using the sum of two Gaussian functions, one of which has a power-law tail to describe the final-state radiation~\cite{Skwarnicki:1986xj}.
While the analytical form is the same, most of the shape parameters are allowed to be different between the positively and negatively charged candidates, to account for possible differences in momentum estimates for positively and negatively charged signal candidates. 
To stabilise the fitting procedure, the width ratio of the two Gaussian functions is shared between the positively and negatively charged signal candidates. 
Apart from the power-law exponent, which is fixed from a fit to all data combined, all parameters are free to vary in the fits. 
To describe the background, a first-order polynomial for each of the two charges is used. 

The fitted signal yields per charge are expressed in terms of the raw asymmetry and the total yield, such that the raw asymmetry and its uncertainty can be determined directly as a parameter in the fit. 
With the same fit parametrisation, per-candidate weights calculated using the \sPlot~\cite{Pivk:2004ty} technique are assigned to statistically subtract the background from the kinematic distributions used in subsequent steps of the analysis. 
In a few kinematic regions, mostly at high values of $\eta$, the charmed-meson yields are too low to properly study the raw asymmetry and its corrections. 
For these regions, no measurements are reported.

\begin{figure}[]
\centering
\includegraphics[width=0.49\textwidth]{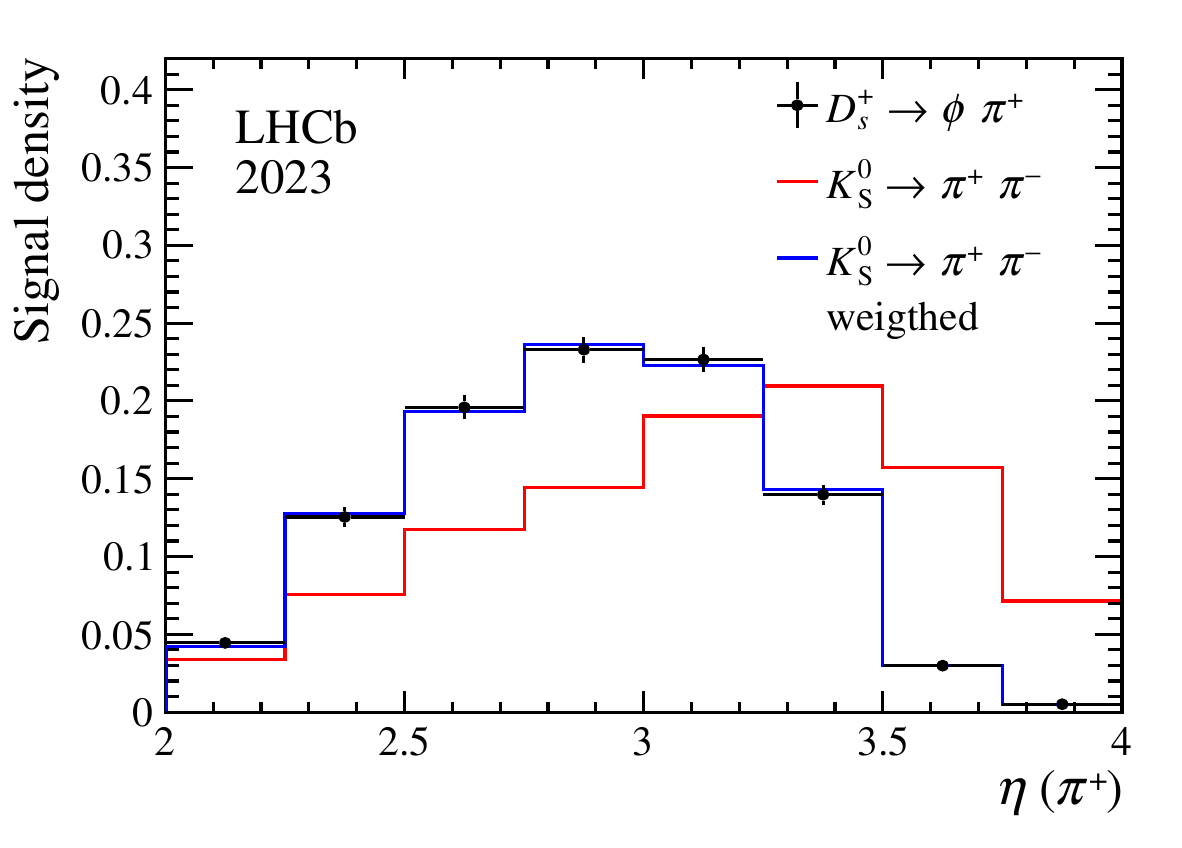}
\includegraphics[width=0.49\textwidth]{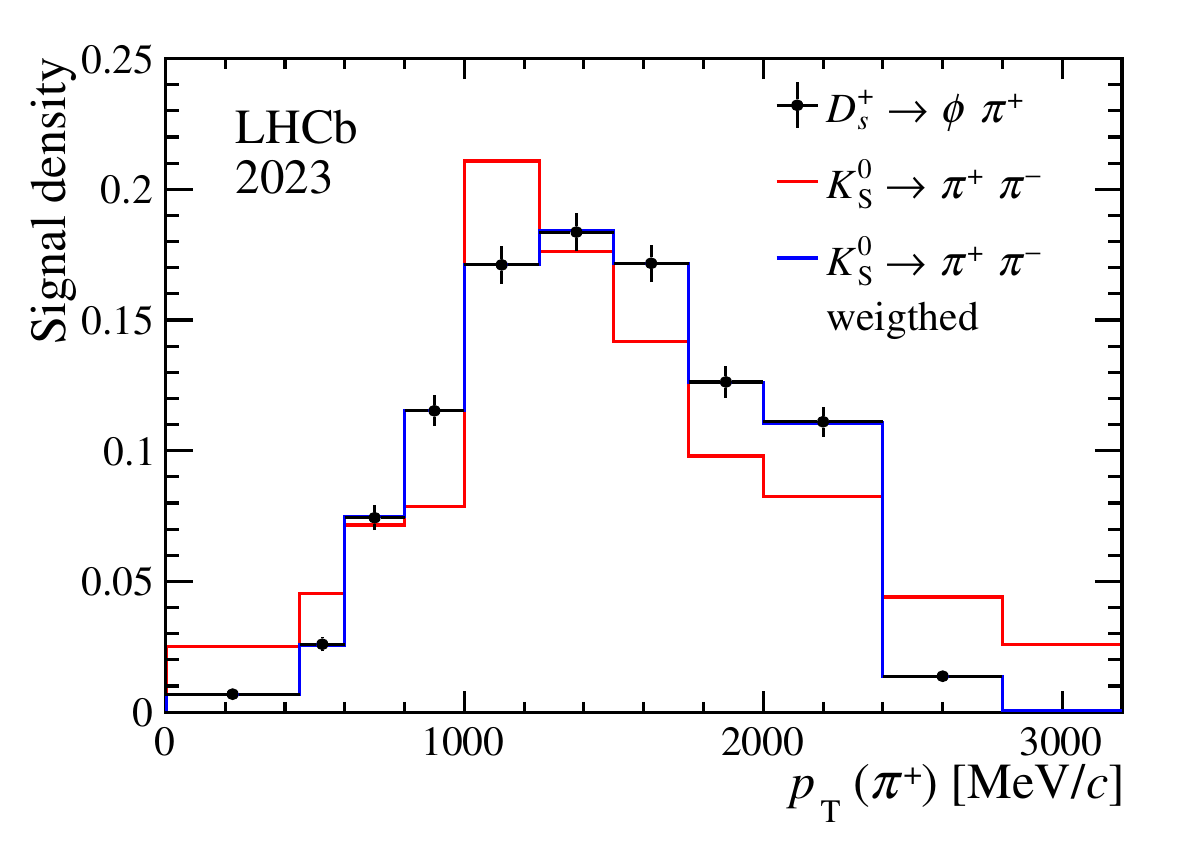} 
\caption{Normalised background-subtracted (left) $\eta$ and (right) $\pt$ distributions of the probe pion from $\KS\to\pip\pim$ decays, and the pion produced in $\Dsp\to\phi\pip$ decays for $\Dsp$ mesons with $2.0 < \eta < 3.0$ and $2.5 < \pt < 4.7 \gevc$. The distributions after applying the weights are shown as well. Only data collected in 2023 are shown. 
}\label{fig:weighting_ks_before_after_weighting}
\end{figure}

Based on Eq.~\ref{eq:Araw_sec}, the nuisance asymmetries $\arec(\pip)$ and $\apid(\pip)$ are determined to obtain the production asymmetry.
Since $\arec(\pip)$ is expected to depend on the kinematics of the pion, a procedure is used to equalise the kinematic distribution of the probe pion in the control sample to that of the pion in signal decays, which assigns weights to the $\KS$ candidates. 
Before determining these weights, the $\pip\pim$ invariant-mass distribution of the control channel, containing $\KS\to\pip\pim$ candidates, is fitted and background-subtracted kinematic distributions are obtained with the \sPlot method. Then, the weights are determined in two steps.
\begin{enumerate}
    \item The binned, two-dimensional, background-subtracted (normalised) distribution of the $\pt$ and $\eta$ of the signal pion is divided by the corresponding distribution of the probe pion. The resulting factors are used as per-candidate weights for each of the probe pions.
    \item Using the previously obtained weights for the probe pion, the same procedure is repeated with the two-dimensional distribution in $\eta$ and the azimuthal angle, $\varphi$.
\end{enumerate}
This weighting strategy, which is repeated for each of the kinematic regions of the charmed meson, is chosen as it provides good stability in the kinematic regions with few signal candidates. 
An example of the agreement between the calibration and signal sample after the complete weighting procedure is presented in Fig.~\ref{fig:weighting_ks_before_after_weighting}. 

After assigning the per-candidate weights, the control sample is divided into a matched sample, containing $\KS\to\pip\pim$ decays for which an extension of the probe VELO track to the rest of the tracking system is found, and an unmatched sample, which contains the remaining candidates. The efficiency is defined as the ratio of matched candidates to the total number of both matched and unmatched candidates.
The sample is also divided based on the charge of the probe pion, providing four categories in total. 
A binned maximum-likelihood fit is then performed on the weighted $\pip\pim$ invariant-mass distributions in each of these four categories to estimate the pion reconstruction efficiency and its charge asymmetry, $\arec(\pip)$. 
Both are included as parameters in the fit, following Eq.~\ref{eq:Adet_pi}. 
In this fit, the signal is described using a sum of two Gaussian functions, one of which features power-law tails on both sides of the peak. 
The background is described by an exponential function. No relation is assumed between the background shapes or yields between the different categories. 
The magnitudes of the resulting pion reconstruction asymmetry range from $0.04\%$ to $2.81\%$, with the largest $\arec(\pip)$ value observed for the highest $\eta(\DsDp)$ region. 

The additional nuisance asymmetry due to the PID requirements on the pion, $\apid(\pip)$, is determined using $\Dstarp\to\Dz(\to\Km\pip)\pip$ decays, following a similar approach to that used for $\arec(\pip)$. 
The $\Dstarp$ data are first assigned per-candidate weights to match the kinematic distributions of the $\DsDp$-decay pion with the same weighting procedure as used for $\arec(\pip)$, and are subsequently split by charge. 
These data are further split into two categories, one with pions passing the PID requirements, and one with those that do not. 
The PID asymmetries are determined through unbinned maximum-likelihood fits to the weighted $\Delta m$ distributions including these asymmetries as free parameters. 
The resulting $\apid(\pip)$ magnitudes range from $0.02\%$ to $1.44\%$, with the largest value observed for the highest $\eta(\DsDp)$ region. 

\subsection{Neutral charmed mesons}
\begin{figure}[tb]
\centering
\includegraphics[width=0.49\textwidth]{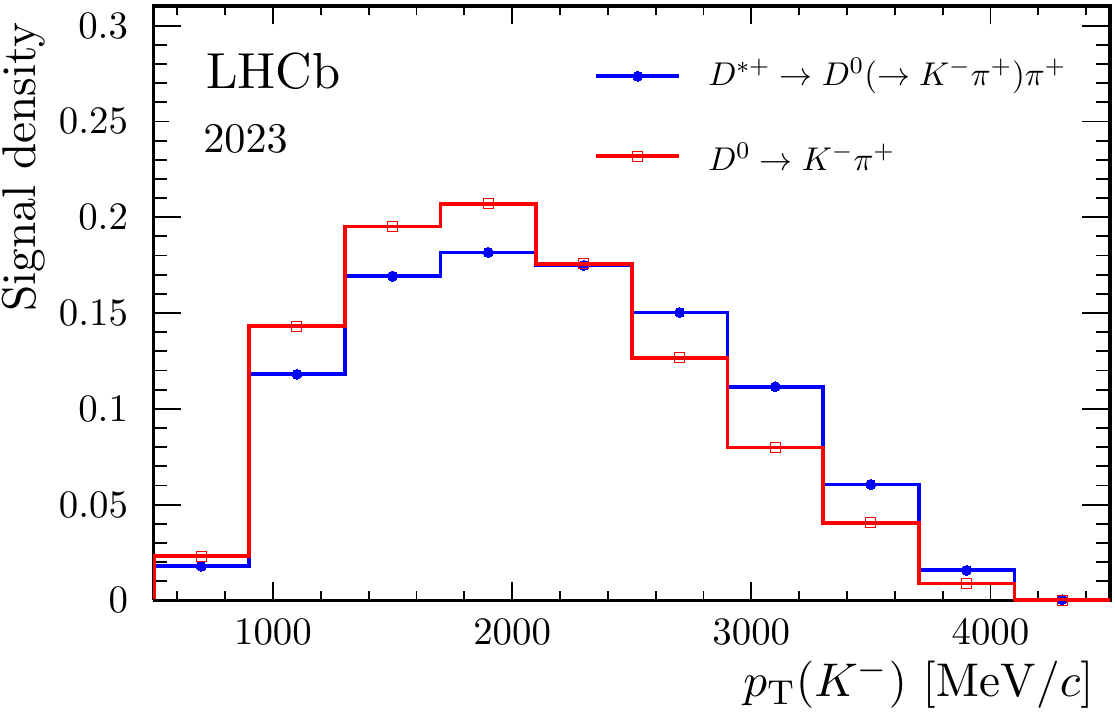}\hfil
\includegraphics[width=0.49\textwidth]{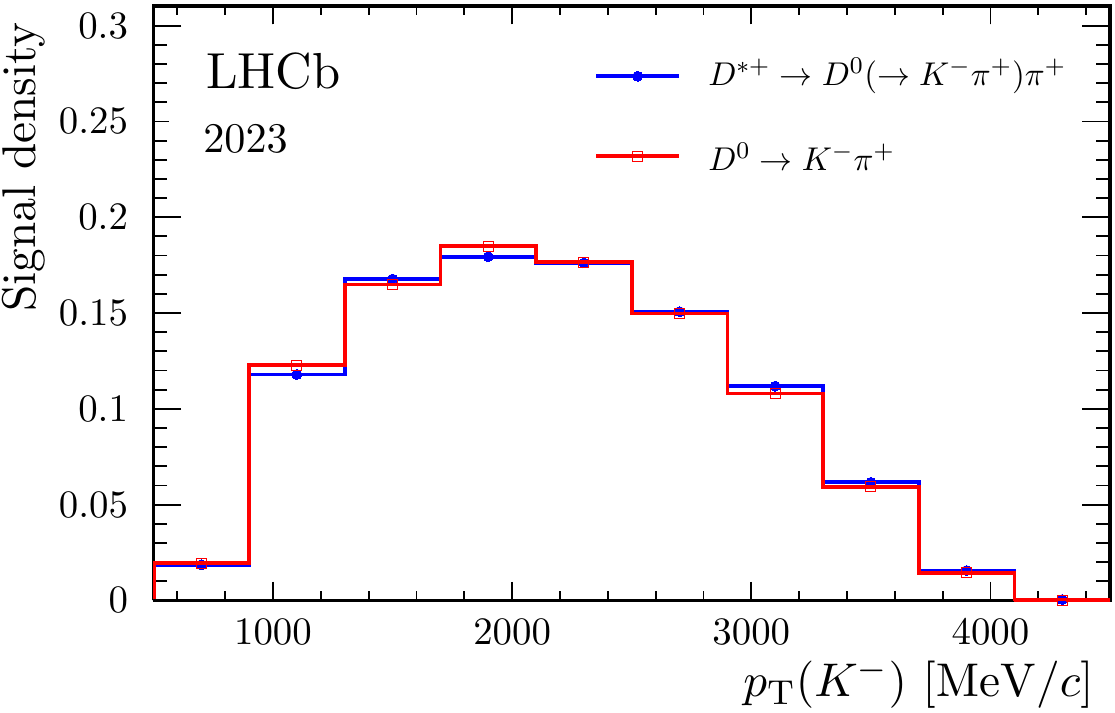}\\
\includegraphics[width=0.49\textwidth]{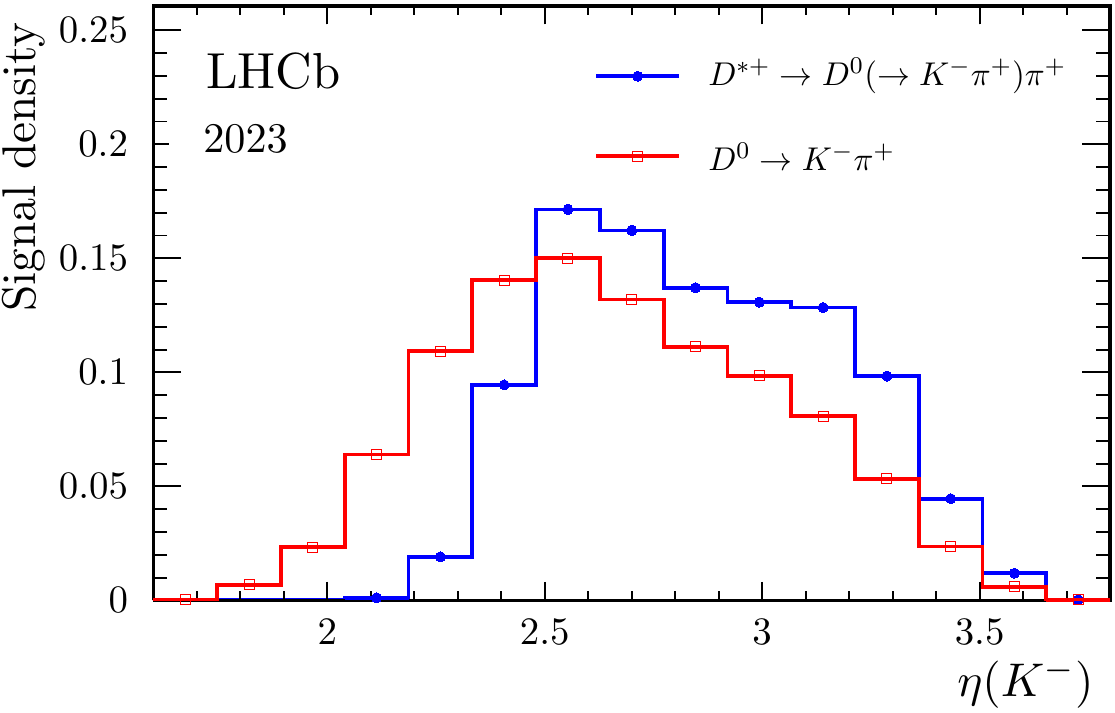}\hfil
\includegraphics[width=0.49\textwidth]{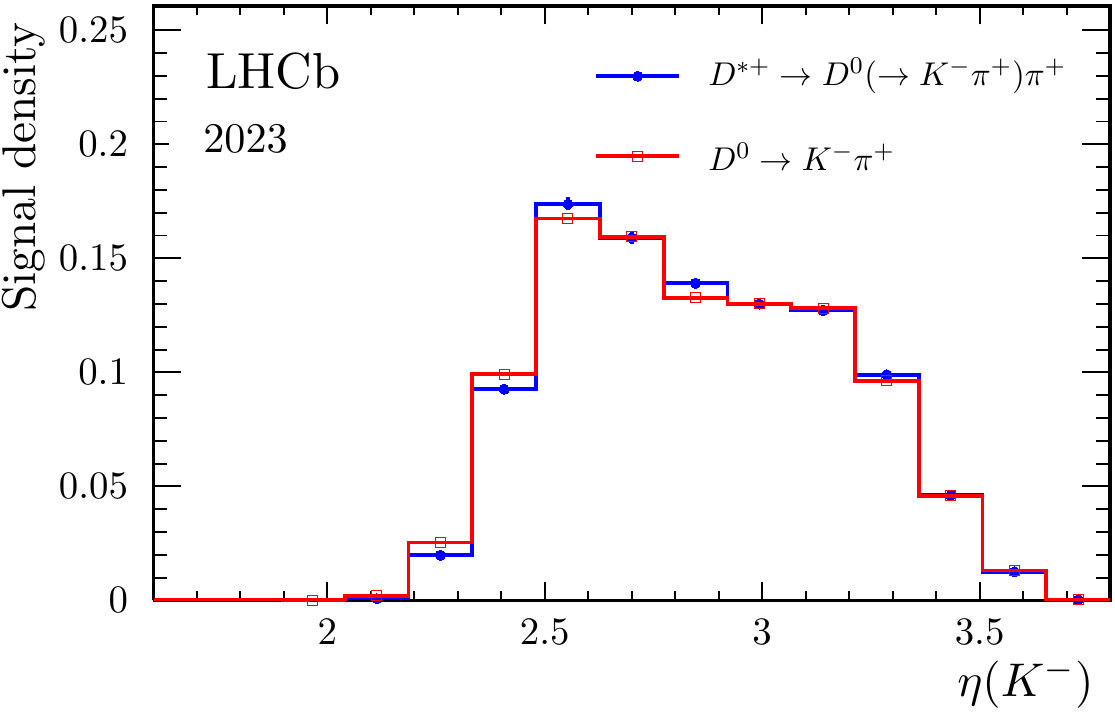}\\
\caption{Normalised background-subtracted (top) $\pt$ and (bottom) $\eta$ distributions of the kaon from $\Dz\to\Km\pip$ decays and $\Dstarp\to\Dz(\to\Km\pip)\pip$ decays for $\Dz$ mesons with $2.0 < \eta < 3.0$ and $2.5 < \pt < 4.7$ \gevc. The distributions are shown (left) before and (right) after applying the kinematic weights. Only data collected in 2023 are shown.}\label{fig:weighting_dz_km_example}
\end{figure}

Differently from the charged modes, the size of the calibration samples, in particular that of \mbox{$\Dstarp\to\Dz(\to\Km\Kp)\pip$} decays, limits the statistical precision on the measurement of $\Ap(\Dz)$. 
In this case, instead of matching the kinematic distributions of the calibration data samples to those of the signal decays, 
the kinematic weighting procedure, needed to cancel the nuisance asymmetries, assigns weights to the tagged and untagged $\Dz\to\Km\pip$ candidates.
These weights are determined by dividing the background-subtracted, three-dimensional $(\pt, \eta, \varphi)$ distribution of the $\Dz$ meson in \mbox{$\Dstarp\to\Dz(\to\Km\Kp)\pip$} decays, by that observed for the tagged and untagged \mbox{$\Dz\to\Km\pip$} decays.
Too sparsely populated regions in $\Dz$ kinematics are removed in all samples to further stabilise the weighting procedure.
While the weights only consider the $\Dz$ meson kinematics, the resulting kinematic distributions of the kaon and the pion agree well between the tagged and untagged $\Dz\to\Km\pip$ decays, as well as the distributions of the \Dstarp and the pion accompanying the \Dz meson between the two tagged decays. 
Examples of this agreement are presented in Fig.~\ref{fig:weighting_dz_km_example}. 

After the weighting procedure, the raw asymmetries of the signal and calibration channels are determined in each kinematic region of the charmed meson through binned least-squares fits. The fits are carried out simultaneously to the meson and antimeson candidates, with the raw asymmetry included as a parameter. The fits to the signal channel are done on the $m(\Km\pip)$ distributions, while those to the $\Dstarp$ decays are done on the $\Delta m$ distributions. 
In all of the fits, the signal is described using a Johnson $S_{U}$ function~\cite{Johnson:1949zj}, with shape parameters shared between the meson and antimeson candidates, except for the peak positions, which are free to vary independently. 
To describe the background in the $m(\Km\pip)$ distribution, an exponential function with free scale parameter is used. 
The background in the $\Delta m$ distributions is modelled using the empirical {\tt RooDstD0BG} function from the \roofit software package~\cite{roofit,ROOT}. 
The parameters describing the background in the $\Delta m$ distributions are shared between the $\Dstarp$ and $\Dstarm$ candidates.

Using the measured raw asymmetries, \adet(\Km\pip) is calculated as described in Eq.~\ref{eq:adetkpi}. This value is then subtracted from the raw asymmetry of the signal channel to determine $\AP(\Dz)$. To account for the correlation between the tagged and untagged samples, statistical uncertainties are estimated using a bootstrapping technique~\cite{efron:1979}. The full analysis procedure is repeated for each bootstrap sample.

The resulting magnitude for the raw asymmetries range from $0.01\%$ to $3.9\%$ across all bins and data samples, with the largest value observed for the highest $\eta(\Dz)$ region. This range serves as an indication for the typical size of the required corrections due to nuisance asymmetries.

\section{Secondary charm production}
\label{sec:secondaries}
\begin{figure}[tb]
\centering
\includegraphics[width=0.49\textwidth]{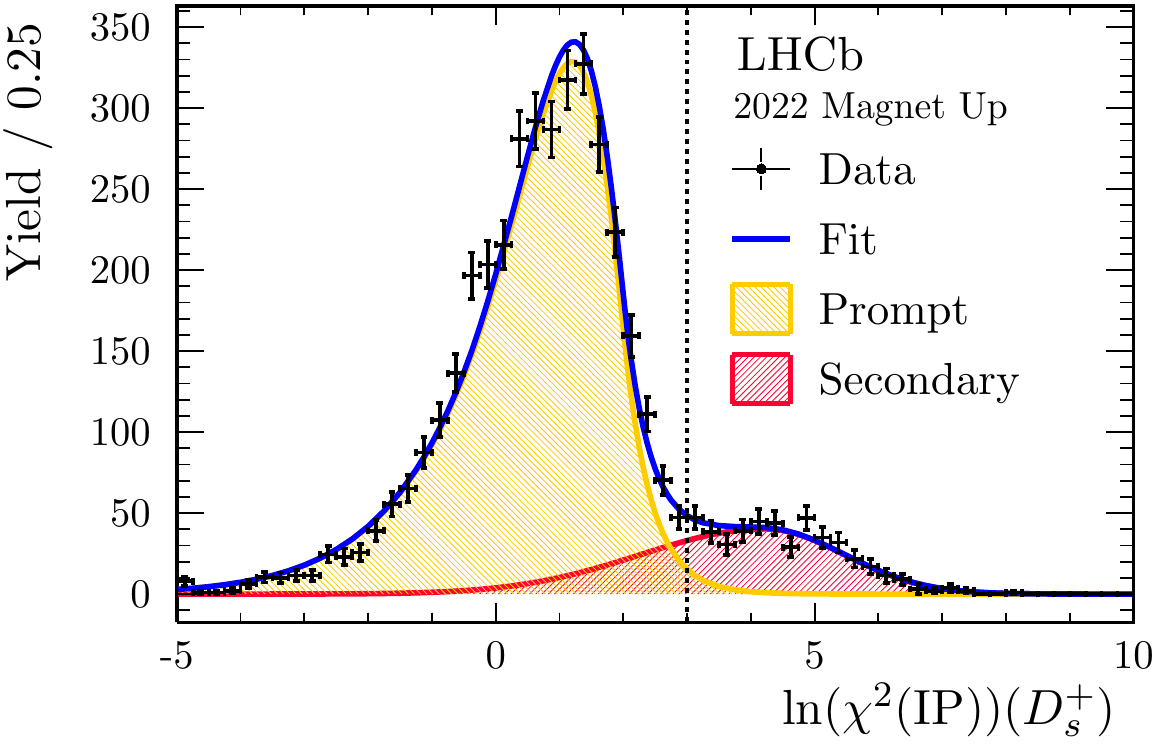}
\includegraphics[width=0.49\textwidth]{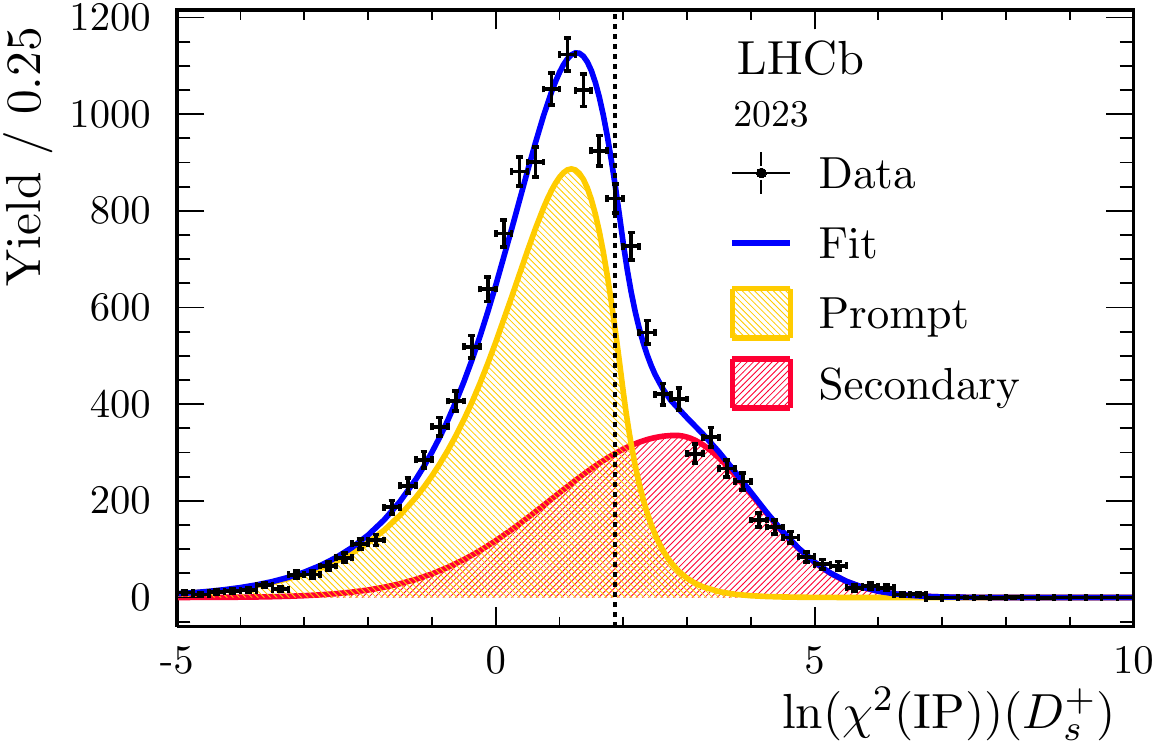}\\

\includegraphics[width=0.49\textwidth]{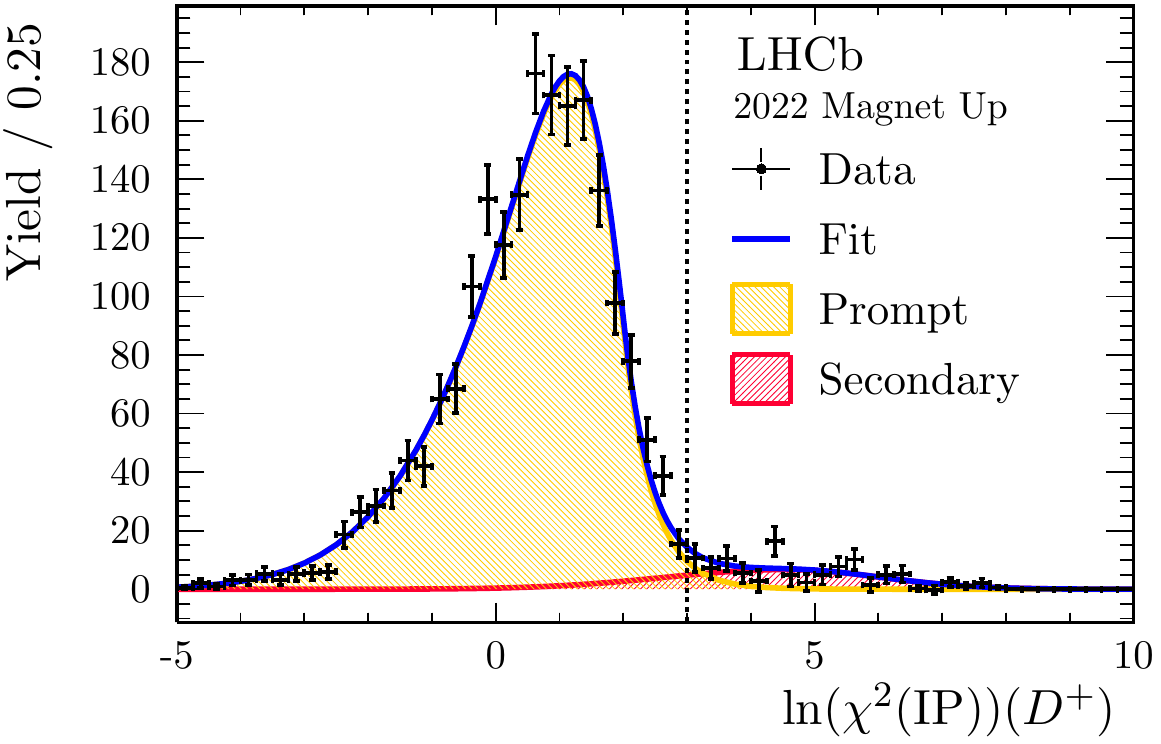}
\includegraphics[width=0.49\textwidth]{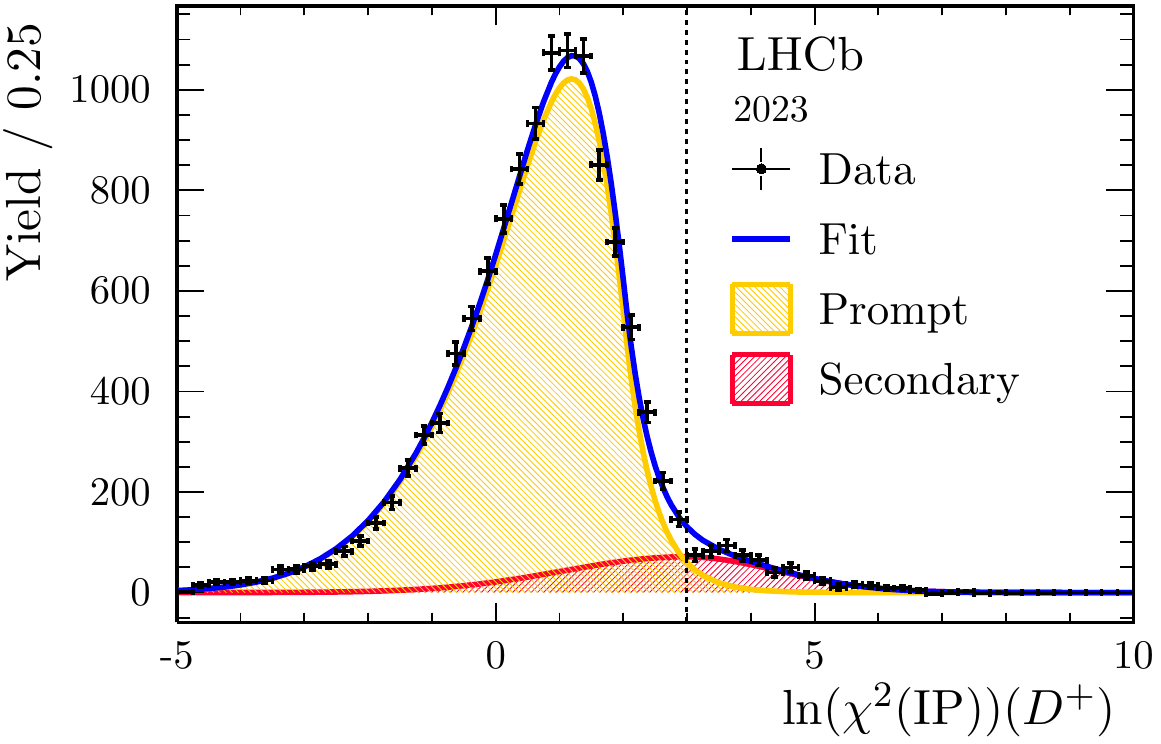}

\includegraphics[width=0.49\textwidth]{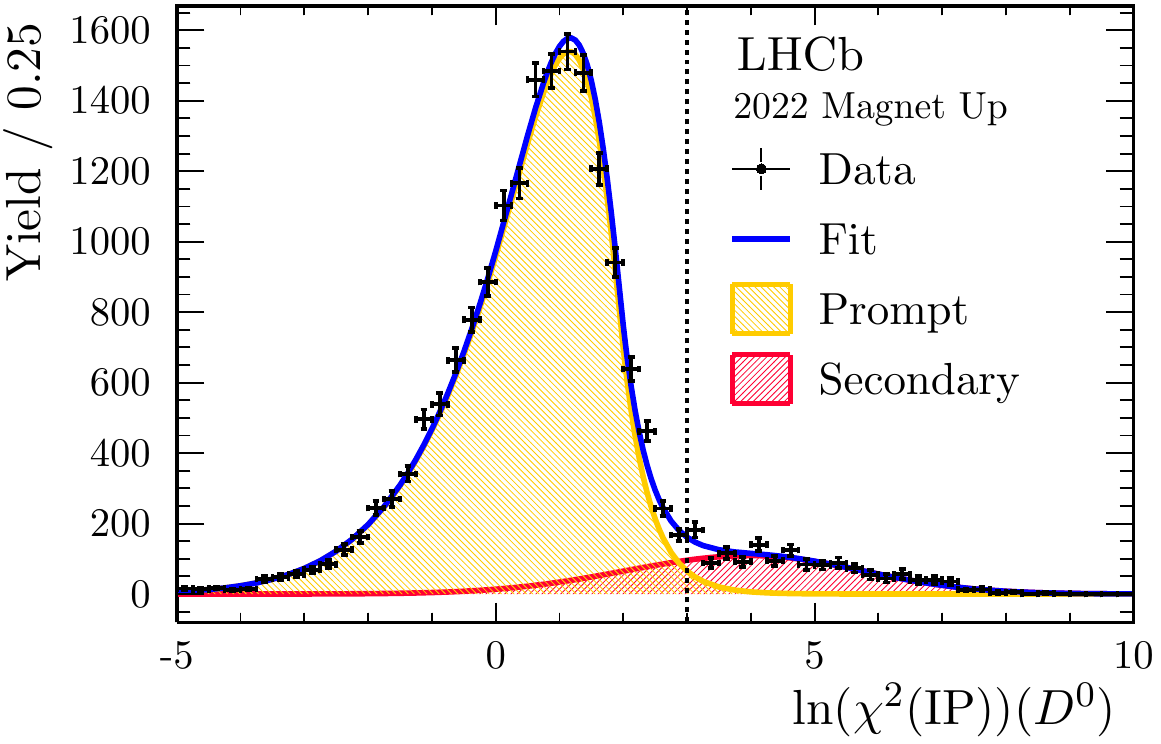}
\includegraphics[width=0.49\textwidth]{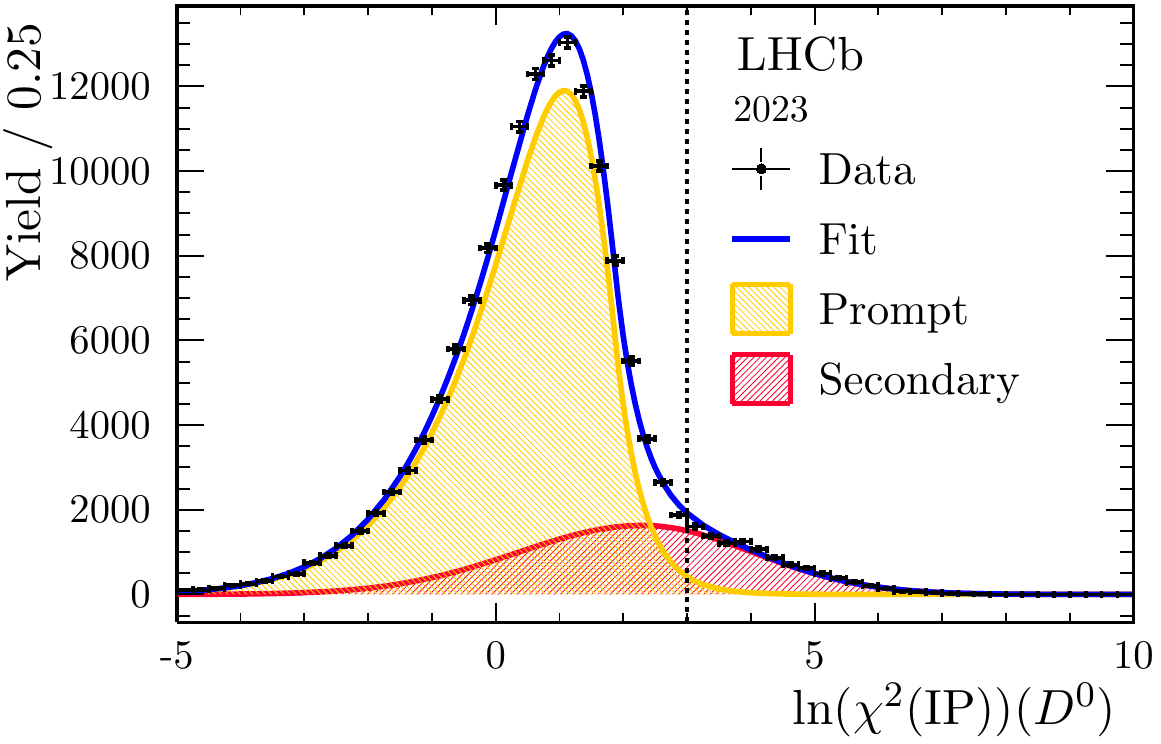}

\caption{Background-subtracted distributions of $\ln(\chisqip)$ for the (top) $\Dsp$, (middle) $\Dp$ and (bottom) $\Dz$ mesons in (left) 2022 and (right) 2023 data. All plots refer to the kinematic region $2.0 < \eta < 3.0$ and $6.5 < \pt < 8.5$ \gevc of the charmed-meson candidates. The dashed line marks the maximal $\ln(\chi^2(\text{IP}))$ allowed by the offline selection.
}\label{fig:seconaries_examples}
\end{figure}

Contributions from charmed mesons produced in \bquark-hadron decays are removed by correcting the measured raw asymmetry by the term in Eq.~\ref{eq:Araw_sec},
\begin{equation}
\label{eq:DeltaAsec}
    \Delta A_{\text{sec}} = \fsec \left[\Aprod(\Xc)- \Asec (\Xc)\right].
\end{equation}
This correction is evaluated for each kinematic region and data sample by considering the distributions of the observable $\ln{(\chisqip)}$ of the meson and antimeson candidates which provide information on both the fraction \fsec~and the asymmetry difference $\Aprod(\Xc)- \Asec (\Xc)$.
For these studies the $\chisqip$ requirement on the charmed-meson candidate is removed, allowing for an unbiased consideration of the secondary contributions. 
Simultaneous unbinned maximum-likelihood fits are performed to the background-subtracted distributions of the meson and antimeson signal candidates to determine the corrections.
In these fits, promptly produced charmed-meson candidates are modelled using a Bukin function~\cite{bukin2007fittingfunctionasymmetricpeaks}, while the secondary component is described using a Gaussian function that has different widths on the left and right sides of the peak.
The peak position of these functions, which depends on the detector resolution, is left free to vary in the fits to data, while the remaining parameters are fixed to values obtained from simulated prompt and secondary decays.
Examples of these fits are shown in Fig.~\ref{fig:seconaries_examples}. 
The yields of prompt and secondary charmed mesons, split per charge, are included as free parameters in the fit, from which the asymmetry difference, \mbox{$\Aprod(\Xc) - \Asec (\Xc)$}, and $\fsec$ are inferred. 

The IP resolution for data recorded in 2023 is particularly impacted by the open VELO configuration, leading to a higher contamination from $\bquark$-hadron decays with respect to 2022 data. 
The largest contributions are found for $\Dsp$ mesons, in line with the expectations from $b$-hadron-decay branching fractions and $\bquark$-hadron production cross-sections~\cite{PDG2024,LHCb-PAPER-2017-037,LHCb-PAPER-2018-050,LHCb-PAPER-2015-041}. 
On average, the fractions of secondary decays are $5.4\%$, $1.6\%$ and $4.1\%$ for $\Dsp$, $\Dp$ and $\Dz$ mesons, respectively.
It is verified that the secondary fraction is stable as a function of the time within a fill in 2022, and thus not affected by the movement of the VELO detector. 

The values for \mbox{$\Aprod(\Xc) - \Asec (\Xc)$} are independent of the detector configuration and are determined primarily using data recorded in 2022, after verifying that there is no difference between the datasets. 
The resulting values of \mbox{$\Aprod(\Xc) - \Asec (\Xc)$} from the different datasets are combined for the $\Dsp$ mesons to increase their statistical precision.
~As the observed secondary contribution of $\Dp$ mesons is too small to make a reliable estimate of the related $\bquark$-hadron production asymmetry, \mbox{$\Aprod(\Dp) - \Asec (\Dp)$} is instead conservatively estimated as $(0 \pm 5)\%$. 
This bound on the systematic uncertainty covers the measured relevant production asymmetries at $\sqs=7$ and $8\tev$~\cite{LHCb-PAPER-2012-026,LHCb-PAPER-2016-062}, and the expectation at 13.6\tev. 

The corrections $\Delta A_{\text{sec}}$ are mostly found to be compatible with zero within three standard deviations when considering statistical uncertainties only. 
Their values remain below one-third of the corresponding statistical uncertainty of $\Aprod$.

\section{Cross-checks and systematic uncertainties}
\subsection{Cross-checks}
Cross-checks are carried out for further scrutiny of the results and deeper understanding of the upgraded \lhcb detector.
None of these cross-checks signalled any shortcomings in the methodology, showing compatibility with $p$-values of at least $5\%$.

\begin{figure}
    \centering
    \includegraphics[width=\linewidth]{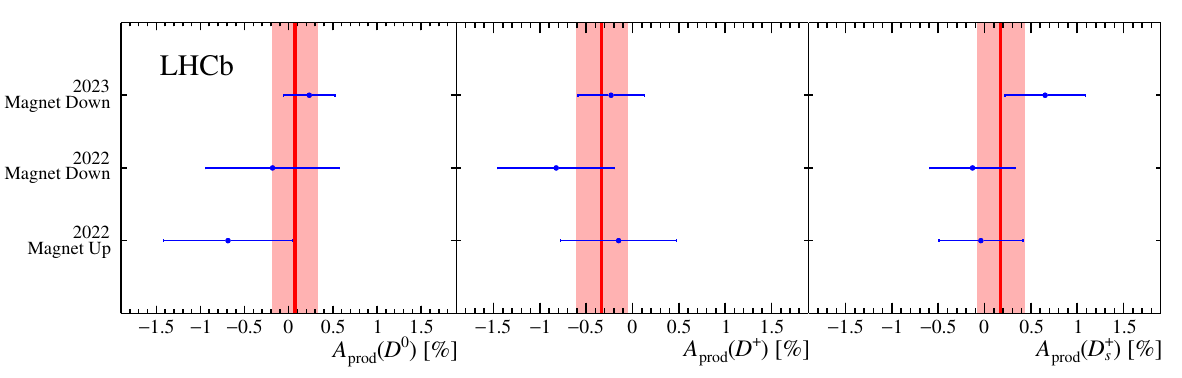}
    \caption{Comparison between the production asymmetries measured separately from the 2023, 2022 \MagDown and 2022 \MagUp data samples. The comparison is shown for the (left)~\Dz meson, (middle)~\Dp meson and (right)~\Dsp meson. The average value and its uncertainty are indicated by the red line and the red band, respectively.}
    \label{fig:samples_compatibility}
\end{figure}

The consistency of the production asymmetries across kinematic regions in the 2022 \MagDown, 2022 \MagUp and 2023 datasets is considered, as these represent different detector conditions. The global $p$-values are found to be 88.4\%, 27.9\%, 57.9\% for the \Dz, \Dsp and \Dp production asymmetries, respectively. The integrated production asymmetries are shown in Fig.~\ref{fig:samples_compatibility} for the three samples and the three decay modes. In addition, the analysis is repeated in even finer intervals of these data-taking periods to evaluate the consistency per dataset. 
The consistency of the $\Dsp$ production asymmetry in the 2022 \MagDown dataset has the weakest compatibility, corresponding to a $p$-value of 5.9\%.

Where possible, the asymmetries in different kinematic regions for the final-state particles are studied. 
In addition, measurements of the $\Dz$ production asymmetry are performed with stricter requirements on the pointing of the charmed mesons to the associated PVs. 
In all of these studies no deviations are found with respect to the baseline analysis approach.  

The $\KS$ mesons, used in the calibration of $\arec(\pip)$, have a longer lifetime than the charmed mesons. Nevertheless, it is confirmed that the typical positions of the pions produced in the $\KS\to\pip\pim$ decays in the tracking system agree with those produced in the charm decays. In addition, $\arec(\pip)$ is determined in different intervals of the $\KS$ decay-vertex longitudinal position, with all values found to be in agreement. 

The \Dz production asymmetry is determined using only events with a single candidate. 
Including events with more candidates changes the integrated production asymmetry by less than one standard deviation of the statistical uncertainty. 

For the data recorded in 2022, it is studied whether the production asymmetries are sensitive to the variation of the position of the VELO across one fill. 
This is achieved by evaluating the variation of the production asymmetry as a function of the time after closure of the \velo. 
The results, included in Appendix~\ref{appendix:crosscheck_consistency}, show no variation in the data that are used. To further test the stability, the baseline result is compared to a result obtained from a dataset with additional selections. 
First, the data-taking periods with the largest variation in the \velo position, which also have a misalignment larger than 10\mum in the $x$-direction and corresponding to about the first 50 minutes of a fill, are removed (about 7\% of the 2022 data). 
Second, about 6\% (3\%) of \Dz ($\DsDp$) candidates close to the upper acceptance of the \velo, where the VELO movement might influence the selection efficiency the most, are rejected by requiring that all final-state particles have $\eta<4.3$. 
No significant deviations are observed, with the lowest $p$-value being 12.1\% for the \Dz production asymmetry measurement in the magnet-down sample.  

\subsection{Systematic uncertainties}
Systematic uncertainties are evaluated per kinematic bin, and per charmed meson. The considered contributions are explained in the following, in order of importance. Table~\ref{tab:summary_of_systematics} presents a summary of these contributions and their size. 

The systematic uncertainties due to the models used in the determination of the secondary component are evaluated by studying the impact of additional degrees of freedom in these fits.
In particular, all the shape parameters fixed from simulation for the prompt and secondary components are left free to vary, one component at a time, and the resulting corrections are compared to the baseline values. 
The largest difference observed with respect to the baseline result is used as an estimate of the systematic uncertainty.
In addition to the fit shapes used to estimate the secondary contribution, differences between the instrumental asymmetries in the promptly produced charmed mesons, and those produced from secondary decays, could also impact this correction. These imperfections are accounted for by multiplying the difference between their kinematic distributions with the observed raw asymmetry in bins of the final-state particles' kinematics, and then summing over all bins. The resulting values are only significant for the $\Dz$ meson, for which these are added to the systematic uncertainty related to the secondary contribution. 

The impact of the choice of fit function on the inferred raw asymmetries of the control and signal channels is studied through a set of pseudoexperiments. 
For each kinematic region, datasets are generated with the same relative fraction and yields of signal and background as observed in data. 
To produce and fit the pseudodata, an alternative and the baseline models are used to perform either of the two steps. 
The alternative model consists of a different function to describe the signal shape, either the sum of three Gaussian functions, of which one includes a low-mass power-law tail, or the same baseline function but with fewer constrained shape parameters. 
The average difference between the production asymmetries measured from the pseudodata samples, and the generated asymmetry, is used as a systematic uncertainty.
When both generating and fitting with the baseline model  no significant difference is observed.
For the charged mesons, the result of the fit is compared to the asymmetry obtained by counting the number of decays of each charge, where the background is subtracted using information from the sidebands and assuming a linear shape. 
The maximum of this difference and the uncertainty obtained by pseudoexperiments is used as a systematic uncertainty.

The systematic uncertainty related to the kinematic-weighting procedure is addressed by estimating the impact of remaining differences in the kinematic distributions between the signal and control modes. 
This is achieved by considering fine-grained one-dimensional projections of the kinematic distributions onto the component with which $\Araw$ shows the largest variation. 
These differences are multiplied by the values of $\Adet$ from the control samples, and summed to evaluate the systematic uncertainty.
For \AP(\DporDsp), the impact of residual correlations among the three components of the momentum are studied by including additional steps in the weighting procedure, using the data sample with the largest variation of $\arec(\pip)$. 
The differences with respect to the baseline strategy are used as systematic uncertainties.

Transitions other than those via the $\Pphi$ meson are expected to be present in the selected $\DporDsp\rightarrow\Km\Kp\pip$ decays, which could introduce a difference  between the $\Km$ and $\Kp$ meson kinematics. 
The additional instrumental asymmetry arising from a difference in the momentum distributions of these kaons is estimated by comparing their distributions and using the raw asymmetry of the $\Dz\to\Km\pip$ decay as a proxy for the kaon detection asymmetry.
The resulting systematic uncertainty is determined to be of the order of $10^{-3}$ and is slightly larger for the $\Dp$ meson than for the $\Dsp$ meson.

The systematic uncertainty associated to the curvature-bias corrections, applied to 2023 data, is estimated by evaluating the raw asymmetries with and without the corrections. 
The observed difference is assigned as the systematic uncertainty.

It is assumed that the flavour of the $\Dz$ meson at production can be inferred directly from the $\Km\pip$ final state. This neglects effects from the doubly Cabibbo-suppressed \mbox{$\Dzb\to\Km\pip$} decays, as well as \mbox{$\Dz$--$\Dzb$} mixing. Using previous experimental results~\cite{HFLAV24}, the impact of the misidentification rate on the production asymmetries is estimated to be below $10^{-4}$, which is assigned as the corresponding systematic uncertainty.

Finally, the uncertainty due to the experimental knowledge on $\ACP$~\cite{LHCb-PAPER-2022-024,LHCb-PAPER-2019-002} is included as part of the systematic uncertainties. 

\begin{table}[!h]
    \caption{\label{tab:summary_of_systematics} Absolute systematic uncertainties on the charm production asymmetries, in units of $10^{-2}$. 
    The values indicate the span across the different kinematic bins. 
    In case of an empty cell (--), the systematic uncertainty is not applicable.
    The statistical uncertainties are presented in Table~\ref{tab:results_aprod}.
    }
        \centering
        \bgroup
        \def\arraystretch{1.25}
        \begin{tabular}{l @{\hskip 2.5em} c @{\hskip 2.2em} c @{\hskip 2.2em} c @{\hskip 1.5em} }
    \multicolumn{4}{c}{Systematic uncertainty~$[10^{-2}]$}\\
        \toprule
        Source \hskip 3.25em & $\Dz$ & $\Dp$ & $\Dsp$ \\
        \cmidrule[0.5pt](r{1.2em}){1-1}
        \cmidrule[0.5pt](l{-0.5em}){2-4}
        Secondary contribution & \makecell{$0.02-0.35$} &\makecell{$0.16-0.41$} & \makecell{$0.03-0.43$}   \\
        Fit models $\Araw$ & \makecell{$0.01-0.05$} &\makecell{$0.08-0.73$} & \makecell{$0.05-0.39$}  \\

	      Residual $\Adet$ & \makecell{$0.04-0.40$} &\makecell{$\leq0.11$} & \makecell{$\leq0.19$}  \\

	      Fit model $\Arec(\pip)$  &  $-$ &\makecell{$\leq0.08$} & \makecell{$\leq0.15$} \\

	      Asymmetries in the $\Km\Kp$ pair  & $-$ & \makecell{$\leq0.12$} & \makecell{$\leq0.09$}  \\
          Curvature bias & \makecell{$\leq0.39$} &\makecell{$\leq0.25$} & \makecell{$\leq0.25$}  \\
Flavour misidentification & \makecell{$0.01$} & $-$ & $-$  \\
	       $\ACP$ & \makecell{0.06} & \makecell{0.05} & $-$  \\
           \midrule
           Total systematic uncertainty & \makecell{$0.08-0.56$} & \makecell{$0.25-0.85$}  &  \makecell{$0.11-0.53$}\\
           Statistical uncertainty & \makecell{$0.68-4.59$} & \makecell{$0.79-1.81$} & \makecell{$0.72-2.59$}\\
           
          \bottomrule
      \end{tabular}
      \egroup
    \end{table}

\section{Results}
\label{sec:results}
\begin{table}[h]
    \caption{\label{tab:results_aprod} Results for the \Dz, \Dp and \Dsp production asymmetries in units of $10^{-2}$, after averaging over the datasets. The first quoted uncertainty is statistical, the second is systematic.}
    \centering
    \bgroup
    \def\arraystretch{1.25}
    \resizebox{\textwidth}{!}{
    \begin{tabular}{l @{\hskip 1.75em} *{4}{r@{ $\pm$ }l } }
    \multicolumn{9}{c}{$\Aprod(\Dz)$ [$10^{-2}$]}\\
    \toprule
    & \multicolumn{8}{c}{$\eta$} \\
    \cmidrule(l{-0.5em}){2-9}
    \pt [\gevc]\hskip 1.25em & \multicolumn{2}{c}{$2.0-3.0$} & \multicolumn{2}{c}{$3.0-3.5$} & \multicolumn{2}{c}{$3.5-4.0$} & \multicolumn{2}{c}{$4.0-4.5$}\\
    \cmidrule[0.5pt](r{1.2em}){1-1}
    \cmidrule[0.5pt](l{-0.5em}){2-9}
    $2.5-4.7$ & $1.92$ & $1.34 \pm 0.36$ & $0.65$ & $0.69 \pm 0.44$ & $-0.91$ & $0.96 \pm 0.16$ & $-3.23$ & $2.13 \pm 0.14$ \\ 
$4.7-6.5$ & $-0.81$ & $0.82 \pm 0.40$ & $0.25$ & $0.68 \pm 0.08$ & $-1.37$ & $1.28 \pm 0.33$ & \multicolumn{2}{c}{--} \\
$6.5-8.5$ & $1.15$ & $0.80 \pm 0.19$ & $0.63$ & $0.82 \pm 0.09$ & $-2.84$ & $2.51 \pm 0.49$ & \multicolumn{2}{c}{--} \\
$8.5-25.0$ & $-0.17$ & $0.62 \pm 0.16$ & $-0.14$ & $0.96 \pm 0.32$ & $3.27$ & $4.59 \pm 0.56$ & \multicolumn{2}{c}{--} \\
       \bottomrule
\\
    \multicolumn{9}{c}{$\Aprod(\Dp)$ [$10^{-2}$]}\\     
          \toprule
    & \multicolumn{8}{c}{$\eta$} \\
    \cmidrule(l{-0.5em}){2-9}
    \pt [\gevc]\hskip 1.25em & \multicolumn{2}{c}{$2.0-3.0$} & \multicolumn{2}{c}{$3.0-3.5$} & \multicolumn{2}{c}{$3.5-4.0$} & \multicolumn{2}{c}{$4.0-4.5$}\\
    \cmidrule[0.5pt](r{1.2em}){1-1}
    \cmidrule[0.5pt](l{-0.5em}){2-9}
        $2.5-4.7$&$-1.13$& $1.05\pm0.46$&$-0.19$& $0.86\pm0.42$&$-0.33$& $0.98\pm0.26$&$-2.14$& $1.78\pm0.37$\\
        $4.7-6.5$&$-0.78$& $0.79\pm0.25$&$-0.47$& $0.82\pm0.48$&$-1.25$& $1.14\pm0.47$&
        \multicolumn{2}{c}{--}
        
        \\
        $6.5-8.5$&$0.38$& $0.87\pm0.45$&$-0.84$& $1.00\pm0.40$&$-0.10$& $1.81\pm0.85$&
        \multicolumn{2}{c}{--}
        
        \\
        $8.5-25.0$&$0.56$& $0.86\pm0.45$&$0.91$& $1.24\pm0.34$&
        \multicolumn{2}{c}{--}&\multicolumn{2}{c}{--}
        \\
      \bottomrule
\\
    \multicolumn{9}{c}{$\Aprod(\Dsp)$ [$10^{-2}$]}\\     
    \toprule
    & \multicolumn{8}{c}{$\eta$} \\
    \cmidrule(l{-0.5em}){2-9}
    \pt [\gevc]\hskip 1.25em & \multicolumn{2}{c}{$2.0-3.0$} & \multicolumn{2}{c}{$3.0-3.5$} & \multicolumn{2}{c}{$3.5-4.0$} & \multicolumn{2}{c}{$4.0-4.5$}\\
    \cmidrule[0.5pt](r{1.2em}){1-1}
    \cmidrule[0.5pt](l{-0.5em}){2-9}
        $2.5-4.7$&$1.19$& $1.17\pm0.35$&$0.65$& $0.84\pm0.45$&$-1.34$& $0.83\pm0.18$&$0.86$& $1.41\pm0.27$\\
        $4.7-6.5$&$-0.54$& $0.83\pm0.23$&$1.64$& $0.76\pm0.21$&$1.00$& $0.88\pm0.13$&$0.70$& $1.91\pm0.43$\\
        $6.5-8.5$&$0.20$& $0.85\pm0.38$&$-0.56$& $0.87\pm0.15$&$-0.37$& $1.27\pm0.21$&
        \multicolumn{2}{c}{--}\\
        $8.5-25.0$&$0.72$& $0.72\pm0.11$&$-1.81$& $0.97\pm0.18$&$0.71$& $2.59\pm0.53$&
        \multicolumn{2}{c}{--}\\
      \bottomrule
\end{tabular}
    }
      \egroup
    \end{table}
    
\begin{figure}[]
\centering
\includegraphics[width=0.6\textwidth]{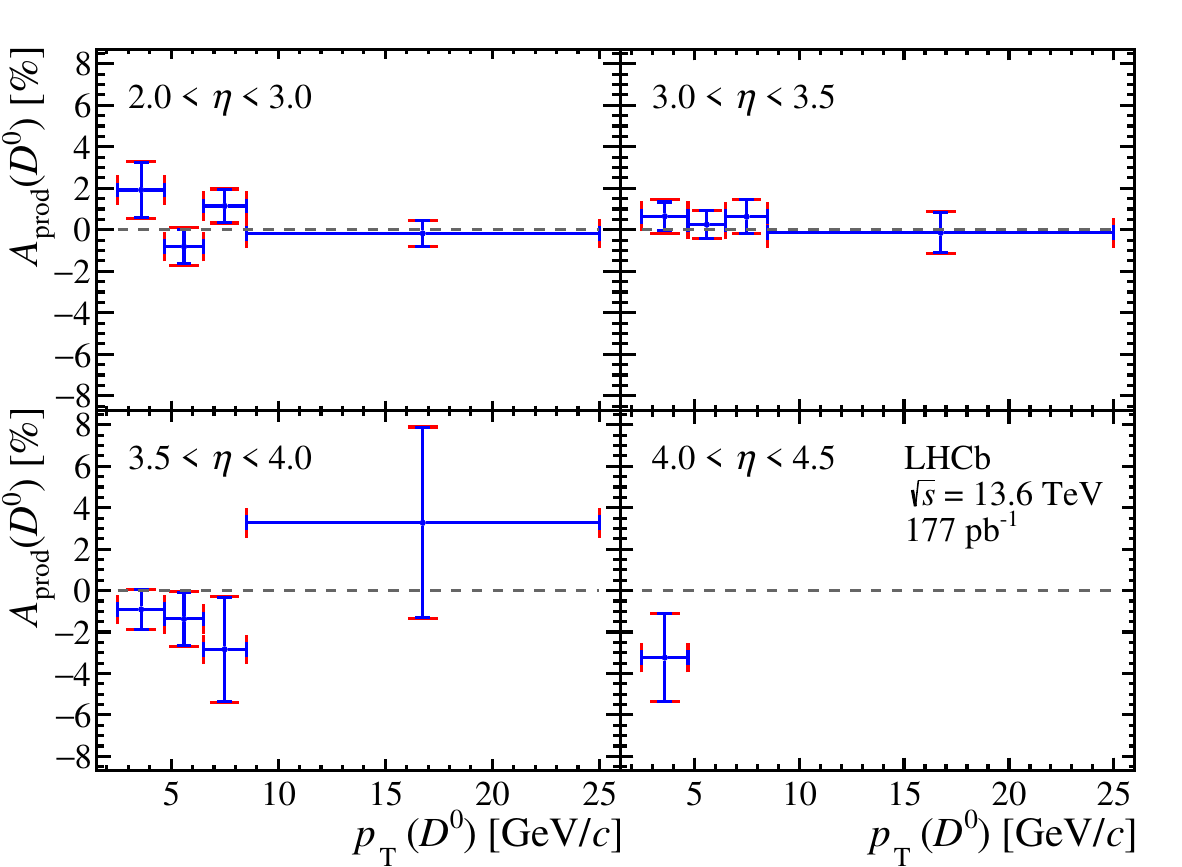}
\includegraphics[width=0.6\textwidth]{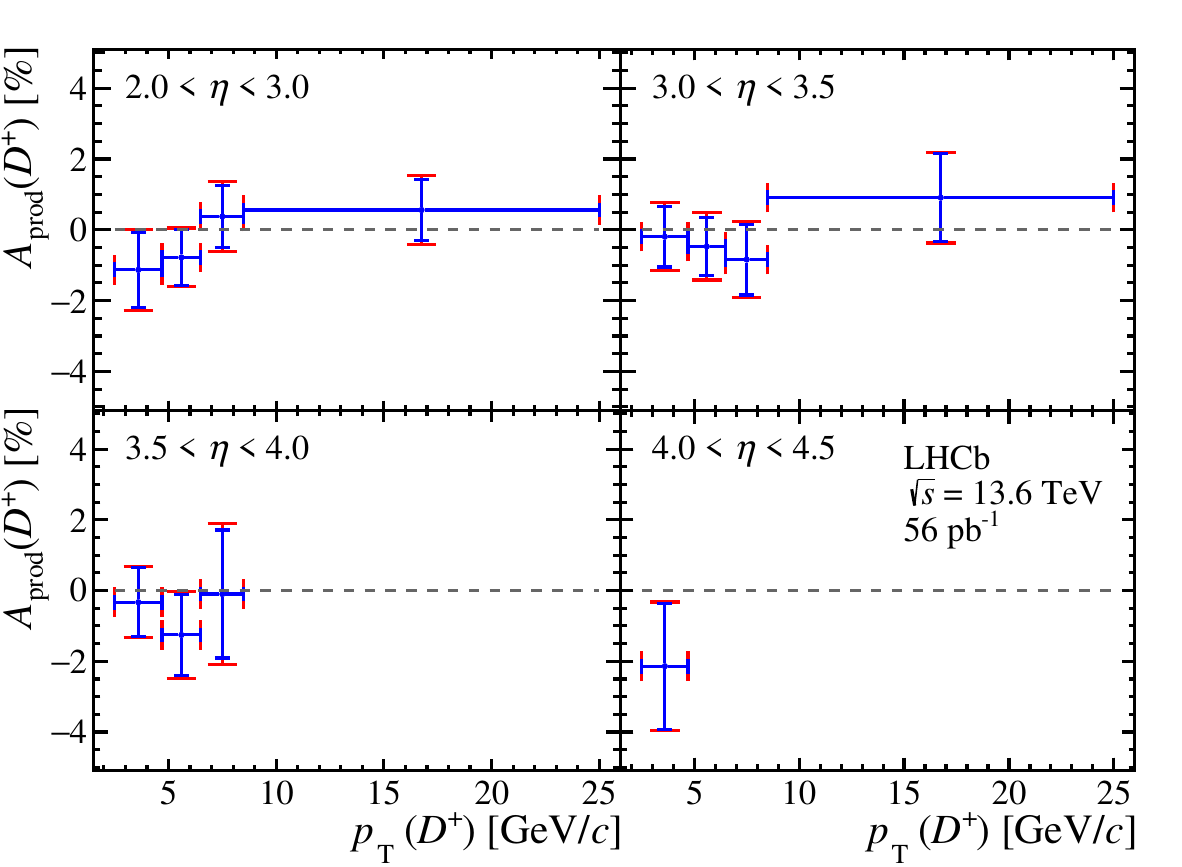}
\includegraphics[width=0.6\textwidth]{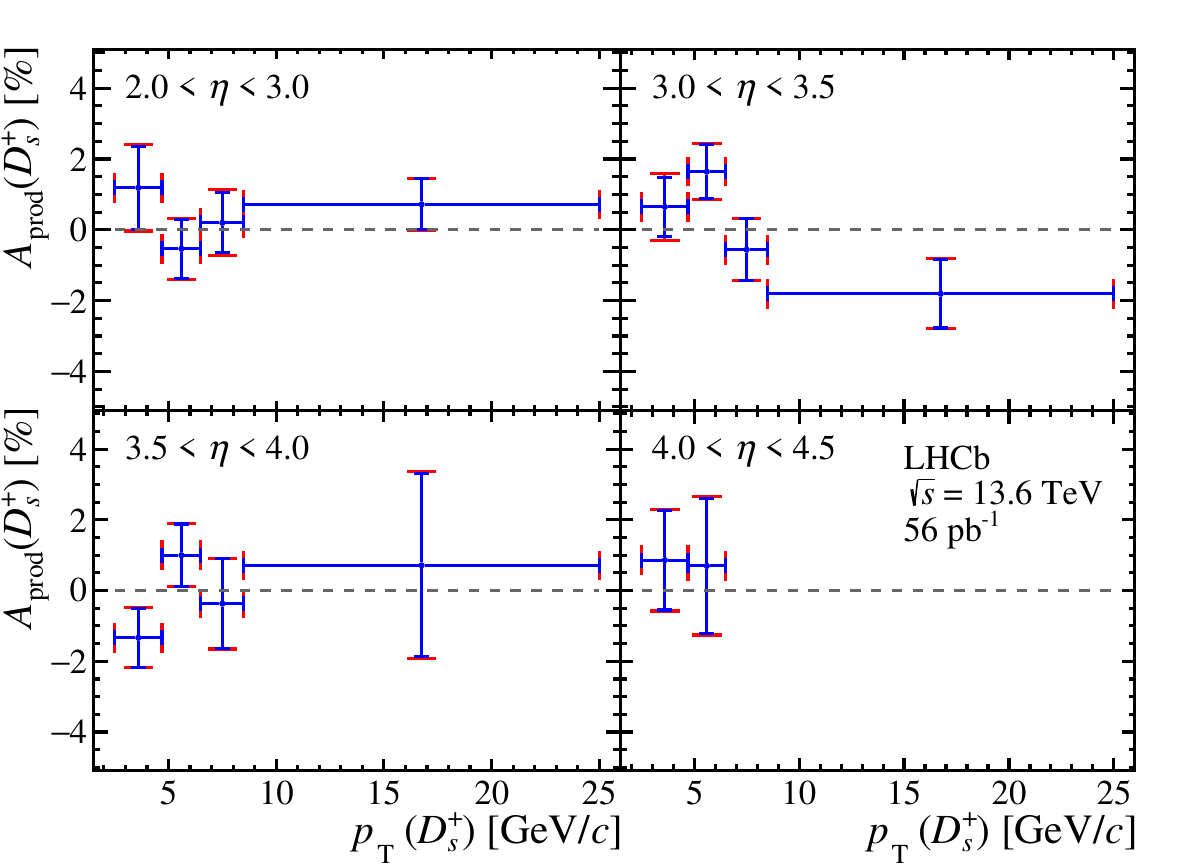}

\caption{\label{fig:results_plot}Production asymmetries for the indicated charmed mesons after averaging over the datasets. The blue error bars represent the statistical uncertainty, and the red ones represent the total uncertainty.}
\end{figure}

Table~\ref{tab:results_aprod} presents the production asymmetries in intervals of meson \pt and $\eta$ for the $\Dz$, $\Dp$ and $\Dsp$ mesons. A graphical representation of the results is given in Fig.~\ref{fig:results_plot}.

While the datasets used to calibrate the instrumental asymmetries are shared between the two charged charmed mesons and kinematic bins, the total uncertainty is found to be dominated by that of the (uncorrelated) raw asymmetries. Consequently, there is only a negligible correlation between the various presented results, at the subpercent level. 

Most of the values are compatible with symmetric production of mesons and antimesons with no dependence on their kinematics within one standard deviation. The corresponding global $p$-values for their compatibility with the absence of any production asymmetry are {$45.3\%$} for the $\Dz$ meson, $77.6\%$ for the $\Dp$ meson, and $21.7\%$ for the $\Dsp$ meson. 

\begin{figure}[]
\centering
\includegraphics[width=0.49\textwidth]{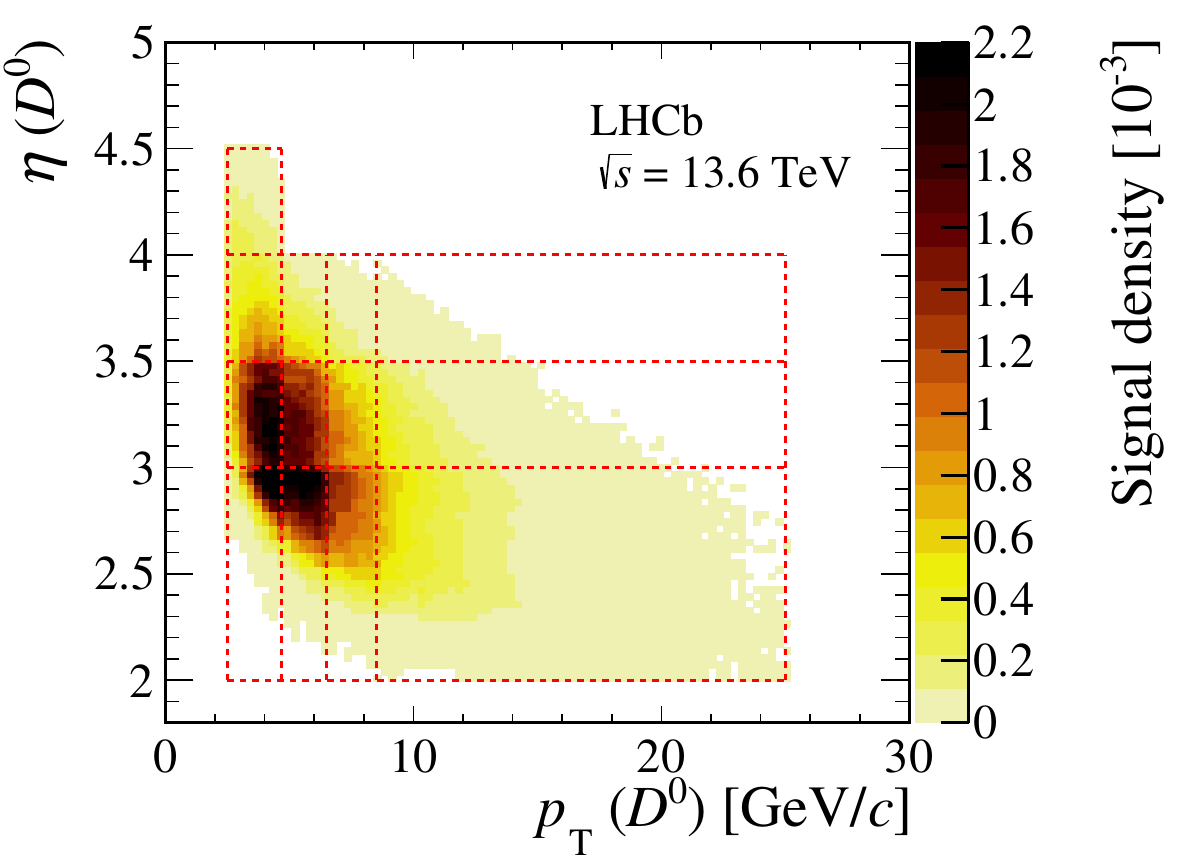}
\includegraphics[width=0.49\textwidth]{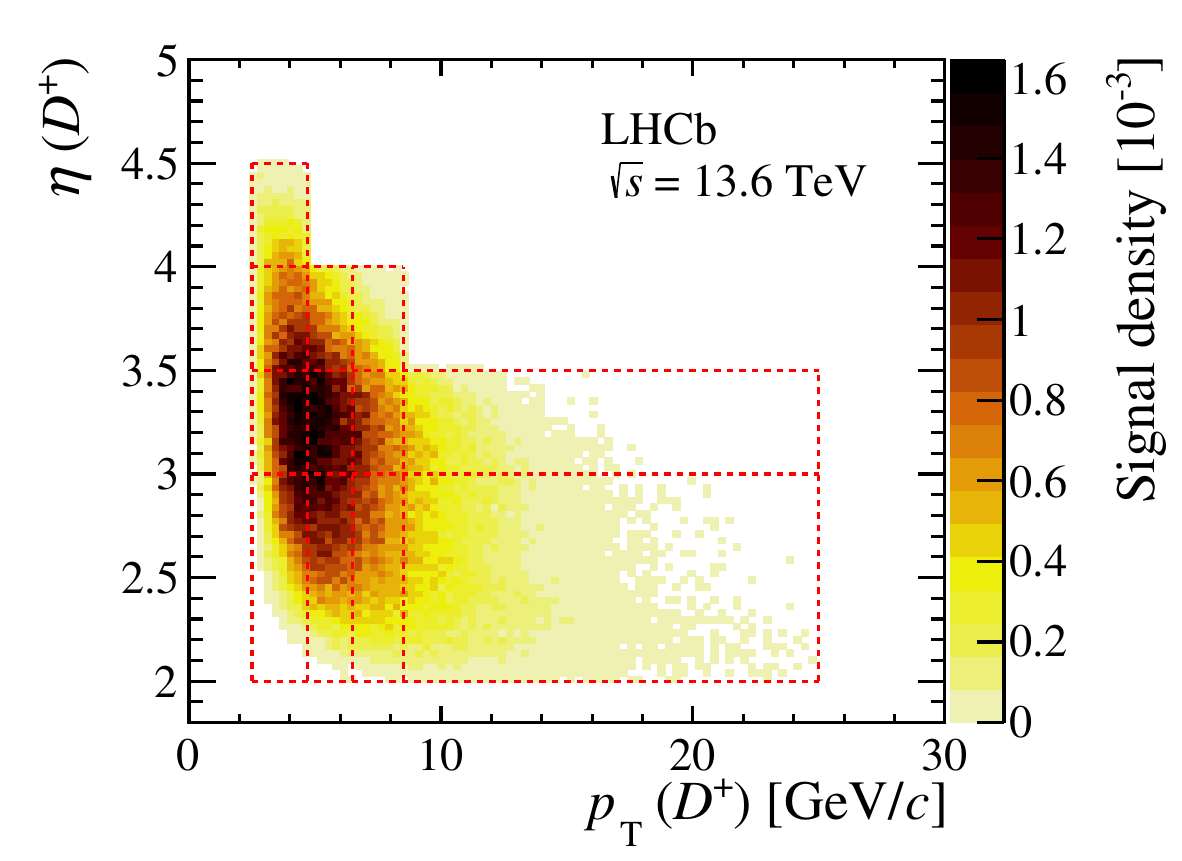} 
\includegraphics[width=0.49\textwidth]{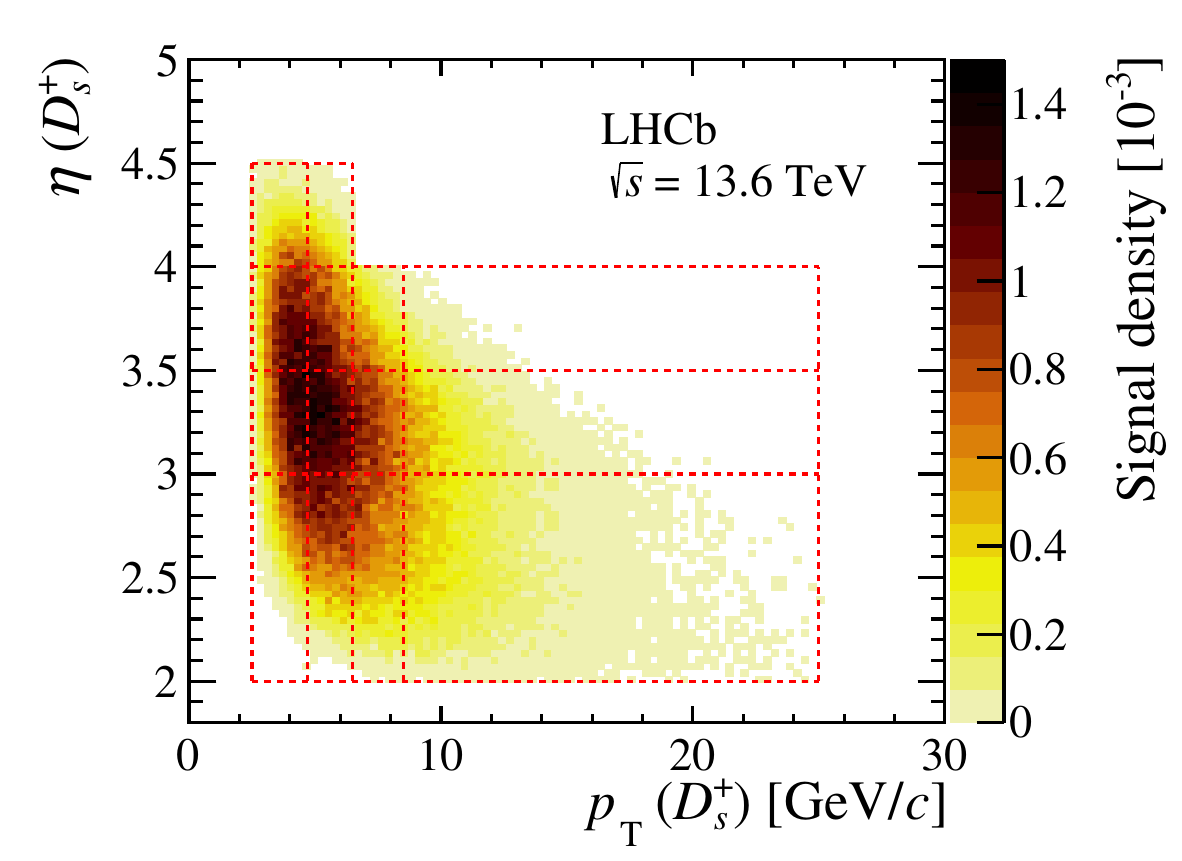}
\caption{\label{fig:charm_kinematic_distributions}Observed $(\pt, \eta)$ distributions for the indicated charmed mesons, not corrected for the instrumental efficiency, after combining all datasets. The dashed red lines indicate the kinematic bins.}
\end{figure}

To illustrate typical production asymmetries relevant for charmed-meson decays at the LHCb experiment, the asymmetries observed across the kinematic intervals are averaged, accounting for the observed $\pt$ and $\eta$ distributions (Fig.~\ref{fig:charm_kinematic_distributions}), not corrected for the instrumental efficiency. The results are:
\begin{align*}
    \Aprod(\Dz) &= (\phantom{-}0.07\pm 0.26\stat \pm 0.10 \syst) \% ,\nonumber \\
    \Aprod(\Dp) &= 
    ( {-0.33} \pm 0.29 \stat \pm 0.14 \syst) \% ,
    \nonumber \\
    \Aprod(\Dsp) &= 
    (\phantom{-}0.18 \pm 0.26 \stat \pm 0.08 \syst) \%
    .
\end{align*}

Previous measurements of the $\Dp$ \cite{LHCb-PAPER-2012-026} and $\Dsp$ \cite{LHCb-PAPER-2018-010} production asymmetries at the LHC showed no significant dependence on either $\pt$ or $\eta$. 
This allows a comparison of the averaged production asymmetries. The averaged results obtained at 13.6\tev are compatible with those at lower centre-of-mass energies, with a combined $p$-value of $8.5\%$.

\begin{figure}[]
\centering
\hspace{0.5cm} 
\includegraphics[width=0.84\textwidth]{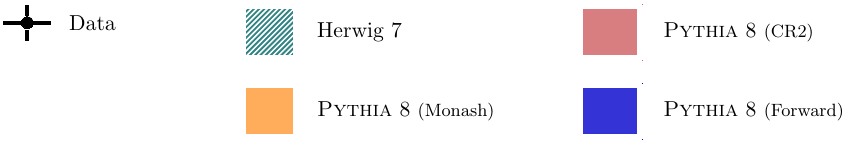} 
\hspace{0.5cm} 
\\
\includegraphics[width=0.49\textwidth]{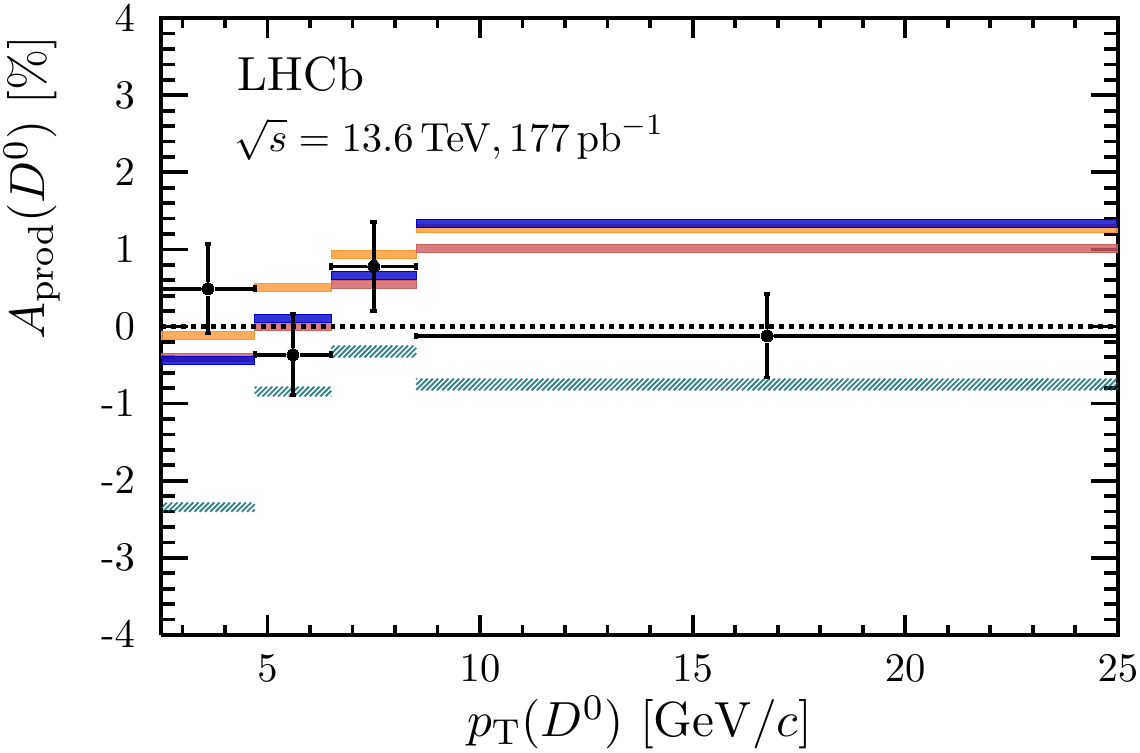}
\includegraphics[width=0.49\textwidth]{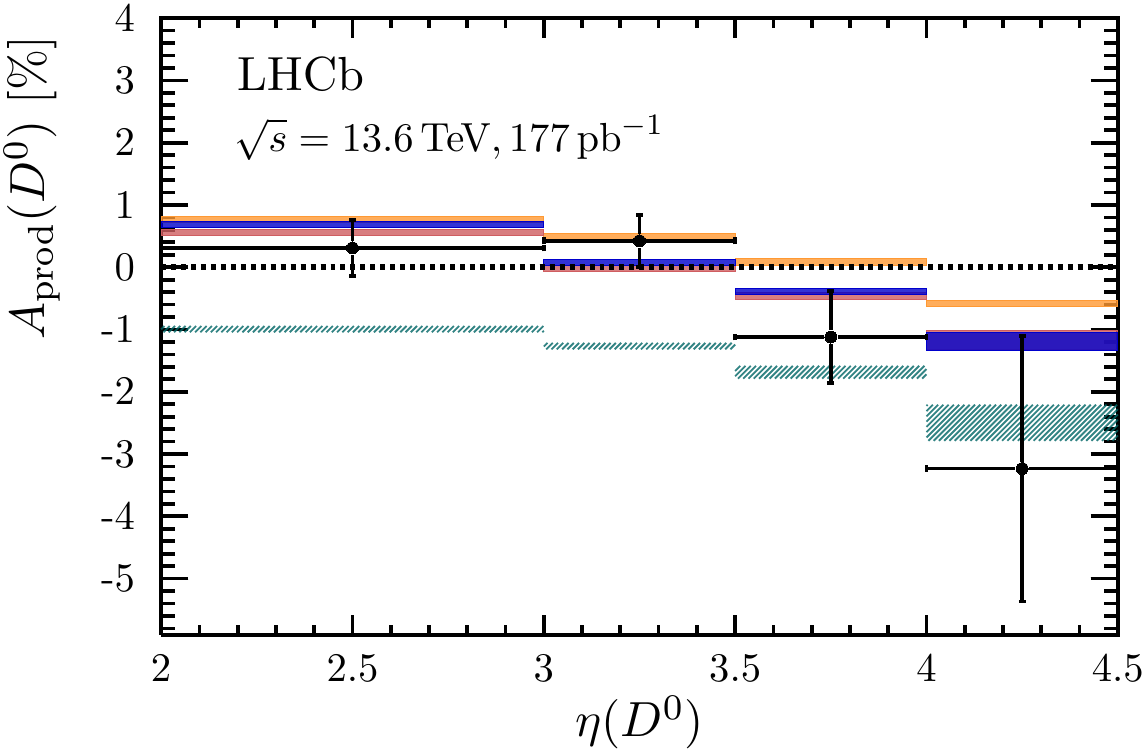} \\
\vspace{1.0em}
\includegraphics[width=0.49\textwidth]{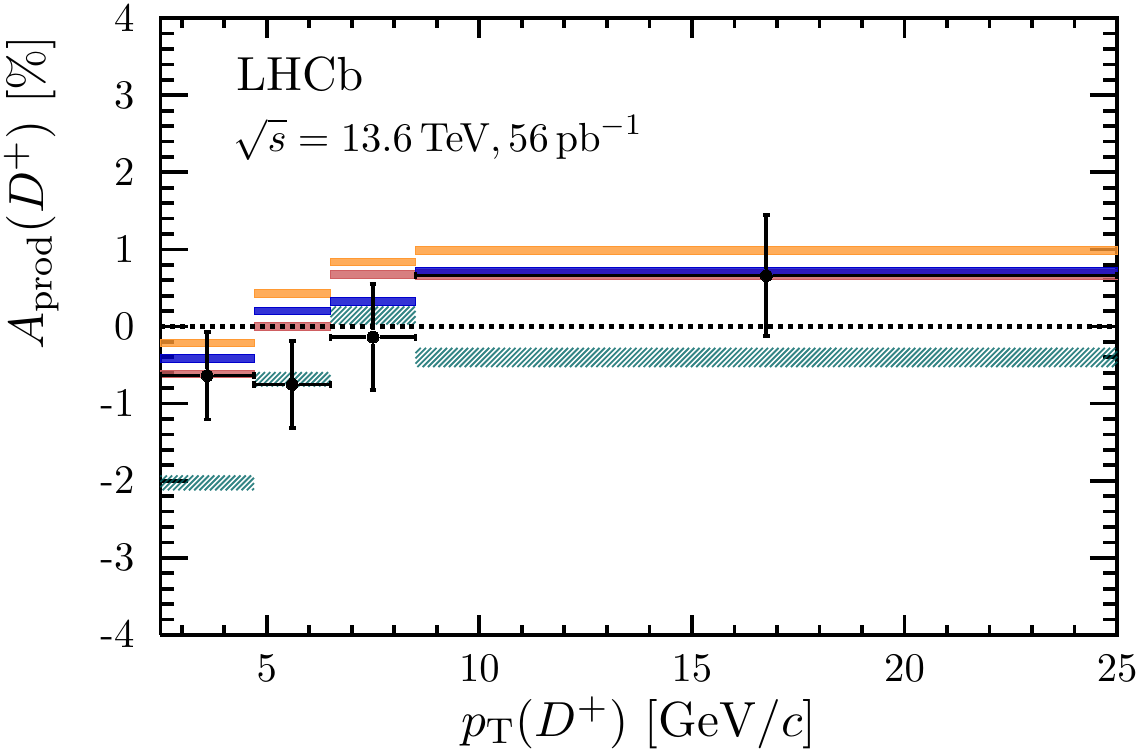}
\includegraphics[width=0.49\textwidth]{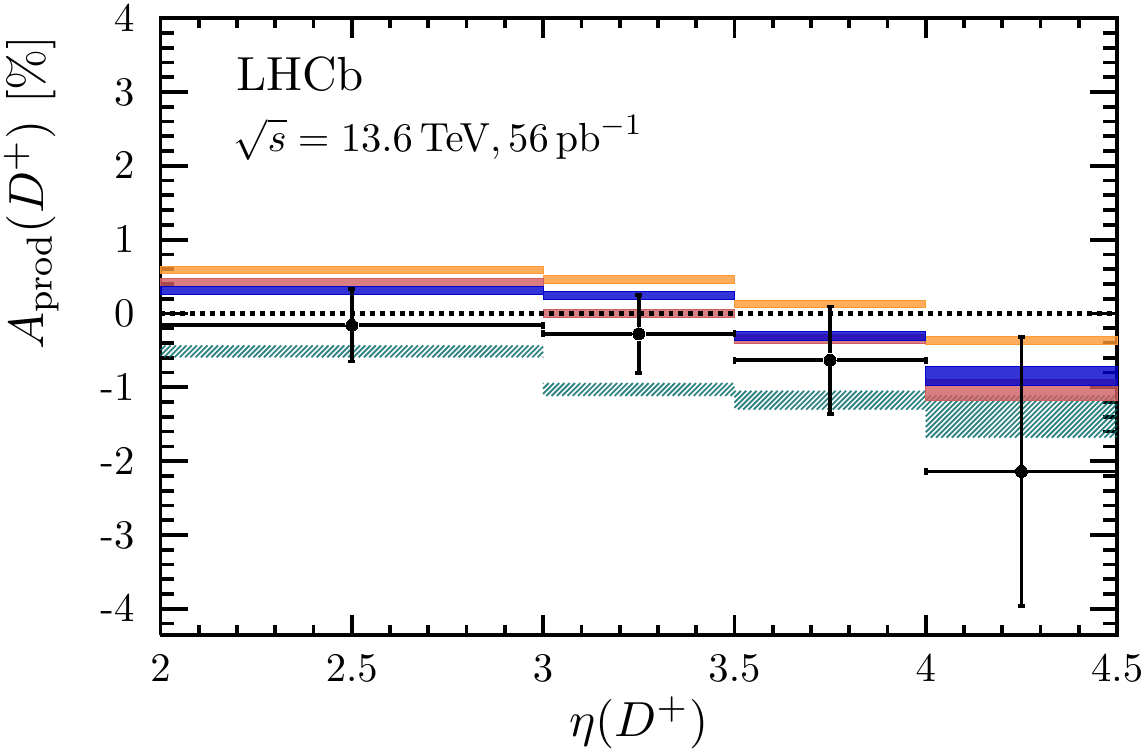} \\
\vspace{1.0em}
\includegraphics[width=0.49\textwidth]{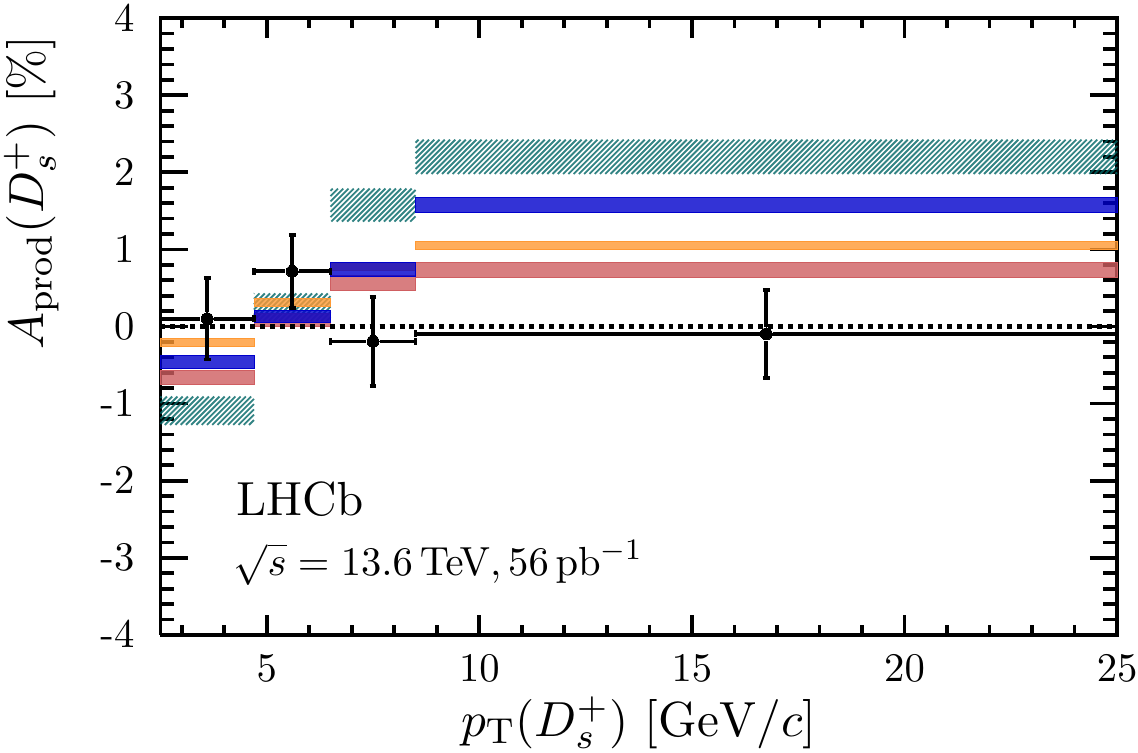}
\includegraphics[width=0.49\textwidth]{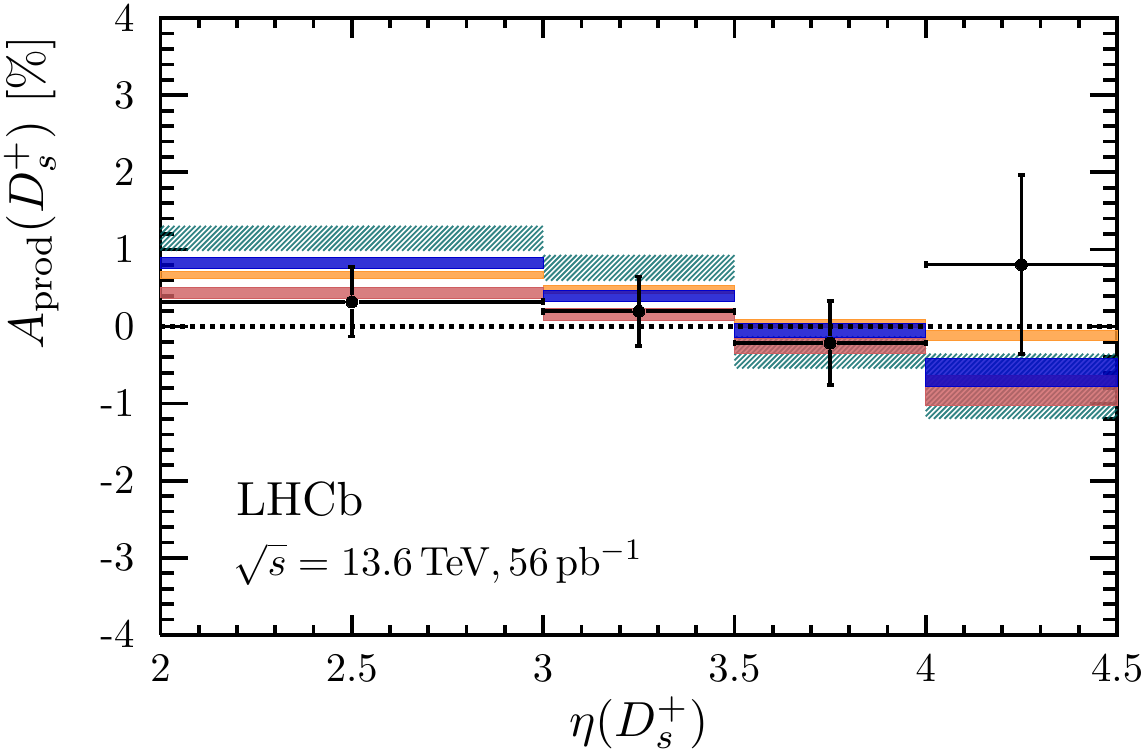} \\

\caption{Comparisons between the production asymmetries predicted with different event generators and the data. 
}\label{fig:predictions_result}
\end{figure}

Figure~\ref{fig:charm_kinematic_distributions} shows the fine-grained kinematic distributions of the charmed mesons that can be used to interpret the presented measurements. The projections of the measurements onto the $\pt$ or $\eta$ axes, shown in Fig.~\ref{fig:predictions_result}, are compared to the predictions from the event generator \mbox{\pythia8}~\cite{Bierlich:2022pfr}, based on a colour-string model, and the event generator \herwigseven~\cite{Bellm:2015jjp}, based on a cluster model. 
For \pythia~8, version 8.310 is used with three different tunings: the Monash tune~\cite{Skands:2014pea}, which is the standard, the ``QCD-inspired'' beam-remnant and colour-reconnection model (CR2)~\cite{Christiansen:2015yqa,Argyropoulos:2014zoa}, and a recent variant of the QCD-inspired model, where the beam-remnant parameters are tuned to the available LHCf data (Forward)~\cite{Fieg:2023kld} . 
The details of the parameters are presented in Appendix~\ref{appendix:pythia_parameters}. 
For \herwigseven, version 7.3.0 is used, which implements the recent developments for the decays of excited mesons and baryons~\cite{Masouminia:2023zhb, Hoang:2024zwl} and the colour-reconnection model. 
The events are generated using a minimum-bias configuration, and those with a charmed meson are retained. 
For all generators, only the charmed hadrons that are not produced in $\bquark$-hadron decays are considered. A per-candidate weighting is used to match the kinematic distributions of the charmed mesons in data, which are presented in Fig.~\ref{fig:charm_kinematic_distributions}. This step is required to account for the detector acceptance, reconstruction and selection efficiency.

Table~\ref{tab:chi2_comparison_generators} presents the compatibility $\chi^{2}$ values between the production asymmetries obtained from the event generators and the data. The predictions follow similar trends between the different charmed mesons. Similar to what was recently seen for $\bquark$-baryon data~\cite{LHCb-PAPER-2021-016}, the charmed-meson data favours the QCD-inspired colour-reconnection model from \pythia8 over the Monash tune. 
The forward-physics tuning, however, does not lead to a better description of the data at high $\pt$. 
The results from \herwigseven~show a similar trend as those from \pythia8, but are shifted to lower values, and are consistently disfavoured at low $\pt$. At higher values of $\pt$, the level of agreement between the data and these predictions improves. 
All considered event generators predict a positive production asymmetry at high $\pt$ for the $\Dsp$ meson, including \herwigseven, which is not observed by this analysis and other measurements at LHCb~\cite{LHCb-PAPER-2018-010}. 
A dedicated tuning of the hadronisation parameters, and a further study of their relation to charm-anticharm quark production are well motivated by these measurements at high $\pt$.

\begin{table}
\caption{\label{tab:chi2_comparison_generators}Compatibility $\chi^{2}$ between the different event generators and the data for the various projections. In each of these, the number of degrees of freedom is four.}
\centering
 \bgroup
 \def\arraystretch{1.25}
 \begin{tabular}{l *{6}{d{3.3}}}
 \toprule
 Generator 
 & 
 \mc{$\pt(\Dz)$} & \mc{$\eta(\Dz)$} & 
 \mc{$\pt(\Dp)$} & \mc{$\eta(\Dp)$} & 
 \mc{$\pt(\Dsp)$} & \mc{$\eta(\Dsp)$} \\
\midrule
\herwigseven             & 29.3 & 25.1 & 7.8 & 3.2 & 27.5  & 6.0  \\
\pythia8 (Monash)    & 10.5 & 5.3  & 7.1  & 6.3  & 7.8    & 1.8 \\
\pythia8 (CR2)           & 7.3   & 3.1   & 3.2 & 2.2  & 7.4    & 2.0 \\
\pythia8 (Forward)   & 10.6 & 3.2  & 3.5 & 2.6  & 13.5 & 2.9 \\
 \bottomrule
 \end{tabular}
 \egroup
 \end{table}

\section{Conclusion}
Doubly differential measurements of \Dz, \Dp and \Dsp production asymmetries in proton-proton collisions at $\sqrt{s}=13.6\tev$ are presented. %
Within the statistical precision, the measurements are consistent with the absence of production asymmetries, and with previous measurements at $\sqs=7~\text{or}~8\tev$. 
The achieved precision on the measurements of $\AP(\Dsp)$ and $\AP(\Dp)$ with the early Run 3 dataset of \lhcb, corresponding to integrated luminosities of $56\invpb$, is comparable to that obtained using the 2011 dataset, corresponding to $1 \invfb$ of integrated luminosity, highlighting the much improved efficiency of the LHCb detector in Run 3.
Comparisons to predictions from different tunes of the \herwigseven~and \pythia~8 event generators are also included. 
Their results are not always able to reproduce the observed dependence on charmed-meson $\pt$.
The presented measurements are the first to use data of the \lhcb Run 3 detector. 
While the precision achieved using the data collected in 2022 and 2023 is statistically dominated, this work demonstrates the reliability of data-driven techniques in challenging detector conditions and the potential for future asymmetry measurements with the upgraded LHCb detector during Run 3 of the LHC. 
The presented results can already serve as valuable input for further studies of $\CP$ violation in the charm sector.
 
%
%
%
%
\section*{Acknowledgements}
\noindent We express our gratitude to our colleagues in the CERN
accelerator departments for the excellent performance of the LHC. We
thank the technical and administrative staff at the LHCb
institutes.
We acknowledge support from CERN and from the national agencies:
ARC (Australia);
CAPES, CNPq, FAPERJ and FINEP (Brazil); 
MOST and NSFC (China); 
CNRS/IN2P3 (France); 
BMBF, DFG and MPG (Germany); 
INFN (Italy); 
NWO (Netherlands); 
MNiSW and NCN (Poland); 
MCID/IFA (Romania); 
MICIU and AEI (Spain);
SNSF and SER (Switzerland); 
NASU (Ukraine); 
STFC (United Kingdom); 
DOE NP and NSF (USA).
We acknowledge the computing resources that are provided by ARDC (Australia), 
CBPF (Brazil),
CERN, 
IHEP and LZU (China),
IN2P3 (France), 
KIT and DESY (Germany), 
INFN (Italy), 
SURF (Netherlands),
Polish WLCG (Poland),
IFIN-HH (Romania), 
PIC (Spain), CSCS (Switzerland), 
and GridPP (United Kingdom).
We are indebted to the communities behind the multiple open-source
software packages on which we depend.
Individual groups or members have received support from
Key Research Program of Frontier Sciences of CAS, CAS PIFI, CAS CCEPP, 
Fundamental Research Funds for the Central Universities,  and Sci.\ \& Tech.\ Program of Guangzhou (China);
Minciencias (Colombia);
EPLANET, Marie Sk\l{}odowska-Curie Actions, ERC and NextGenerationEU (European Union);
A*MIDEX, ANR, IPhU and Labex P2IO, and R\'{e}gion Auvergne-Rh\^{o}ne-Alpes (France);
Alexander-von-Humboldt Foundation (Germany);
ICSC (Italy); 
Severo Ochoa and Mar\'ia de Maeztu Units of Excellence, GVA, XuntaGal, GENCAT, InTalent-Inditex and Prog.~Atracci\'on Talento CM (Spain);
SRC (Sweden);
the Leverhulme Trust, the Royal Society and UKRI (United Kingdom).

 
%

%
%
%
%
%
%

%
\clearpage
\section*{Appendices}
\appendix

\section{PYTHIA 8 parameters}
\label{appendix:pythia_parameters}
Table~\ref{tab:generators_pythia_parameters} presents all parameter values used in the different \pythia8 configurations.

 \begin{table*}[!b]
 \centering
 \caption{Parameters for the \pythia8 event generator for the various tunes considered. When the parameter is not specified, or a dash is written, the default value is used.
 \label{tab:generators_pythia_parameters}
}
\resizebox{1\columnwidth}{!}{
 \begin{tabular}{lccc}
 \toprule
 \pythia8 parameter &  Monash~\cite{Skands:2014pea} & CR2~\cite{Christiansen:2015yqa,Argyropoulos:2014zoa} & \thead{Forward \\ Physics~\cite{Fieg:2023kld}}\\
\midrule
 \texttt{SoftQCD:nonDiffractive}       &   on & on & on \\
 \texttt{PhaseSpace:pTHatMin} [\gev]      &    0.4        & 0.4 & 0.4\\
 \texttt{Beams:eCM} [\tev]                              &13.6  & 13.6    & 13.6 \\
 \texttt{Beams:AllowMomentumSpread}                              &on  & on    & on \\
 
 \texttt{BeamRemnants:remnantMode}                              & -- & 1    & 1\\
 \texttt{BeamRemnants:dampPopcorn}                              & -- & --    & 0\\
 \texttt{BeamRemnants:hardRemnantBaryon}                              & -- & --    & on\\
 \texttt{BeamRemnants:aRemnantBaryon}                              & -- & --    & 0.36\\
 \texttt{BeamRemnants:bRemnantBaryon}                              & -- & --    & 1.69\\
 \texttt{BeamRemnants:primordialKTsoft}                              & -- & --    & 0.58\\
 \texttt{BeamRemnants:primordialKThard}                              & -- & --    & 1.8\\
 \texttt{BeamRemnants:halfScaleForKT}                              & -- & --    & 10\\
 \texttt{BeamRemnants:halfMassForKT}                              & -- & --    & 1\\
 \texttt{BeamRemnants:primordialKTremnant}                 & -- & --    & 0.58\\
 \texttt{MultiPartonInteractions:pT0Ref}                       & --  & 2.15  & 2.15\\

 \texttt{ColourReconnection:mode}                    &0	& 1   & 1  \\ 
 \texttt{ColourReconnection:allowDoubleJunRem}                       & --  & off  & off\\
 \texttt{ColourReconnection:allowJunctions}                       & --  & on  & on\\
 \texttt{ColourReconnection:junctionCorrection}                       & --  & 1.2  & 1.2 \\
 \texttt{ColourReconnection:m0}                       & --  & 0.3  & 0.3 \\

  \texttt{StringZ:aLund}                       & --  & 0.36  & 0.36\\
  \texttt{StringZ:bLund}                       & --  & 0.56  & 0.56\\

 \texttt{StringFlav:probQQtoQ}                       & --  & 0.078  & 0.078\\
 \texttt{StringFlav:probStoUD}                       & --  & 0.2 & 0.2\\
 \texttt{StringFlav:probQQ1toQQ0join}                       & --  & 0.0275  & 0.0275\\
 \bottomrule
 \end{tabular}
}
\end{table*}
\clearpage

\section{Invariant-mass distributions per dataset}
\label{appendix:mass_plots}
\begin{figure}[b]
\centering
\includegraphics[width=0.49\textwidth]{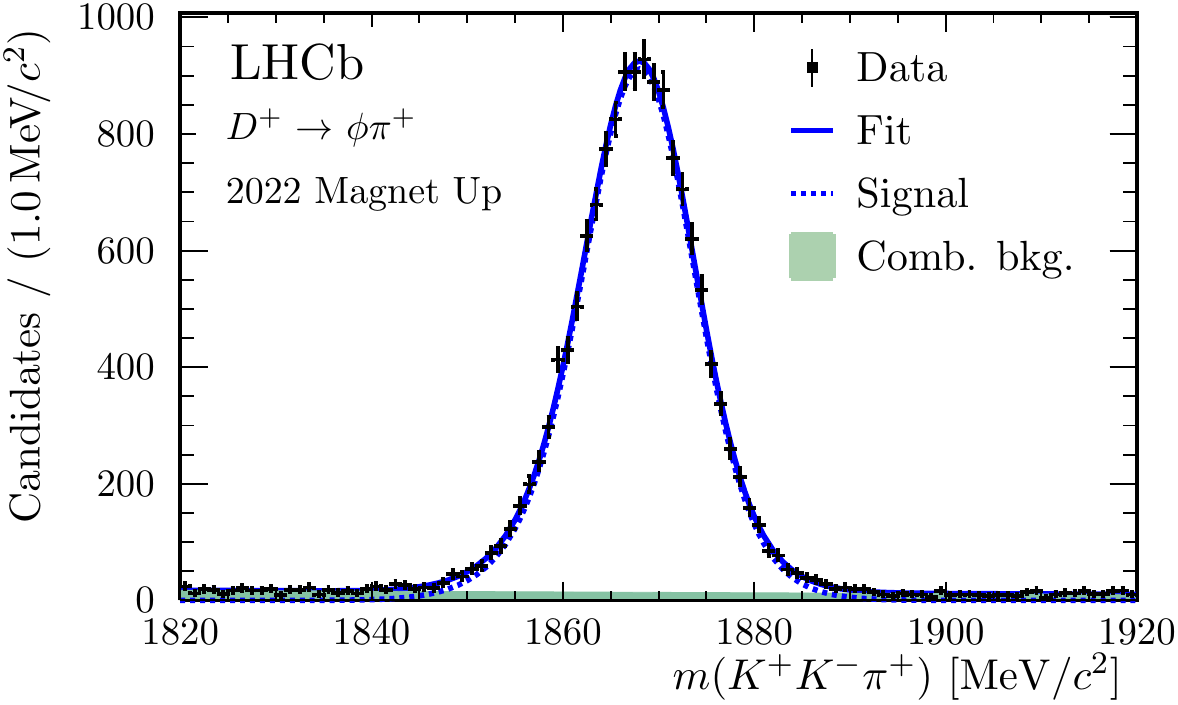}
\includegraphics[width=0.49\textwidth]{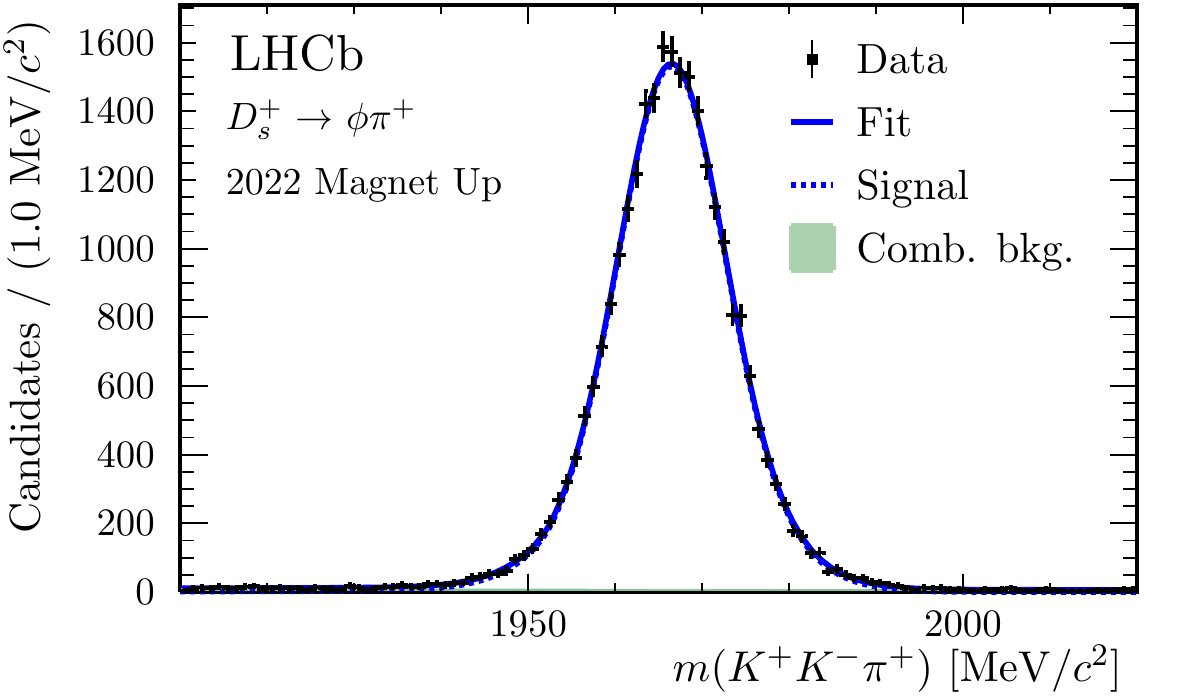}\\

\includegraphics[width=0.49\textwidth]{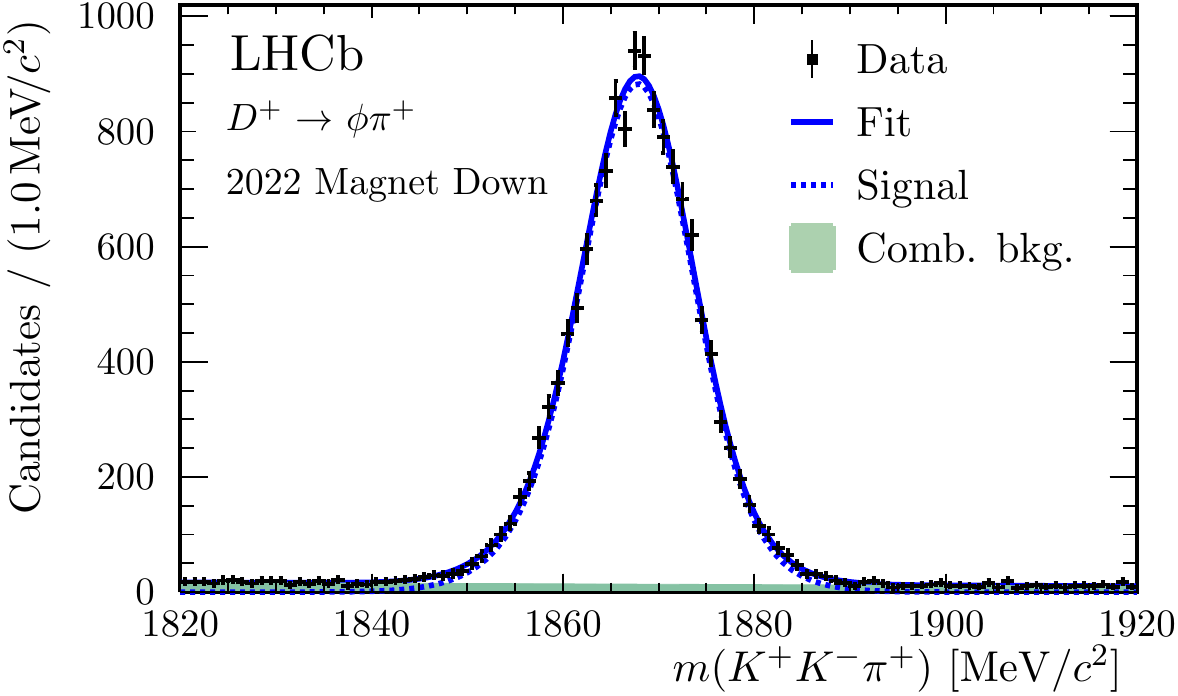}
\includegraphics[width=0.49\textwidth]{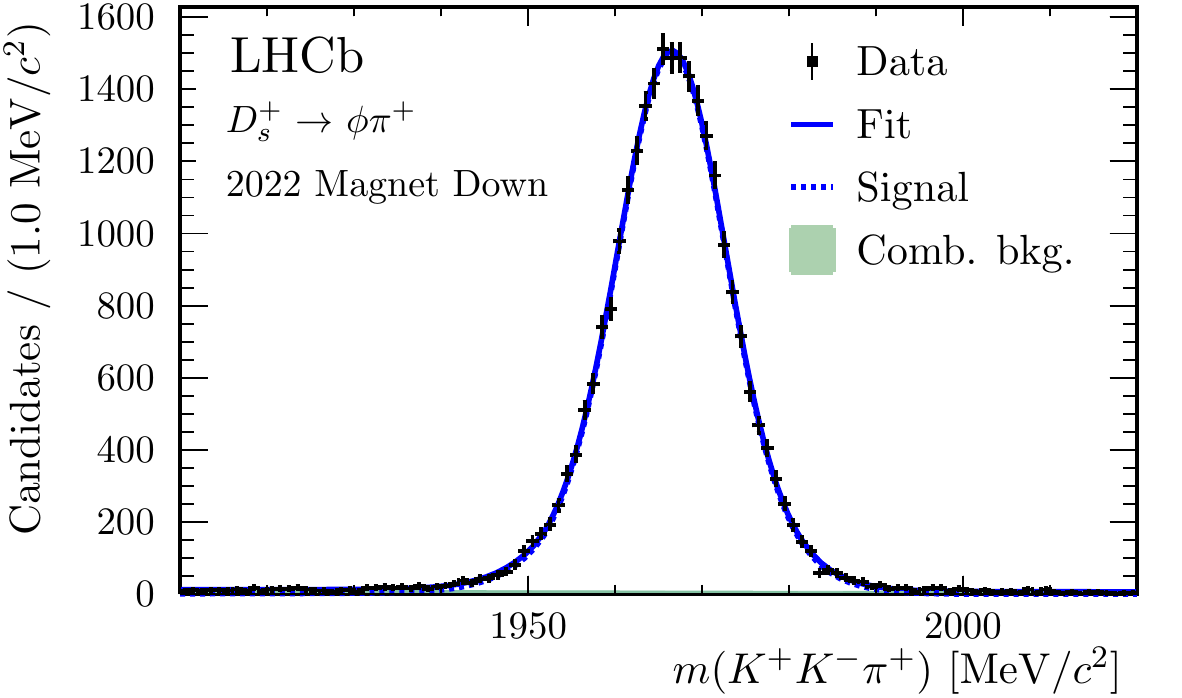} \\

\includegraphics[width=0.49\textwidth]{figs/Fig1a.pdf}
\includegraphics[width=0.49\textwidth]{figs/Fig1b.pdf} \\

\caption{Invariant-mass distributions of $\Kp\Km\pip$ combinations, $m(\Kp\Km\pip)$, for positively charged (left) $\Dp$ and (right) $\Dsp$ candidates for (top) 2022 \MagUp, (centre) 2022 \MagDown, and (bottom) 2023 data.
}\label{fig:appendix_selection_dsp_dp_mass_peaks_all_years}
\end{figure}

\begin{figure}[]
\centering
\includegraphics[width=0.49\textwidth]{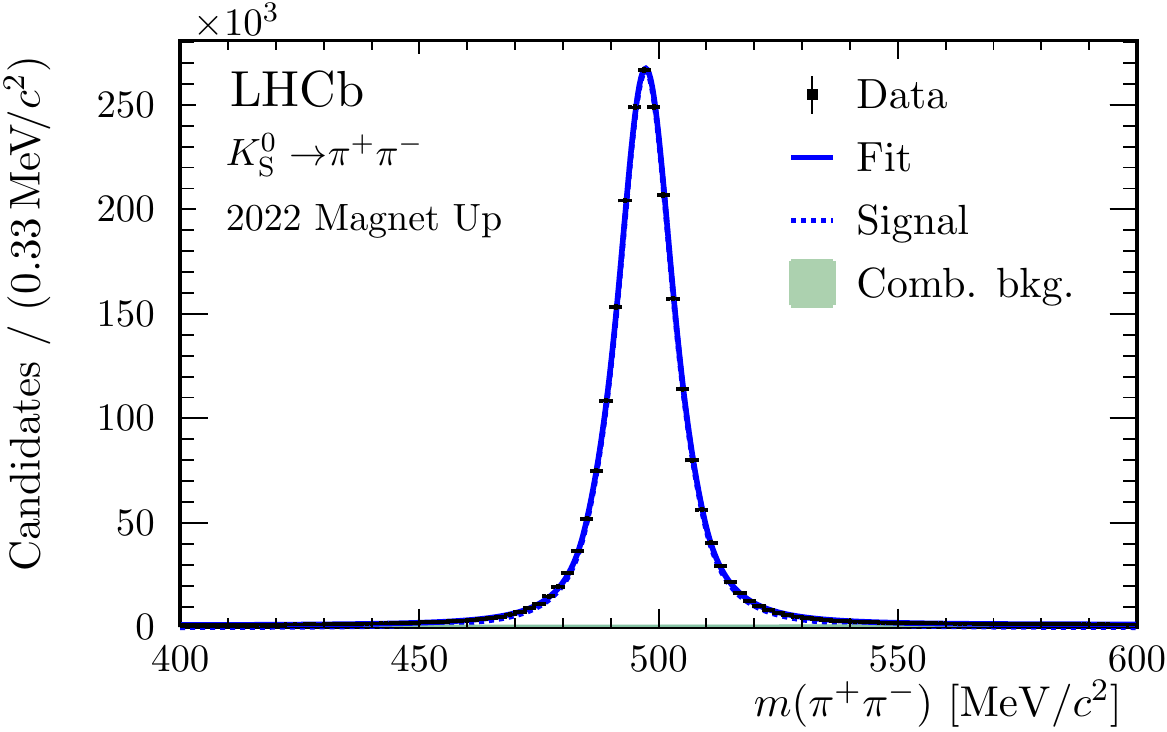}
\includegraphics[width=0.49\textwidth]{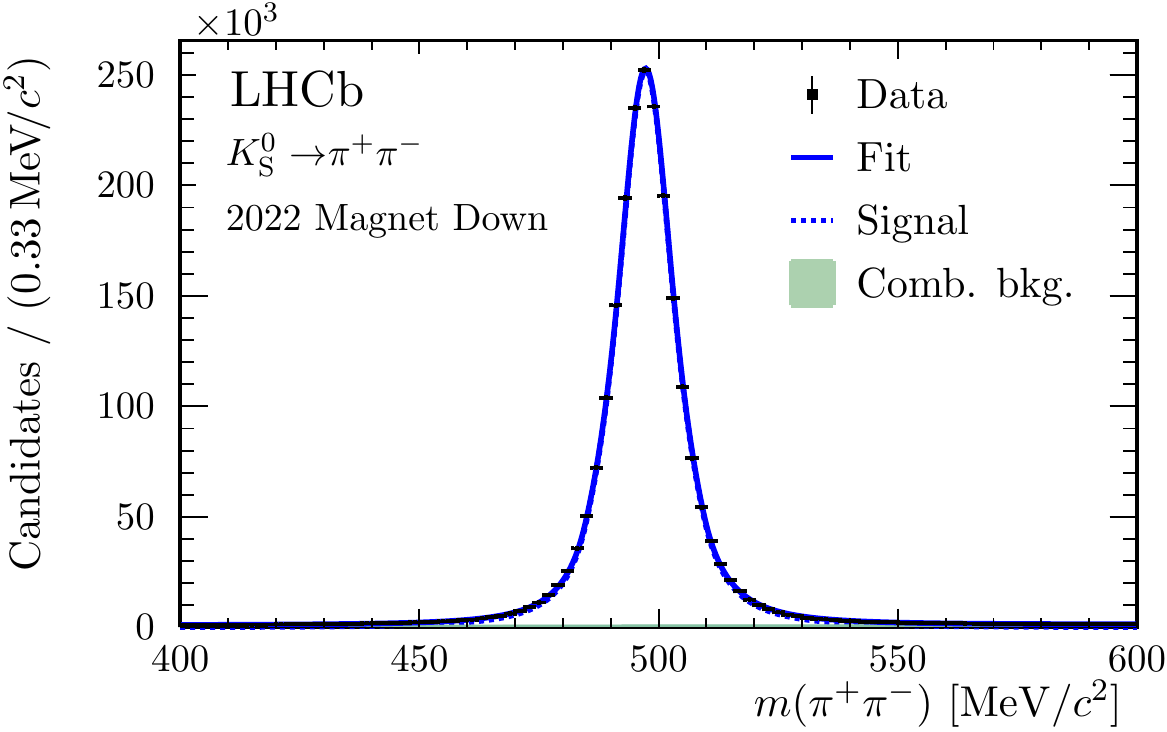}\\
\includegraphics[width=0.49\textwidth]{figs/Fig2.pdf}

\caption{Invariant-mass distributions of $\pip\pim$ combinations, $m(\pip\pim)$, for the $\KS\to\pip\pim$ candidates for (top left) 2022 \MagUp, (top right) 2022 \MagDown, and (bottom) 2023 data.
}\label{fig:appendix_selection_ks_mass_peaks_all_years}
\end{figure}

Figure~\ref{fig:appendix_selection_dsp_dp_mass_peaks_all_years} presents the invariant-mass distributions of the $\Kp\Km\pip$ combination for the $\Dsp$ and $\Dp$ candidates, per dataset. Figure~\ref{fig:appendix_selection_ks_mass_peaks_all_years} presents the invariant-mass distributions of $\pip\pim$ pairs for the $\KS$ candidates, used to estimate $\arec(\pip)$, for  different datasets. 

\begin{figure}[h]
\centering
\includegraphics[width=0.49\textwidth]{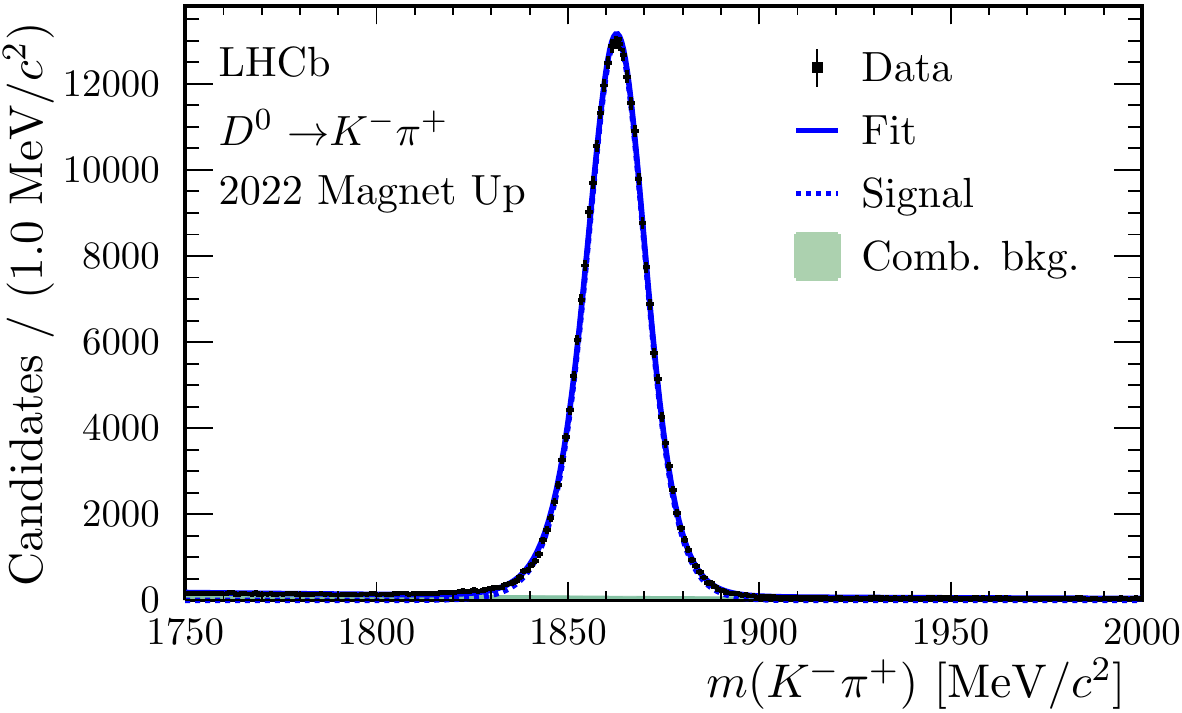}
\includegraphics[width=0.49\textwidth]{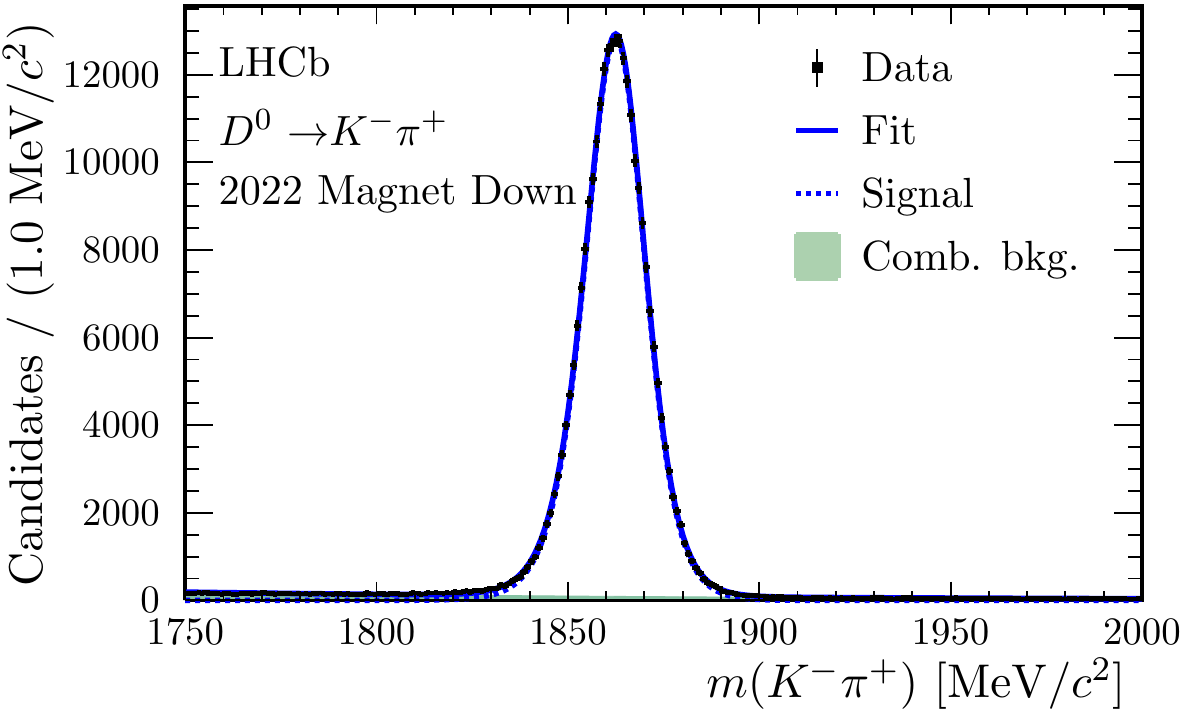}\\

\includegraphics[width=0.49\textwidth]{figs/Fig3.pdf}\\

\caption{Invariant-mass distributions of $\Km\pip$ combinations, $m(\Km\pip)$, for $\Dz$ candidates in the (top-left) 2022 \MagUp, (top-right) 2022 \MagDown, and (bottom) 2023 data.  Data from the charge-conjugated $\Dzb$ decays are not included in this figure.
}\label{fig:appendix_selection_dz_mass_peaks_all_years}
\end{figure}

\begin{figure}[h]
\centering
\includegraphics[width=0.49\textwidth]{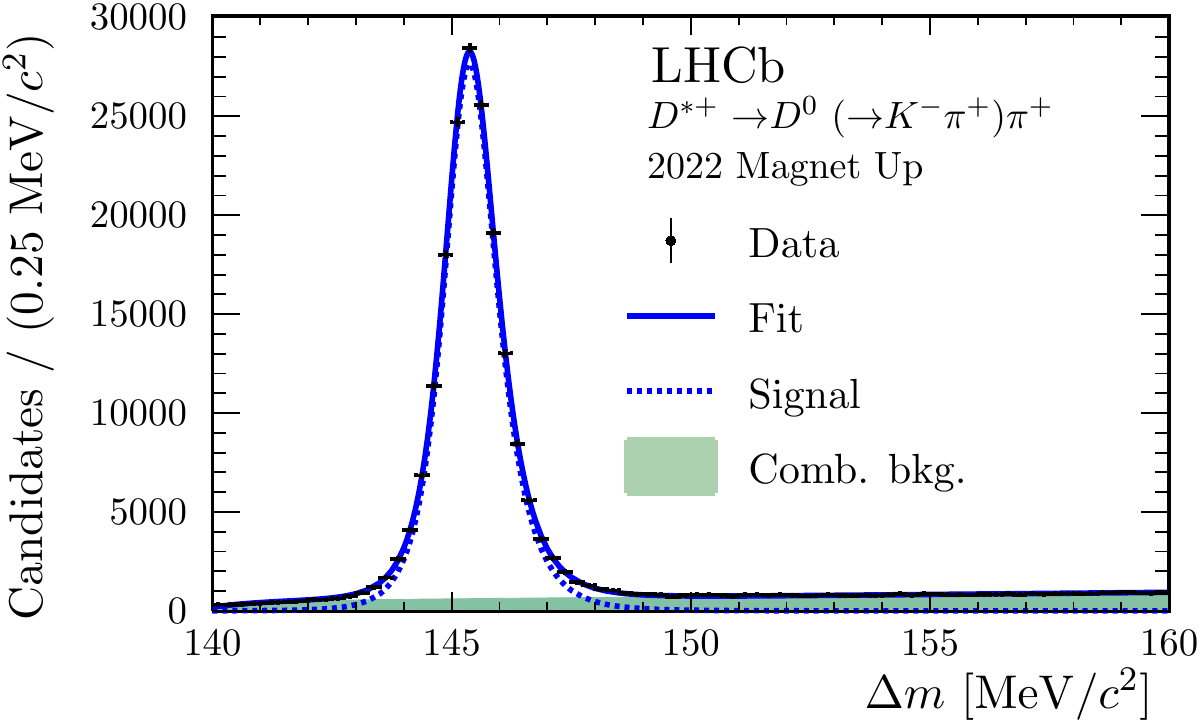}
\includegraphics[width=0.49\textwidth]{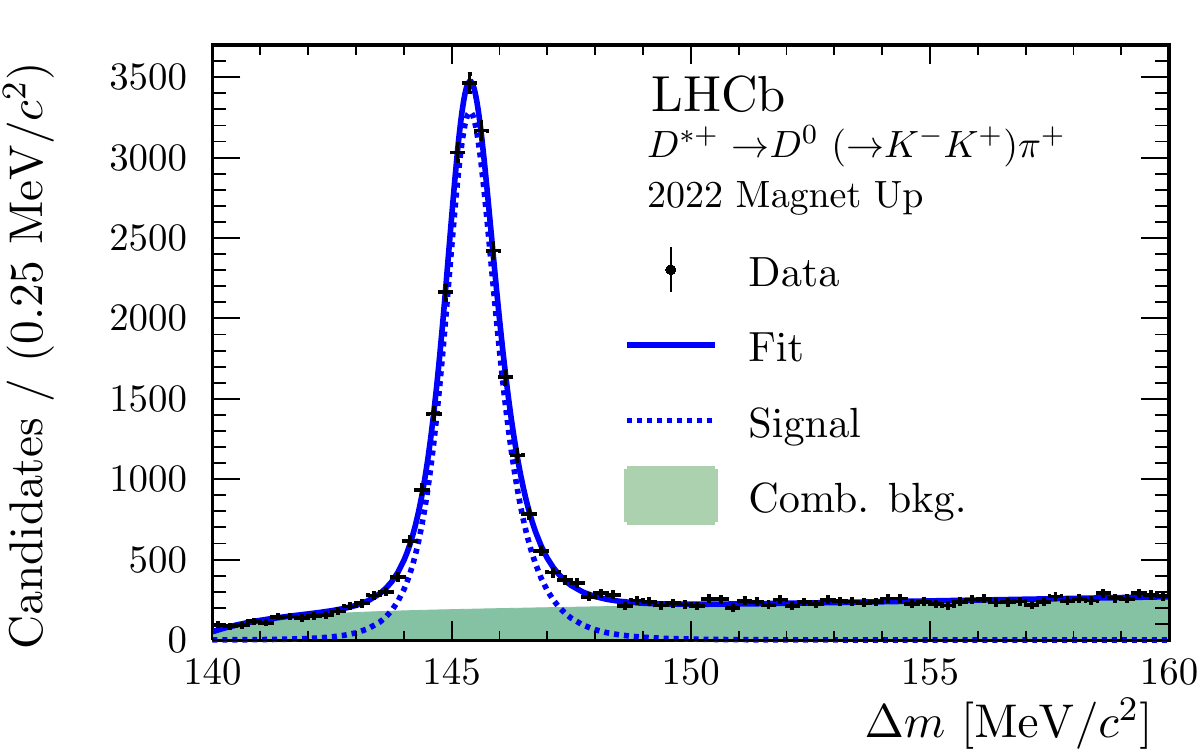}\\

\includegraphics[width=0.49\textwidth]{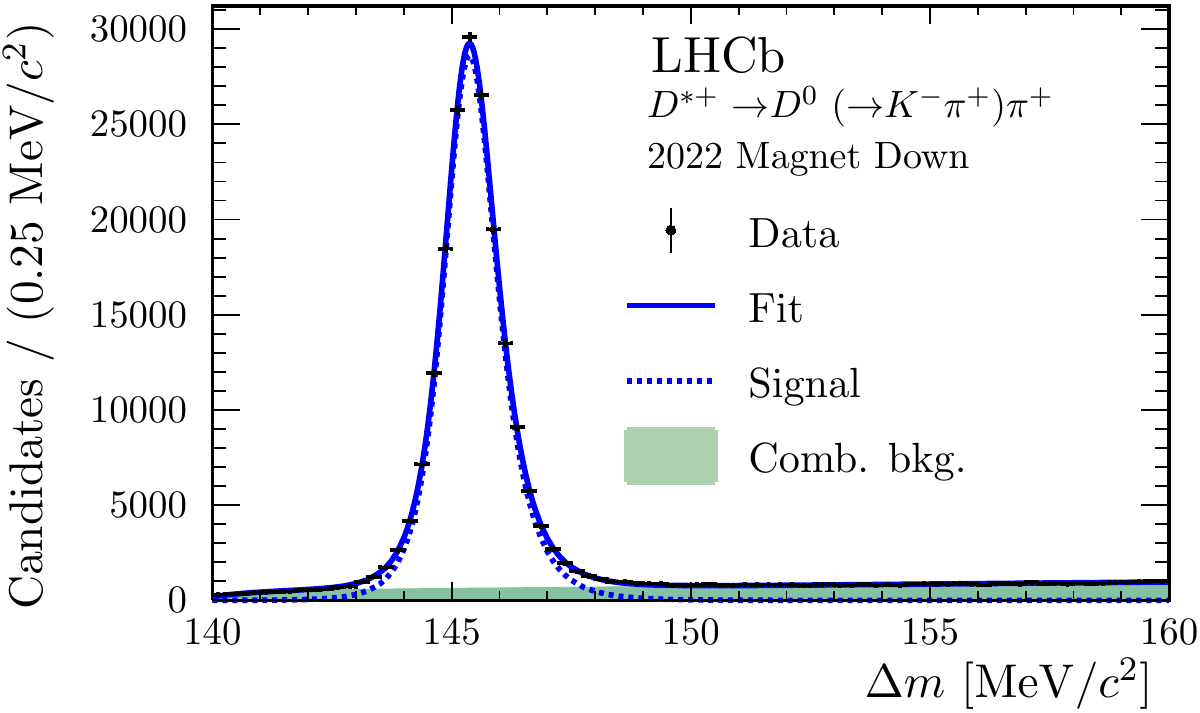}
\includegraphics[width=0.49\textwidth]{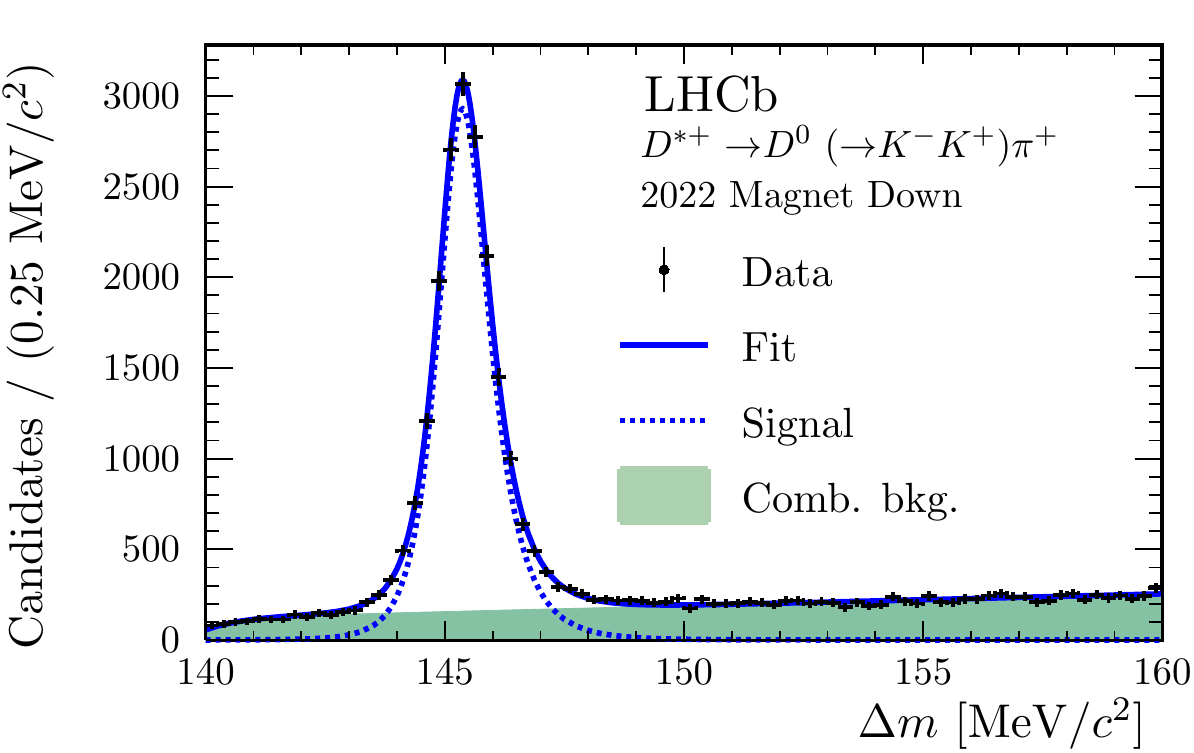}\\

\includegraphics[width=0.49\textwidth]{figs/Fig4a.pdf}
\includegraphics[width=0.49\textwidth]{figs/Fig4b.pdf}\\

\caption{Distributions of the difference in invariant mass of the reconstructed \Dstarp and $\Dz$ candidates ($\Delta m$), for candidates from the positively charged (left) $\Dstarp\to\Dz(\to\Km\pip)\pip$ and (right) $\Dstarp\to\Dz(\to\Kp\Km)\pip$ decays, and for different data samples.
}\label{fig:appendix_selection_dstar_mass_peaks_all_years}
\end{figure}

The $m(\Km\pip)$ distributions for the $\Dz\to\Km\pip$ candidates are presented in Fig.~\ref{fig:appendix_selection_dz_mass_peaks_all_years} for  different datasets. Similarly, Fig.~\ref{fig:appendix_selection_dstar_mass_peaks_all_years} presents the $\Delta m$ distributions of the $\Dstarp$ decays, used as control modes to measure the $\Dz$-meson production asymmetry, for different datasets.

\clearpage
\section{Production asymmetries in different subsamples of the datasets}
\label{appendix:crosscheck_consistency}
\begin{figure}[h]
    \centering
    \includegraphics[width=0.5\linewidth]{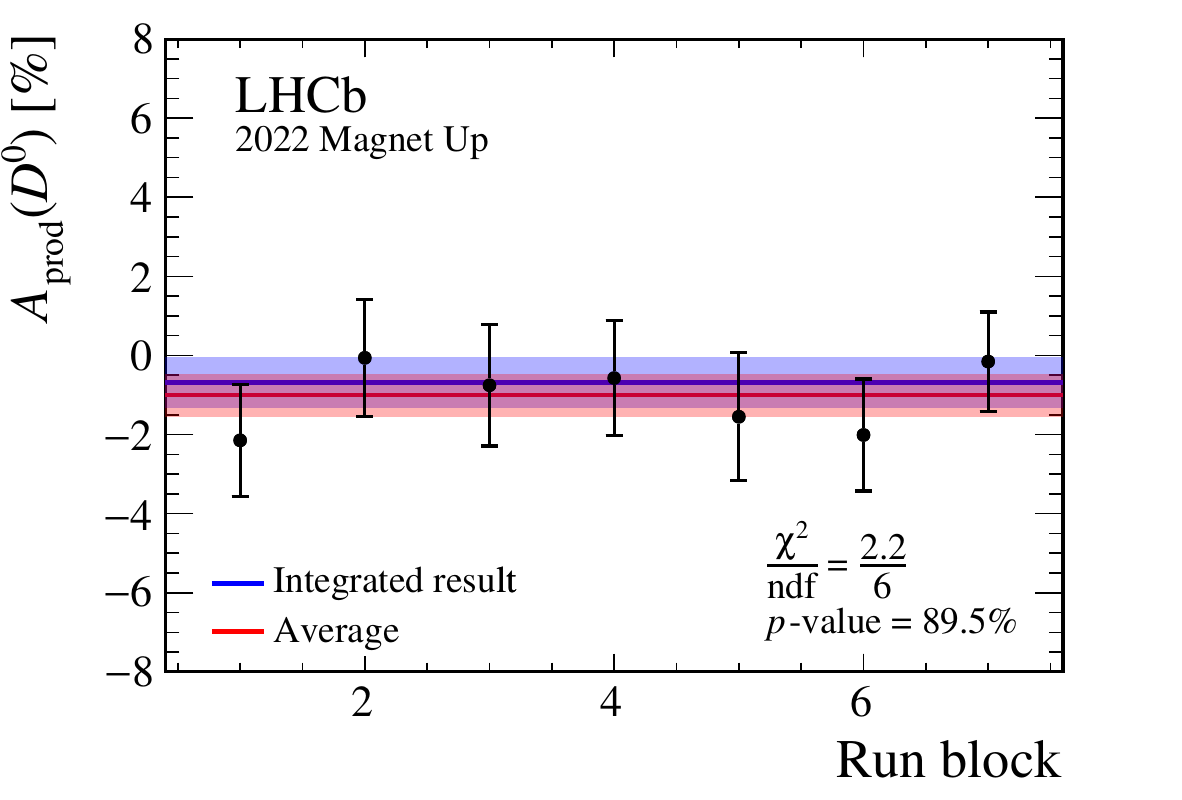}\hfill
    \includegraphics[width=0.5\textwidth]{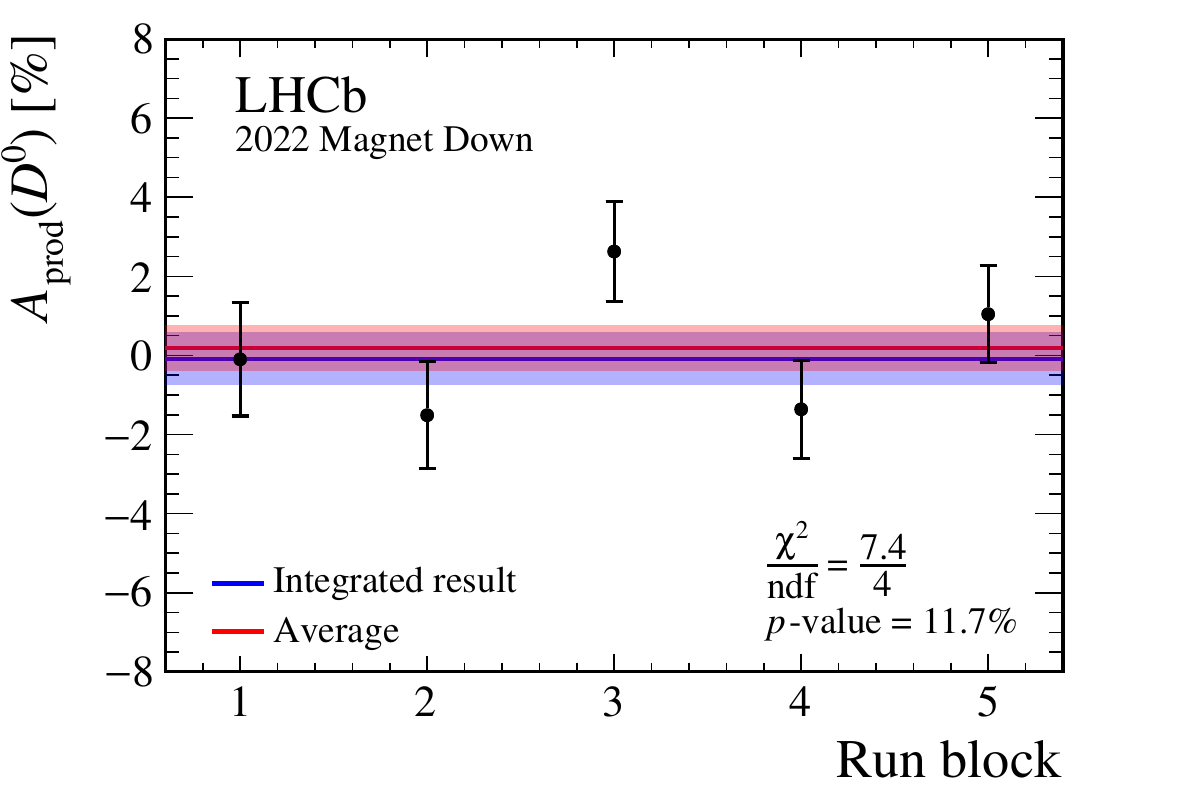}\\
    \includegraphics[width=0.5\textwidth]{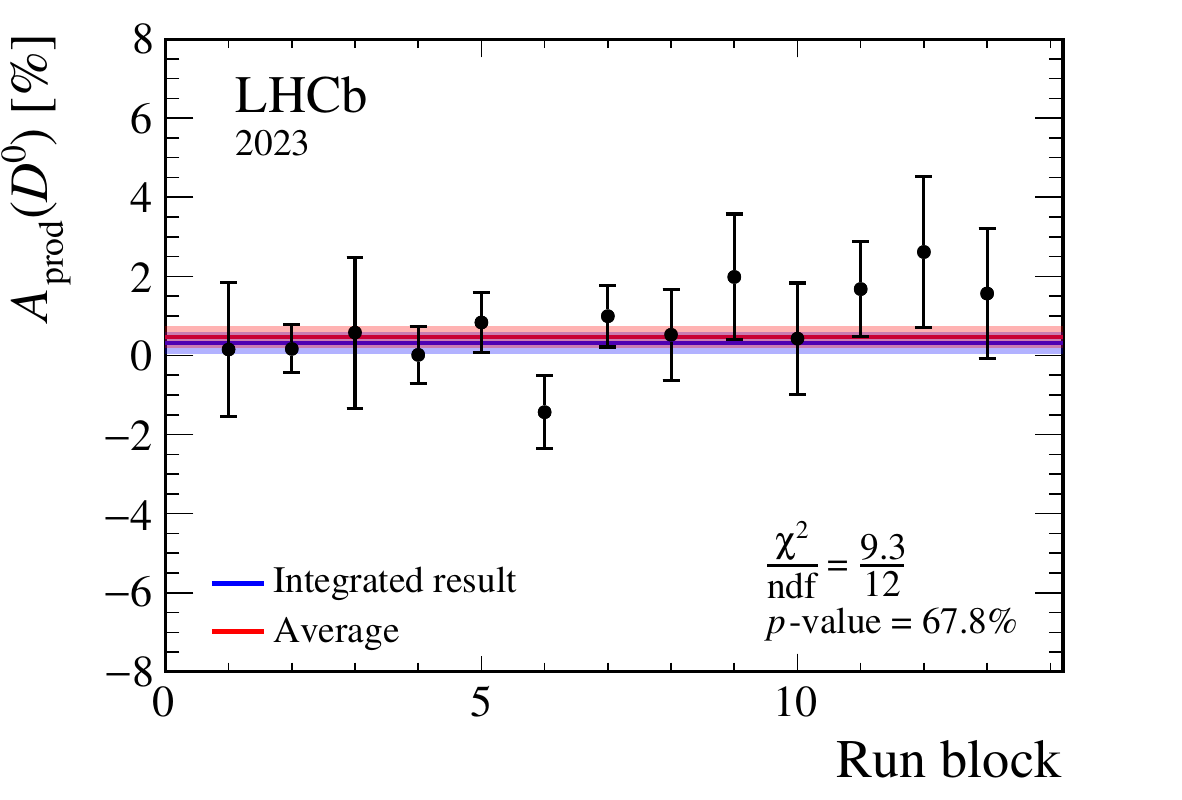}
    \caption{Comparison of the \Dz production asymmetry in different subsamples (Run blocks) of the different datasets. The check for 2022 (left) \MagUp and (right) \MagDown are shown in the top row, while the check for 2023 data is presented in the bottom row. The plots show the average computed among the subsample's measurements together with compatibility $\chi^2$, degrees of freedom and $p$-values and also the integrated results obtained in each subsample with the baseline strategy. The coloured bands represent the uncertainties of the integrated results and the averages.}
    \label{fig:RunBlock_D0}
\end{figure}

\begin{figure}[h]
    \centering
    \includegraphics[width=0.5\linewidth]{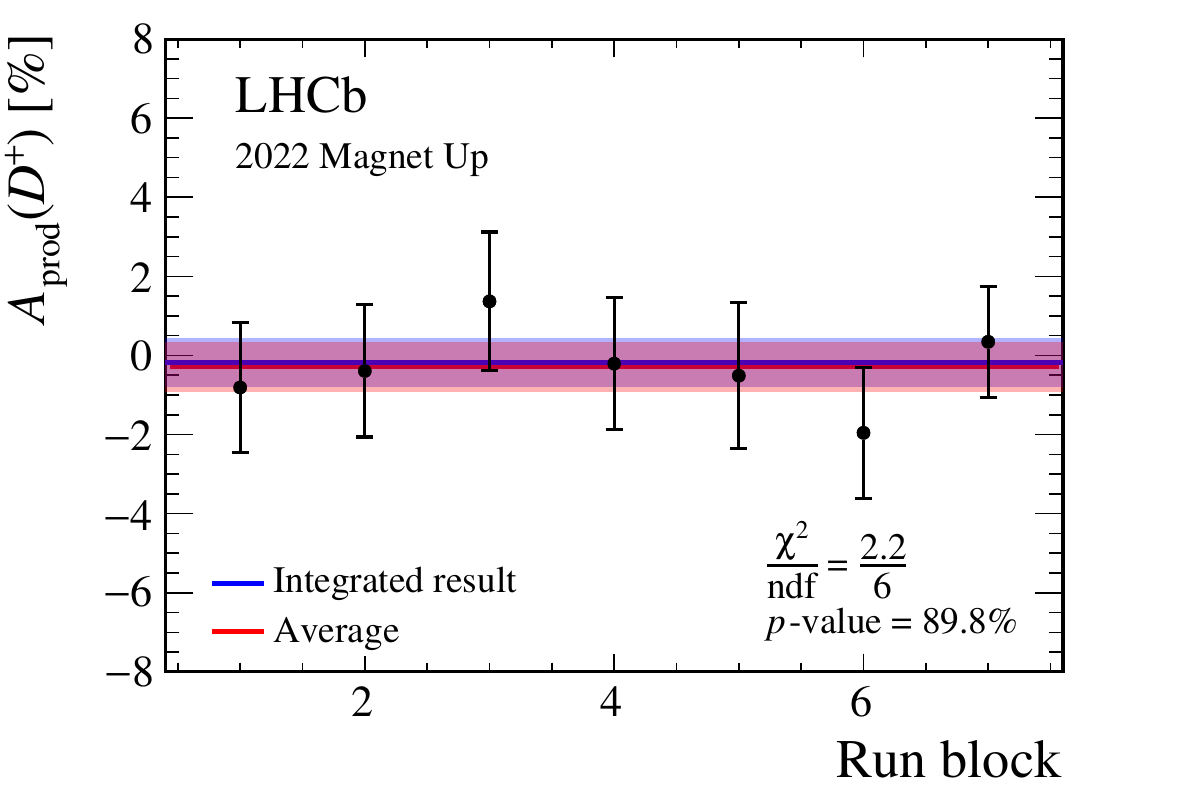}\hfill
    \includegraphics[width=0.5\textwidth]{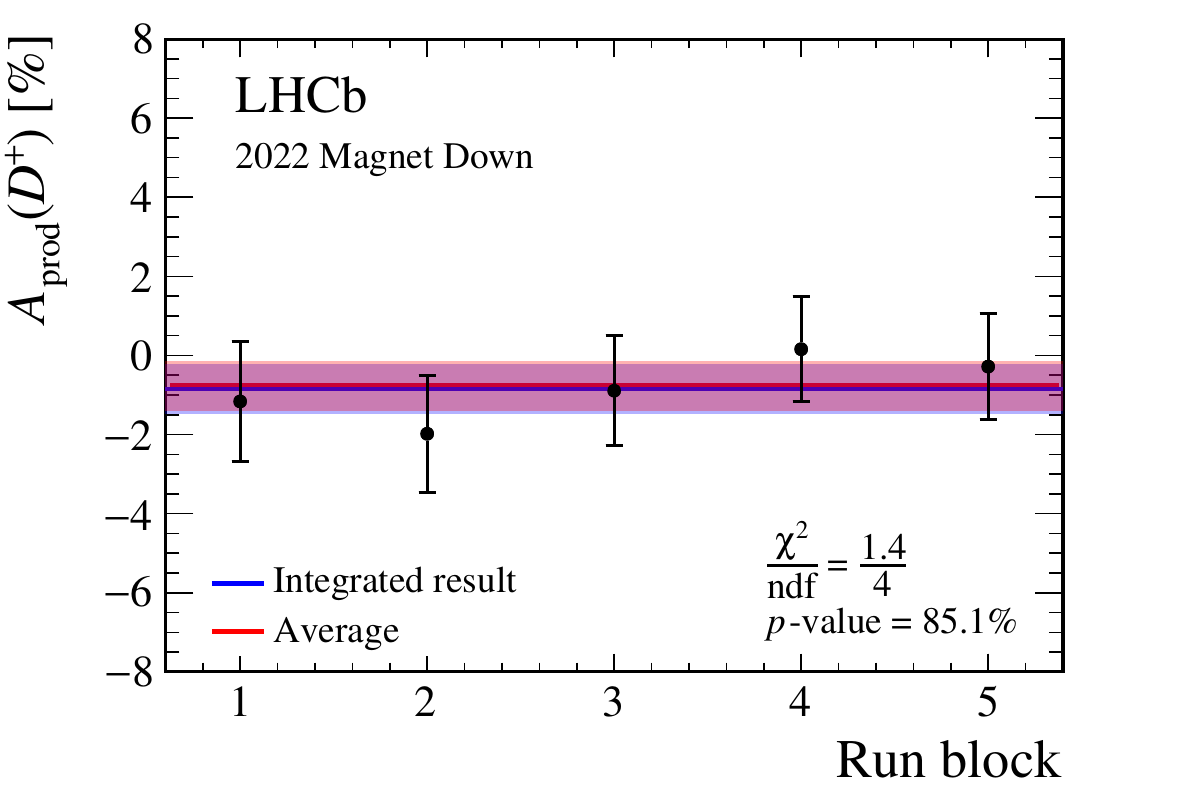}\\
    \includegraphics[width=0.5\textwidth]{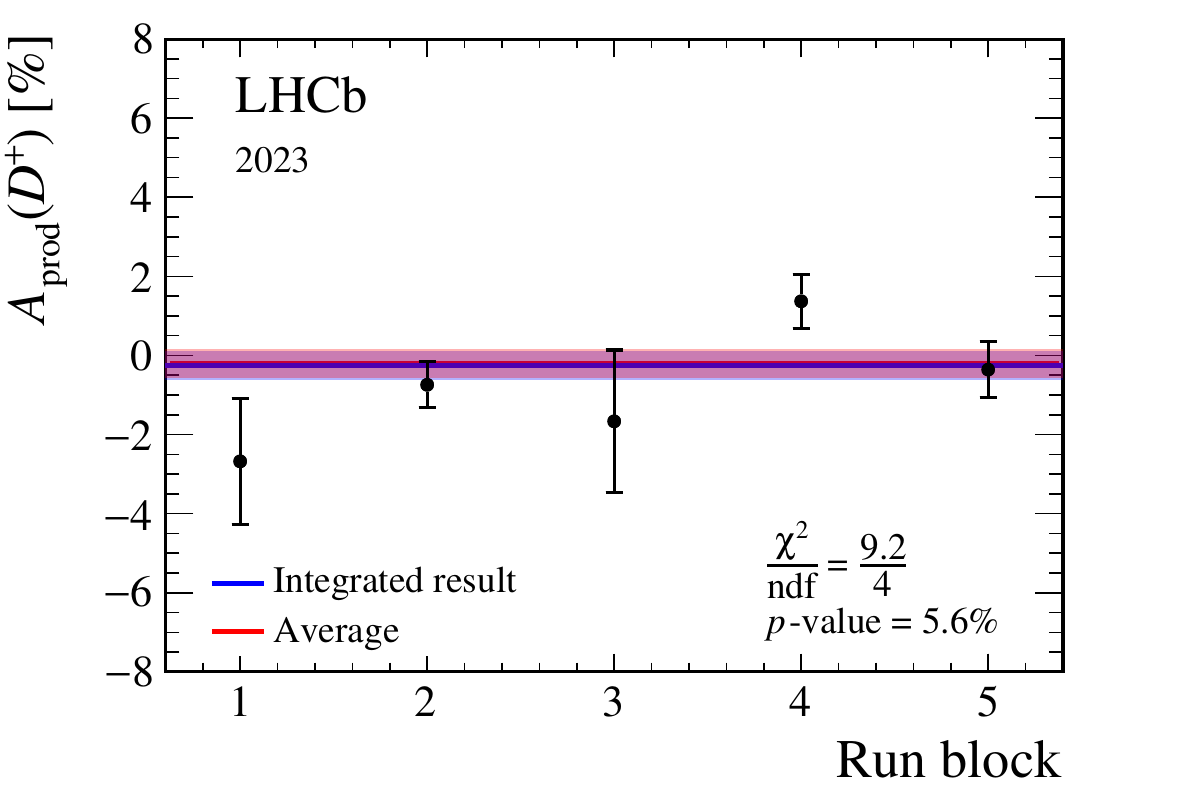}
    \caption{Comparison of the \Dp production asymmetry in different subsamples (Run blocks) of the different datasets. The check for 2022 (left) \MagUp and (right) \MagDown are shown in the top row, while the check for 2023 data is presented in the bottom row. The plots show the average computed among the subsample's measurements together with compatibility $\chi^2$, degrees of freedom and $p$-values and also the integrated results obtained in each subsample with the baseline strategy. The coloured bands represent the uncertainties of the integrated results and the averages.}
    \label{fig:RunBlock_Dp}
\end{figure}

\begin{figure}[h]
    \centering
    \includegraphics[width=0.5\linewidth]{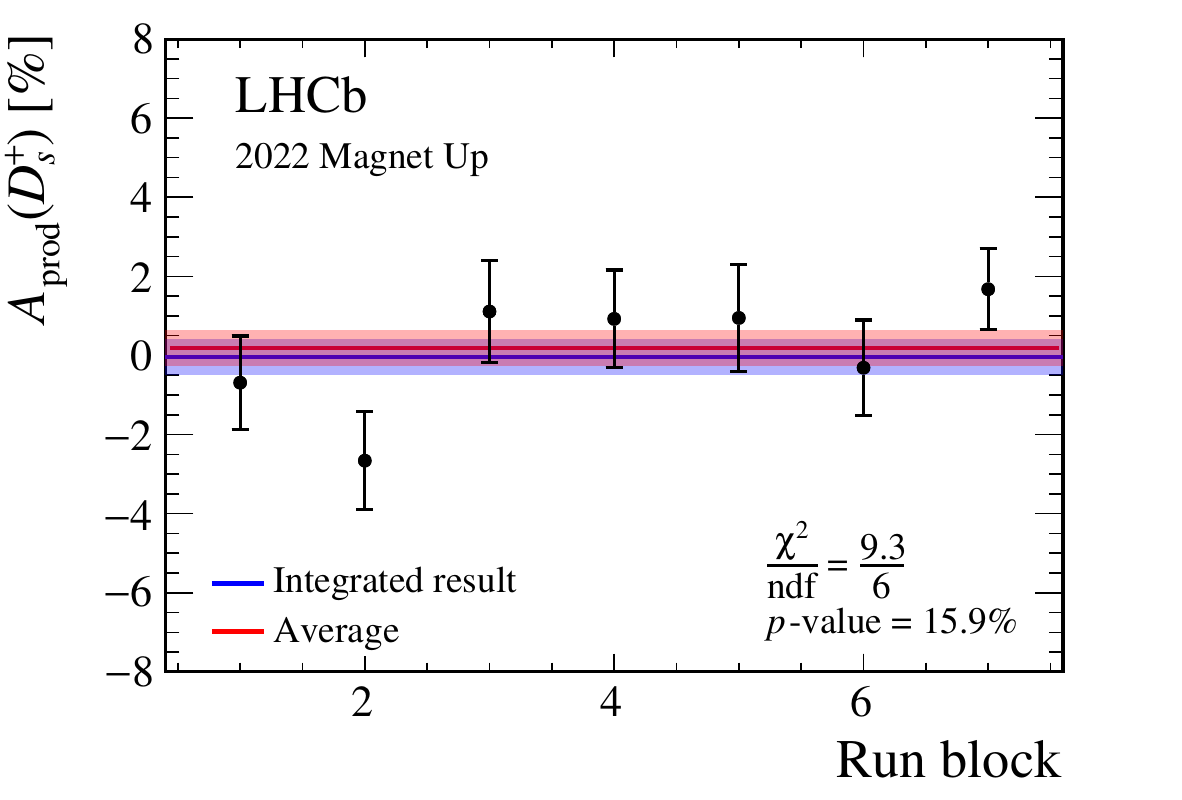}\hfill
    \includegraphics[width=0.5\textwidth]{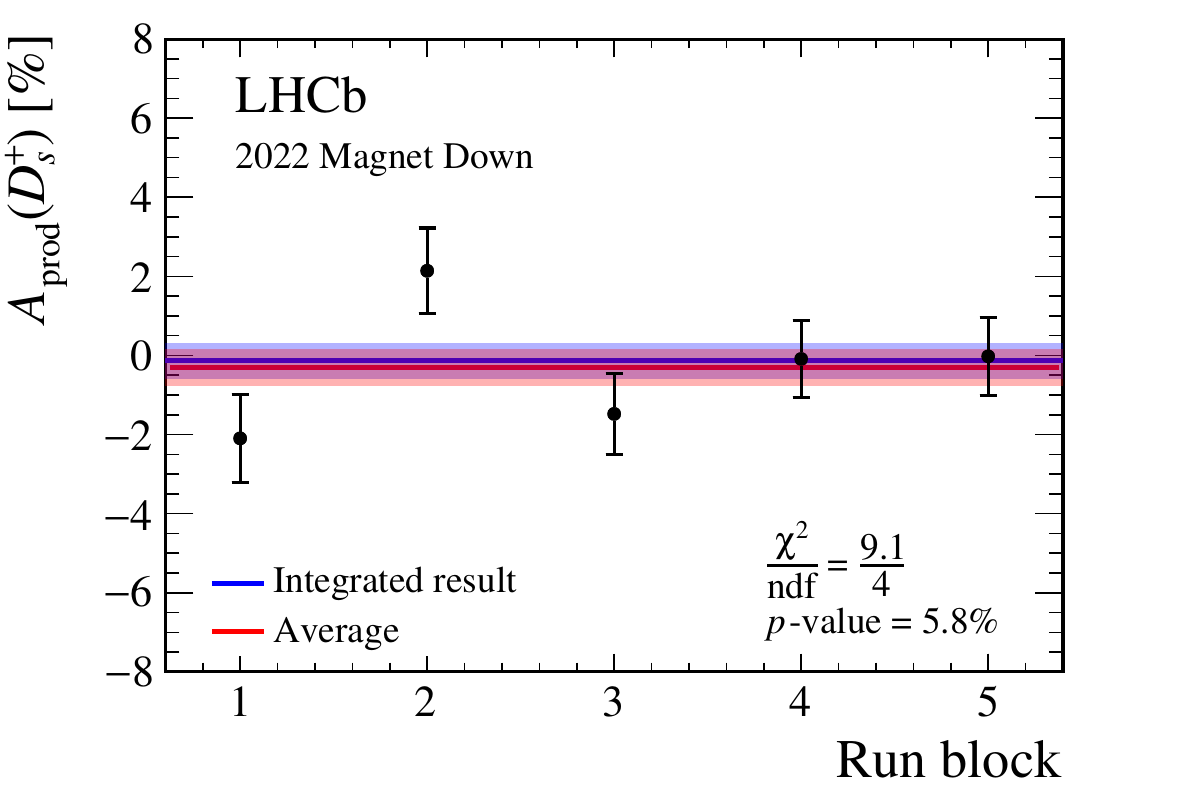}\\
    \includegraphics[width=0.5\textwidth]{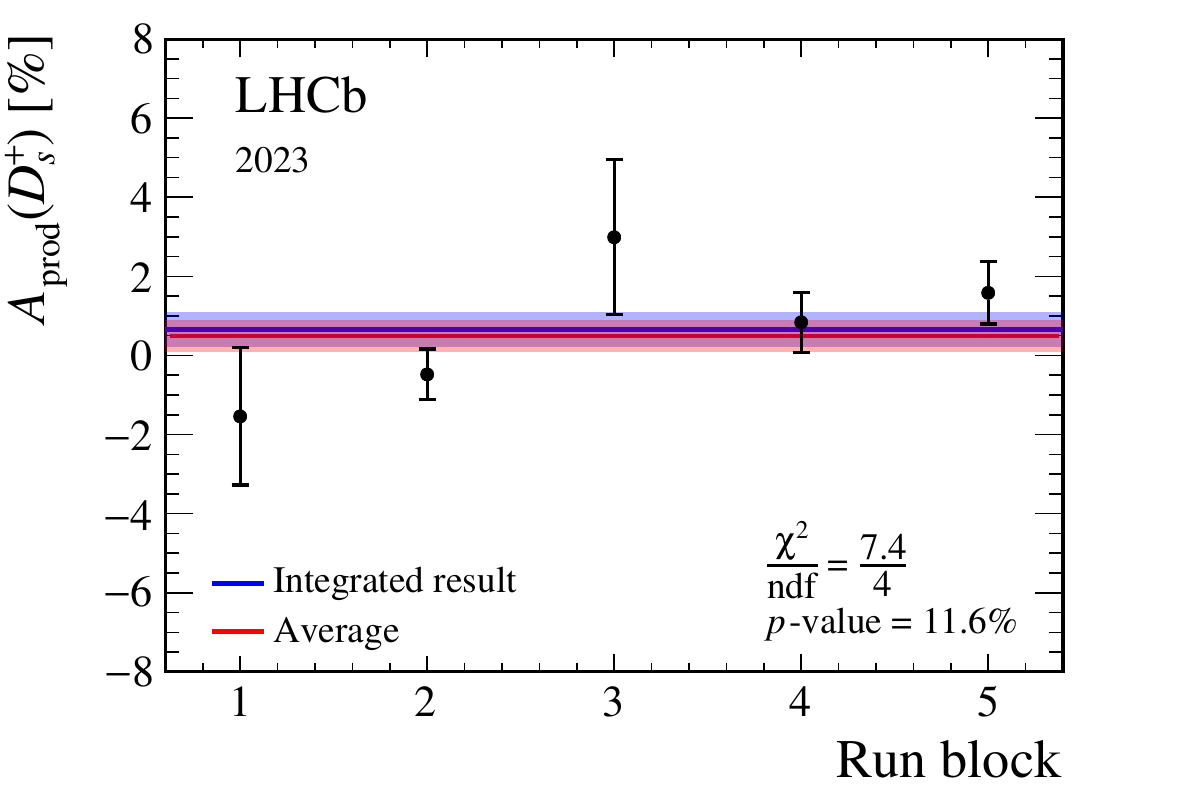}
    \caption{Comparison of the \Dsp production asymmetry in different subsamples (Run blocks) of the different datasets. The check for 2022 (left) \MagUp and (right) \MagDown are shown in the top row, while the check for 2023 data is presented in the bottom row. The plots also show the average values and the integrated results, corresponding to the results obtained in each subsample with the baseline strategy. The coloured bands represent the uncertainties of the integrated results and the averages.}
    \label{fig:RunBlock_Ds}
\end{figure}

\begin{figure}[h]
    \centering
    \includegraphics[width=\linewidth]{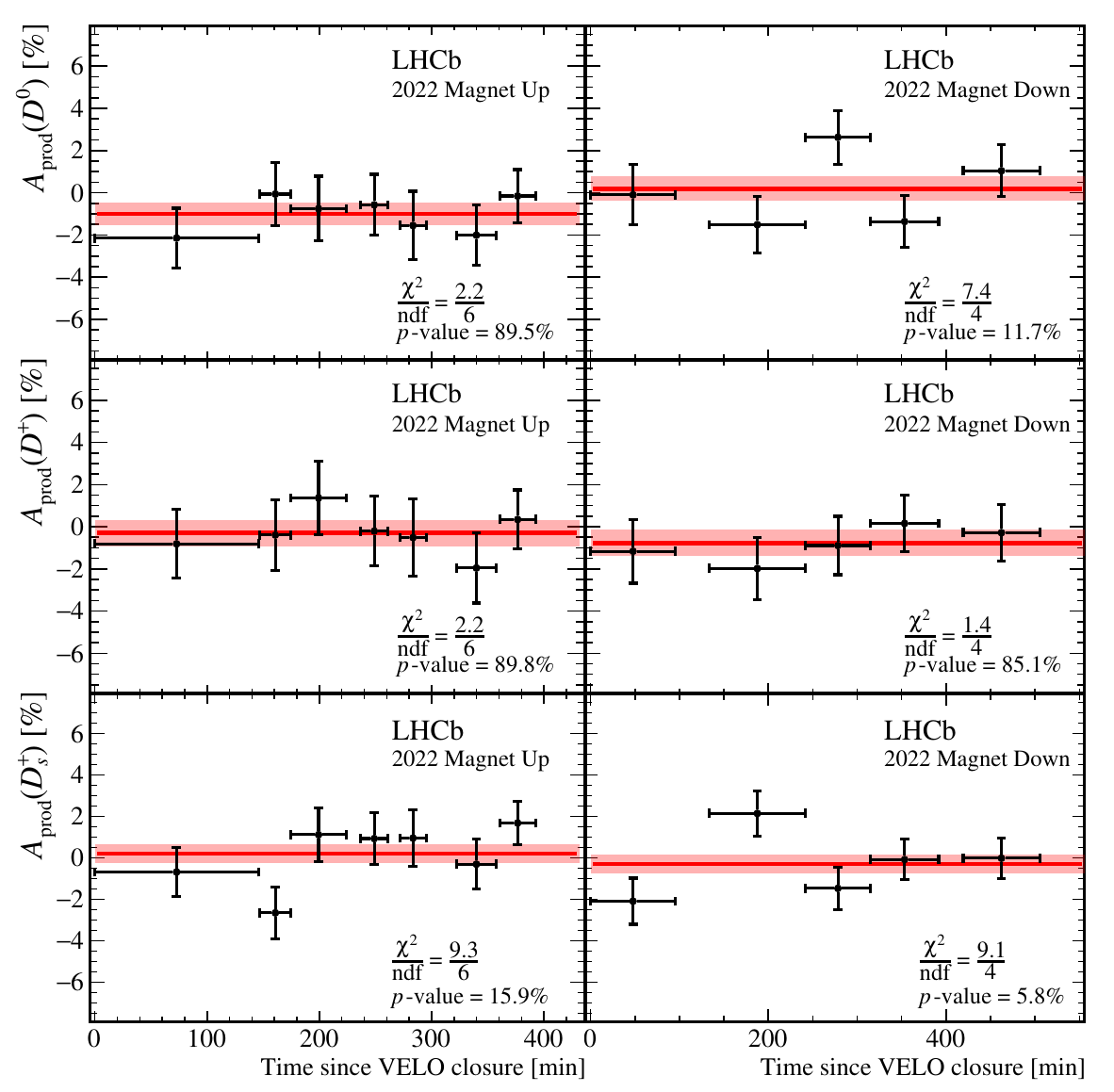}
    
    \caption{Production asymmetry measurements in bins of time and magnet polarity after the \velo closure for the (top) \Dz meson, (middle) \Dp meson and (bottom) \Dsp meson in the 2022 (left) \MagUp and (right) \MagDown datasets. The average value and its uncertainty are indicated by the red line and the red band, respectively. Gaps between the bins indicate that no data from these periods is utilised.}
    \label{fig:crosscheck_time_after_closing}
\end{figure}

Figures~\ref{fig:RunBlock_D0}, \ref{fig:RunBlock_Dp} and \ref{fig:RunBlock_Ds} show the comparison between the production asymmetries measured in different subsamples (RUNNUMBER blocks) of the three datasets used in the analysis.
On each plot the integrated results obtained through the baseline strategy are also shown for reference.
The compatibility $\chi^2$ shown on the plots refer to the different measurements performed in the RUNNUMBER blocks.
The results from different subsamples are compatible with each other within two standard deviations.
Figure~\ref{fig:crosscheck_time_after_closing} presents these measurements in blocks for 2022 data as a function of the time after closure of the \velo, across an entire fill of the LHC. 

\clearpage

%
%

\addcontentsline{toc}{section}{References}
\bibliographystyle{LHCb}
\ifx\mcitethebibliography\mciteundefinedmacro
\PackageError{LHCb.bst}{mciteplus.sty has not been loaded}
{This bibstyle requires the use of the mciteplus package.}\fi
\providecommand{\href}[2]{#2}

\newpage
%
%
%
\centerline
{\large\bf LHCb collaboration}
\begin
{flushleft}
\small
R.~Aaij$^{38}$\lhcborcid{0000-0003-0533-1952},
A.S.W.~Abdelmotteleb$^{57}$\lhcborcid{0000-0001-7905-0542},
C.~Abellan~Beteta$^{51}$\lhcborcid{0009-0009-0869-6798},
F.~Abudin{\'e}n$^{57}$\lhcborcid{0000-0002-6737-3528},
T.~Ackernley$^{61}$\lhcborcid{0000-0002-5951-3498},
A. A. ~Adefisoye$^{69}$\lhcborcid{0000-0003-2448-1550},
B.~Adeva$^{47}$\lhcborcid{0000-0001-9756-3712},
M.~Adinolfi$^{55}$\lhcborcid{0000-0002-1326-1264},
P.~Adlarson$^{83}$\lhcborcid{0000-0001-6280-3851},
C.~Agapopoulou$^{14}$\lhcborcid{0000-0002-2368-0147}
% ...

{\footnotesize \it

% ...
$^{14}$Universit{\'e} Paris-Saclay, CNRS/IN2P3, IJCLab, Orsay, France\\
% ...
$^{38}$Nikhef National Institute for Subatomic Physics, Amsterdam, Netherlands\\
% ...
$^{47}$Instituto Galego de F{\'\i}sica de Altas Enerx{\'\i}as (IGFAE), Universidade de Santiago de Compostela, Santiago de Compostela, Spain\\
% ...
$^{51}$Physik-Institut, Universit{\"a}t Z{\"u}rich, Z{\"u}rich, Switzerland\\
% ...
$^{55}$H.H. Wills Physics Laboratory, University of Bristol, Bristol, United Kingdom\\
% ...
$^{57}$Department of Physics, University of Warwick, Coventry, United Kingdom\\
% ...
$^{60}$School of Physics and Astronomy, University of Glasgow, Glasgow, United Kingdom\\
$^{61}$Oliver Lodge Laboratory, University of Liverpool, Liverpool, United Kingdom\\
% ...
$^{69}$Syracuse University, Syracuse, NY, United States\\
% ...
$^{83}$Department of Physics and Astronomy, Uppsala University, Uppsala, Sweden, associated to $^{60}$\\
% ...
}
\end{flushleft} 
 
\end{document}